\documentclass[desactivate]{aa}  

\usepackage{graphicx}
\usepackage{txfonts}
\usepackage{xcolor}
\usepackage[colorlinks=true,linkcolor=blue,citecolor=blue]{hyperref}
\usepackage{gensymb}
\usepackage{subfigure}

\newcommand{\silicate}{Mg$_2$SiO$_4$ }

\newcommand{\wasp}{WASP-43 b}

\newcommand{\orcidlink}[1]{\protect\href{https://orcid.org/#1}{\protect\includegraphics[width=8pt]{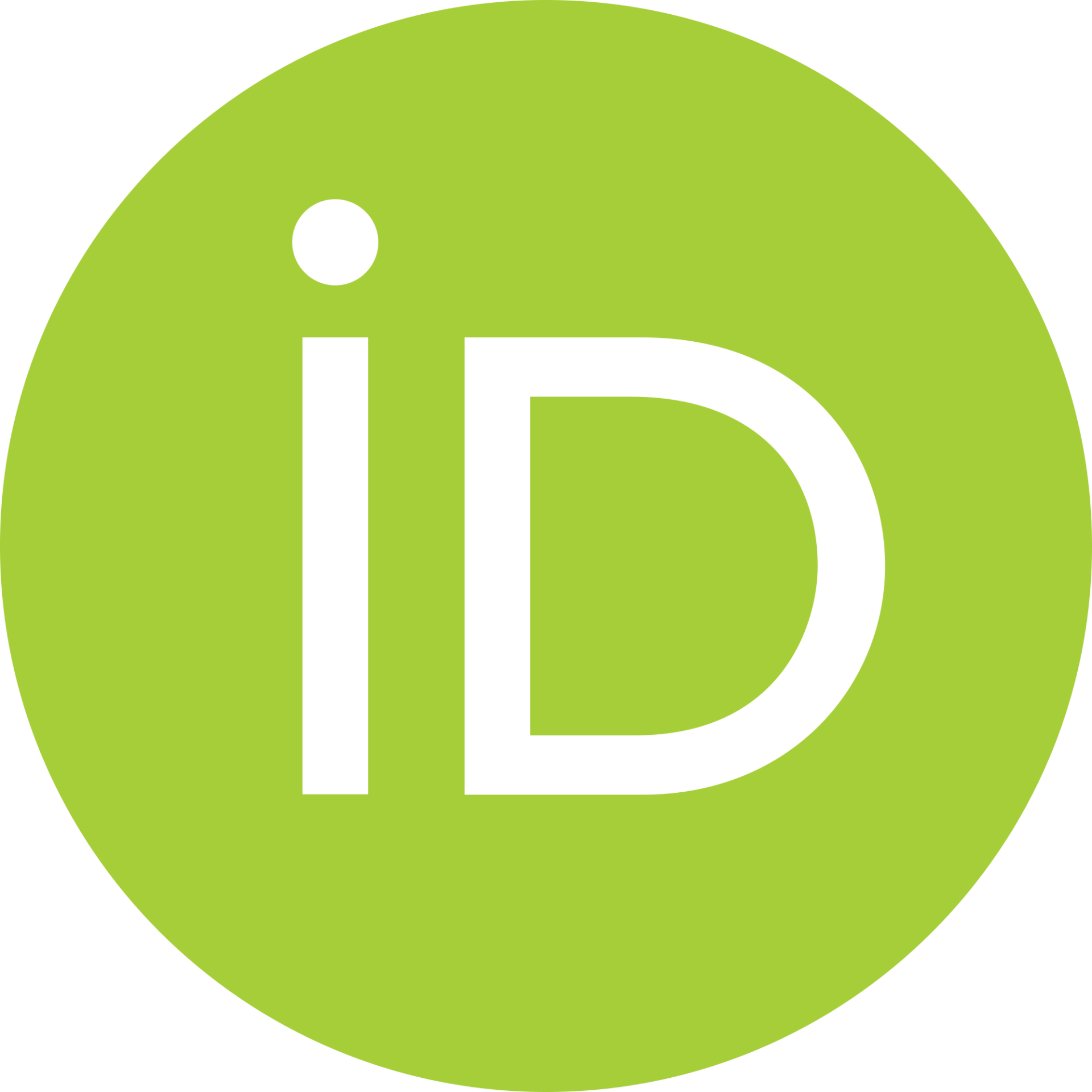}}}

\begin{document} 

   \title{The radiative and dynamical impact of clouds in the atmosphere of the hot Jupiter \wasp}


   \author{L. Teinturier
          \inst{1,2}\thanks{\email{lucas.teinturier@obspm.fr}}\orcidlink{0000-0002-0797-5746} \and B. Charnay \inst{1} \and A. Spiga \inst{2} \and B. Bézard \inst{1}\orcidlink{0000-0002-5433-5661}  \and J. Leconte\inst{3}\orcidlink{0000-0002-3555-480X} \and A. Mechineau \inst{3} \and E. Ducrot \inst{4}\orcidlink{0000-0002-7008-6888} \and E. Millour \inst{2}\orcidlink{0000-0003-4808-9203} \and N. Clément \inst{3} \orcidlink{0000-0001-9928-2981}
          }

   \institute{LESIA, Observatoire de Paris, Université PSL, Sorbonne Université, Université Paris Cité, CNRS, 5 place Jules Janssen, 92195 Meudon, France\\
              \email{lucas.teinturier@obspm.fr}
         \and
            Laboratoire de Météorologie Dynamique, IPSL, CNRS, Sorbonne Université, Ecole Normale Supérieure, Université PSL, Ecole Polytechnique, Institut Polytechnique de Paris, 75005 Paris, France 
        \and Laboratoire d’astrophysique de Bordeaux, Univ. Bordeaux, CNRS, B18N, allée Geoffroy Saint-Hilaire, 33615 Pessac, France
        \and AIM, CEA, CNRS, Université Paris-Saclay, Université Paris Cité, 91191 Gif-sur-Yvette, France
        }

   \date{Received XXX; accepted YYYY}

 
  \abstract
  {Hot Jupiters are tidally locked gaseous exoplanets that exhibit large day--night temperature contrasts. Their cooler nightsides are thought to host clouds, as has been suggested by numerous observations. However, the exact nature of these clouds, their spatial distribution, and their impact on atmospheric dynamics, thermal structure, and spectra is still unclear. }
  {We investigate the atmosphere of \wasp, a short period hot Jupiter recently observed with JWST, to understand the radiative and dynamical impact of clouds on the atmospheric circulation and thermal structure. We aim to understand the impact of different kinds of condensates potentially forming in~\wasp, with various sizes and atmospheric metallicities.}
  {We used a 3D global climate model (GCM) with a new temperature-dependent cloud model that includes  radiative feedbacks coupled with hydrodynamical integrations
  to study the atmospheric properties of \wasp. We produced observables from our GCM simulations and compared them to spectral phase curves from various observations to derive constraints on the atmospheric properties.}
  {We show that clouds have a net warming effect, meaning that the greenhouse effect caused by clouds is stronger than the albedo cooling effect. We show that the radiative effect of clouds has various impacts on the dynamical and thermal structure of \wasp. Depending on the type of condensates and their sizes, the radiative-dynamical feedback will modify the horizontal and vertical temperature gradient and reduce the wind speed. For super-solar metallicity atmospheres, fewer clouds form in the atmosphere, leading to a  weaker feedback. Comparisons with spectral phase curves observed with HST, \textit{Spitzer}, and JWST indicate that \wasp's nightside is cloudy and rule out sub-micron \silicate cloud particles as the main opacity source. Distinguishing between cloudy solar- and cloudy super-solar-metallicity atmospheres is not straightforward, and further observations of both reflected light and thermal emission are needed.}
  {}

\titlerunning{Radiative clouds on \wasp}
\authorrunning{Teinturier et al}
   \maketitle
%

\section{Introduction}

    Hot Jupiters are giant gaseous exoplanets in an extremely close-in orbit around their star. Due to their high mass -- between 0.5 and 13 Jupiter masses ($M_J$) -- and large radius -- 0.8 to 2 Jupiter radii ($R_J$) -- they are relatively easy to detect (more than 500 hot Jupiters have been detected to date). As hot Jupiters are expected to be in a synchronous rotation around their star with orbital periods of less than ten days, they receive an intense instellation on their permanent dayside, but not on their permanent nightside. This extreme contrast between the day and nightsides gives rise to vigorous atmospheric circulation, notably a super-rotating equatorial jet whose speed reaches a few km/s \citep{showman_equatorial_2011}. This powerful equatorial jet is associated with waves \citep{perez-becker_showman_2013,komacek_atmospheric_2016,pierrehumbert_hammond_2019} in hot Jupiters and transports heat from the irradiated dayside to the dark nightside, making the atmosphere an intrinsically 3D system \citep{showman_guillot_2002}. In the last few decades, an ever-growing sample of hot Jupiters suitable for atmospheric characterisation has emerged \citep{tsiaras_2018,edwards_2022}. Using both ground- and space-based telescopes, multiple constraints have been put on their chemistry, atmospheric circulation, and thermal structure. Atmospheric characterisation with transit spectroscopy, eclipse spectroscopy, and phase curves provides insight into the interplay between atmospheric dynamics, atmospheric chemistry, and thermal structure. Of these three, phase curves, and in particular spectral phase curves, yield information on the longitudinal structure of the atmosphere (see \cite{parmentier_exoplanet_2018} for a review and references within). Using the wavelength dependence of the phase curve and its shape yields constraints on the thermal structure and the chemical composition. To date, only a dozen exoplanets have available infrared phase curves. In the next few years, this number will significantly increase, thanks to the unique capabilities of the newly launched JWST. In the next few decades, the PLAnetary Transits and Oscillations of stars Space Telescope (PLATO) and \textit{Ariel} will also provide photometric and spectral phase curves at different wavelengths \citep{tinetti_2018,charnay_survey_2021,rauer_2014}. Infrared phase curves, such as those provided by the \textit{Hubble} Space Telescope (HST), \textit{Spitzer}, and JWST, constrain the redistribution of heat between the day and nightside of the planet, and thus constrain the atmospheric dynamics at play. At shorter wavelengths, optical phase curves probe the reflected light and allow us to place constraints on the cloud distribution and the composition of these clouds \citep{oreshenko_optical_2016,parmentier_transitions_2016}. Such observations have been conducted by the \textit{Kepler} Space Telescope \citep{heng_closed-formed_2021} and by CHEOPS \citep{deline_atmosphere_2022} and will be performed by PLATO once launched. However, discrepancies in reflected light measurement and thermal emission, in the form of geometric and Bond albedo estimates, have arisen and are difficult to reconcile \citep{schwatrz_2015}. Thus, simultaneous observations of reflected light and thermal emission are needed, as will be performed by \textit{Ariel} and its wide spectral coverage, from $0.55$ to $7.8$ $\mathrm{\mu}$m simultaneously. \\
    
    Clouds have been hypothesised to be in the atmosphere of hot Jupiters, though their behaviour is not yet totally understood \citep{stevenson_2016,sing_2016}. They are thought to have a major impact on the shape of transmission spectra \citep{sing_2016}, masking molecular features and hindering a precise determination of molecular abundances. The impact of clouds on phase curves is manifold. If present on the irradiated dayside, they would modify the redistribution of heat \citep{Pont_2013}. Homogeneous clouds would tend to raise the infrared photosphere, increasing the measured contrast and reducing the offset of the phase curve \citep{sudarsky_2003}. Patchy clouds would lead to subtler differences that could only be disentangled by 3D global climate simulations \citep{parmentier_transitions_2016}. \\ 
    
    In this paper we investigate the atmosphere of the hot Jupiter \wasp~\citep{helllier_2011,gillon_2012}. This planet is well known for multiple reasons. First, it is a non-inflated hot Jupiter orbiting a bright, nearby, and quiet K7 star. Second, it orbits its star in 19.5 hours, which is one of the shortest known orbits. This peculiar combination of star properties and the short orbital period makes \wasp~a prime target for phase curve observations with space-based telescopes. \cite{stevenson_thermal_2014} used HST to observe near-infrared (1.1-1.7 $\mu$m) spectroscopic phase curves. As expected, they found a large day--night temperature contrast and an eastward offset of the hotspot relative to the sub-stellar point. They also found that \wasp~poorly redistributes heat and that water is the only chemical species effectively shaping the phase-resolved emission spectra. \cite{stevenson_spitzer_2017} used the \textit{Spitzer} Space Telescope's IRAC instrument in the $3.6$ and $4.5$ $\mu$m channels to further investigate the heat redistribution mechanisms. They find that nightside optically thick clouds were needed to explain their data.

\cite{murphy_lack_2022} observed two additional phase curves of \wasp~with the $4.5$ $\mu$m channel of \textit{Spitzer} spaced in time by several weeks. Using also the data from \cite{stevenson_spitzer_2017}, and thus spanning timescale of years, they found no significant time variability in the atmosphere of \wasp. This indicates that weather is negligible in this particular planetary atmosphere within the precision and time coverage of the present data. Another finding of their study is that clouds are likely present on the nightside of the planet, while the dayside is relatively cloudless. Moreover, the Early Release Science programme of the newly launched JWST \citep{bean_2018} chose \wasp~as a target for a phase curve with the Mid-Infrared Low Resolution Spectrometer (MIRI-LRS) instrument \citep[$5$-$12$ $\mu$m;][]{venot_global_2020,ERSpaper}. Putting together the likely presence of clouds, the remarkable amount of observations, and the absence of detectable weather, \wasp~is the perfect target for studying the radiative and dynamical impact of clouds in the atmosphere of hot, gaseous, giant exoplanets. A few studies have investigated the role of magnetohydrodynamics in the atmospheric dynamics and phase curves for hot Jupiters \citep{rauscher_three-dimensional_2013,rogers_2014} and found that magnetic effects should slow the winds and change the location of the offset of the phase curves. However, as no constraints on the strength and geometry of the magnetic field of exoplanets are currently known, we neglect this effect in our study. \\

    Many modelling studies have attempted to understand the atmospheric circulation and cloudiness of \wasp~\citep{kataria_atmospheric_2015,mendonca_revisiting_2018,mendonca_three-dimensional_2018,carone_equatorial_2020,venot_global_2020,schneider_exploring_2022,murphy_lack_2022,deitrick_2022}, with different assumptions regarding the clouds, the atmospheric chemistry, the atmospheric metallicity, or the radiative transfer. \cite{kataria_atmospheric_2015} find that their cloud-free 5x solar metallicity models best match the observations but still overestimate the nightside flux. \cite{mendonca_revisiting_2018} simulated a cloudy nightside by imposing an extra opacity at these longitudes and reproduced the HST and \textit{Spitzer} data relatively well. They suggest that CO$_2$ might be enhanced on the nightside. \cite{mendonca_three-dimensional_2018} investigated the impact of disequilibrium chemistry of H$_2$O, CO, CO$_2$, and CH$_4$ along their prescribed thick nightside cloud deck. They find that the cloud deck strongly reduces the impact of disequilibrium chemistry. Moreover, they find that for a C/O ratio of 0.5, no differences between equilibrium and disequilibrium chemistry can be detected in the phase-resolved emission spectra. However, for a higher C/O ratio, departure from chemical equilibrium is non-negligible for methane, but the calculations do not match the observed flux. The authors also warn that their model uses double-grey radiative transfer instead of fully and self-consistent multi-wavelength radiative transfer. \cite{carone_equatorial_2020} and \cite{schneider_exploring_2022} used the same model, the expeRT/MITgcm, with double-grey or k-correlated radiative transfer. They simulated a deep, cloud-free atmosphere and show that the deep atmosphere may have a strong influence on the observable atmospheric dynamics of hot Jupiters. \cite{venot_global_2020} used the \cite{parmentier_transitions_2016} model to simulate cloud-free and cloudy atmospheres with different assumptions for the cloud modelling. They find that cloud-free models are unable to reproduce the nightside flux, while cloudy simulations with MnS or MgSiO$_3$ and cloud particles of 1 $\mu$m match the observations better. \cite{murphy_lack_2022} used the  \cite{roman_clouds_2021} model with double-grey radiative transfer and prescribed clouds with radiative feedbacks. Their models include 13 or 8 cloud species, whose formation depends on the local thermal conditions. Finally, \cite{deitrick_2022} used the newly updated version of \texttt{THOR} with multi-wavelength radiative transfer and  multiple scattering.  Their simulations are in agreement with the HST and \textit{Spitzer} observations, especially on the dark nightside. However, and as the authors mention, their cloud modelling is not an attempt to yield realistic conclusions about cloud physics and is uninformative with regard to cloud dynamics. Thus, modelling the coupling between cloud formation and atmospheric dynamics has not yet been performed with the aim of understanding the cloud distribution and feedbacks on the thermal and dynamical structure of \wasp.  \\ 
    
    In this work we adapt the \texttt{generic Planetary Climate Model (generic PCM)} to the case of hot Jupiter atmospheres with a new cloud model that includes temperature-dependent clouds, atmospheric advection with the flow, and radiative feedbacks. The aim of this study is to understand the dynamical and radiative impact of clouds on the atmospheric state and to understand the newly acquired MIRI-LRS phase curve \citep{ERSpaper}. In Section \ref{sec: model} we describe the \texttt{generic PCM}, the cloud model, and the initialisation of our models. In Section \ref{sec: dynamics} we explore the cloudless and cloudy atmospheric circulation of \wasp~and the impact of clouds in solar and 10x solar atmospheres. In Section \ref{sec: observations} we produce spectral phase curves from our models and compare them to observations. We also make predictions for \textit{Ariel} observations. Section \ref{sec: discussion} provides a summary of our results, and Section \ref{sec: conclusion} concludes this work.

\section{The Generic PCM}\label{sec: model}
 The \texttt{Generic PCM}, formerly known as the generic LMDZ global climate model (GCM), has been specifically developed for the study of exoplanets, giant planets, and paleoclimate studies \citep{Wordsworth_2011,charnay_2013,charnay_3d_2015-1,charnay_3d_2015,charnay_formation_2021,leconte_nature_2013,leconte_2013,spiga_global_2020,cabanes_2020,bardet_2022}. The model couples two main modules, the dynamical core (an (hydrodynamical solver) and the physical package (e.g. radiative transfer, cloud formation, sub-grid-scale processes). The LMDZ dynamical core uses a finite difference longitude-latitude grid to solve the primitive hydrostatic equations of meteorology \citep{hourdin_2020}. \cite{spiga_global_2020} made use of a new dynamical core, \texttt{DYNAMICO} \citep{dubos_dynamico-10_2015}, suited for the simulation of gas giants, and applied it to the case of Saturn. Their study was motivated by the small Rossby radius of Saturn, which needs to be resolved with high horizontal resolution. However, the Rossby radius of hot Jupiters is significantly larger than on Saturn. Thus, the traditional core provides high enough horizontal resolution for the study presented in this paper. 

\
In the \texttt{generic PCM}, the physical packages include multiple parameterisations of physical phenomena without assumptions made regarding the type of planets it simulates. We describe this package and the new features developed for this study in Section \ref{physics_package}. Section \ref{model_init} describes the model initialisation procedure.

\subsection{Physical packages}\label{physics_package}
\subsubsection{Existing parameterisations}

In the \texttt{generic PCM}, the physical package includes a radiative transfer scheme based on the k-correlated method and the two-stream equations, including collision-induced absorption of H$_{\rm 2}$-H$_{\rm 2}$ and H$_{\rm 2}$-He and Rayleigh scattering by H$_{\rm 2}$ and He. This scheme is based on the model described in \cite{Toon_1989}. The shortwave and longwave solvers are separated and independent in the model but they overlap in our study, as we study hot planets around relatively cool stars. We used 27 bins in the shortwave-stellar channel and 26 in the longwave-planetary channel, as listed in Table \ref{bins}. Radiative impacts of clouds and aerosols can be taken into account for specific species, and we extend this scheme to work with any kind of aerosol (see Sect. \ref{RGCS}).

The model also includes vertical turbulent mixing, and an adiabatic temperature adjustment, relaxing the temperature profile towards an adiabatic profile for atmospheric regions that are convectively unstable. This last scheme is almost never activated in our simulations, as we find no differences between a simulation with and without this adiabatic adjustment. 

\subsubsection{Inclusion of a scheme for cloud condensation} \label{GCS}

We tackled the issue of generalising the water cycle to any kind of species. The idea behind this development is to take into account the condensation and sublimation of tracers that will mimic the formation and evaporation of clouds. To achieve this goal, we made use of the Clausius-Clapeyron law to compute the saturation vapour pressure P$_{\rm sat}$ for each layer, grid cells, and species, $ i$: 
\begin{equation}
    P_{sat,i} = P_{ref,i}\exp{\Bigg[-\frac{\Delta_{vap}H_i}{R}\Big(\frac{1}{T}-\frac{1}{T_{ref,i}}\Big)-\alpha_i*Z\Bigg]}
,\end{equation}
where P$_{\rm ref,i}$ and T$_{\rm ref,i}$ are a reference pressure (at $Z = 0$) and temperature for species $i$, $\Delta_{\rm vap}$H$_i$ is the enthalpy of vaporisation of species $i$, $\rm \alpha_i$ is a species-dependent coefficient and $Z$ is the logarithm of the atmospheric metallicity. For species used in this study, the value of $\alpha_i$ is taken from either \cite{visscher_atmospheric_2010} or \cite{morley_neglected_2012}. For example, we used for the saturation vapour pressure of Mg and Mn  the equations from Table 2 of \cite{visscher_atmospheric_2010} and Eq. 9 of \cite{morley_neglected_2012}: 
\begin{subequations}\label{eq: psat reference}
    \begin{align}
        \log_{10}{(P_{\rm sat, Mg})} &= 8.25 - 27250/T -[\rm Fe/H] - \log_{10}{X'_{\rm H_2O}} \\ 
        \log_{10}{P_{\rm sat, Mn}} &= 11.532-23810/T - [\rm Fe/H].
    \end{align}
\end{subequations}\\
These equations are used in the model, transcribed with the parameters of Table \ref{tab:psat}. \\
\begin{table}
\caption{Numerical values used to compute the saturation vapour pressure of \silicate and MnS. }             
\label{tab:psat}      
\centering          
\begin{tabular}{|ccccc|  }     
\hline\hline       
Species & $\Delta_{\rm vap}$H [J/mol] & P$_{\rm ref}$  [bar] & T$_{\rm ref}$ [K] & $\alpha$ \\ 
\hline                    
   \silicate & 521700 & 1 & 2303 & 2.3025\\
   MnS & 455800 & 1 & 2064 & 2.3025\\
\hline                  
\end{tabular}
\end{table}
\indent If the saturation vapour pressure of a species is below the partial pressure of the gas in that layer, that layer is saturated and the gaseous species entirely condenses into a solid. Otherwise, an equilibrium is computed between solid and vapour phases, as 
\begin{subequations}
    \begin{align}
        q_{sat,i} &= \epsilon_i \frac{P_{sat,i}}{P-(1-\epsilon_i)P_{sat,i}} \\
        \epsilon_i &= \frac{m_i}{m},
    \end{align}
\end{subequations}
where $\rm q_{\rm sat,i}$ is the specific concentration  of species $i$ at saturation, $\rm m_{\rm i}$ the molecular weight of specie $i$ and $m$ is the mean molecular weight of the atmosphere without including species $i$.  \\
Once clouds are formed in the atmosphere, they can be transported by the dynamics and/or experience sedimentation at a terminal velocity. We did not take coagulation or coalescence into account, and we assumed the particles to be spherical. We did take the latent heat release from cloud condensation into account, but its effect is negligible on the thermal structure. The terminal velocity is computed using a Stokes law corrected for non-linearity \citep{Fuchs1965,ackerman_marley_2001}: 
\begin{equation}\label{eq: terminal_velocity}
    V_f = \frac{2 \beta a^2g(\rho_i-\rho)}{9\eta}
,\end{equation}
with $a$ the particle radius, treated as a free parameter, $g$ the gravity, $\rho_i$ the density of species $i$, $\rho$ the atmospheric density, $\eta$ the atmospheric viscosity, and $\beta$ the dimensionless Cunningham slip factor:
\begin{equation}
    \beta = 1 + K_n\big(1.256+0.4\exp{[-1.1/K_n]}\big)
,\end{equation}
with K$_{\rm n}$ the Knudsen number, given by
\begin{subequations}
    \begin{align}
        K_n &= \frac{\lambda}{a} \\
        \lambda & = \frac{k_B T}{\sqrt{2}\pi d^2P}.
    \end{align}
\end{subequations}
Viscosity is computed as a weighted expression of the viscosity of H$_2$, He and H$_2$O, following \cite{rosner_transport_1986}, \cite{petersen_properties_1970} and \cite{sengers_kamgarparsi_1984}:
\begin{equation}
    \eta = q_{H_2}\Big(2\times10^{-7}T^{0.66}\Big)+q_{He}\Big(1.9 \times 10^{-5}\big(\frac{T}{273.15}\big)^{0.7}\Big)+8 \times 10^{-6}q_{H_2O}
,\end{equation}
with q$_{\rm i}$ the specific concentration of species $i$ and the temperature $T$ in kelvins. Finally, correction for high Reynolds number flow is done based on \cite{clift_motion_1971,charnay_formation_2021}.

The specificity of our scheme compared to the existing literature is that clouds and condensable vapour can be advected horizontally and vertically, and clouds can be vertically sedimented. This transport scheme is based on the `Van-Leer I' finite volume scheme from \cite{hourdin_armengaud_1999}. Moreover, the scheme can handle an unlimited number of tracers at the same time, if computational facilities permit. Thus, cloud formation is directly driven by atmospheric dynamics and thermodynamics. In the case of tidally locked exoplanets with strong day-night temperature contrast, this scheme allows us to test the hypothesis of nightside cloud formation, without further ad hoc assumptions about the atmospheric state or the cloud distribution.

\subsubsection{Inclusion of the radiative effects of clouds}\label{RGCS}
We further extended our cloud condensation scheme by taking cloud radiative effects into account  in the two stream radiative transfer equation. This extension of the cloud condensation scheme allows for coherent radiative feedback of condensable species without assuming grey or double-grey cloud opacities, as done in \cite{komacek_patchy_2022}, \cite{mendonca_revisiting_2018}, and \cite{roman_clouds_2021}. The \texttt{generic PCM} already takes into account the radiative effects of clouds for a few specific aerosols (H$_{\rm 2}$O, CO$_{\rm 2}$, NH$_3$ and H$_{\rm 2}$SO$_{\rm 4}$). Here, we coupled the condensation scheme for any species as described in section \ref{GCS}, with radiative effects, in a consistent way. 

Radiative effects of clouds can be fully described by the extinction efficiency Q$_e$, single scattering albedo $\tilde \omega_0$ and scattering asymmetry parameter g$_0$ in the plane-parallel two-stream framework. For each cloud species, we computed these spectral parameters  offline as a function of wavelength and particle radius, using Mie theory (further assuming sphericity of particles). Then, during run-time, these parameters were re-computed at each time step and for each grid cell, in the atmosphere, based on the particle mean radius $\overline{r}$ (treated as a free parameter) and assuming a log-normal size distribution, following \cite{madeleine_revisiting_2011}:
\begin{equation}
n(r) = \frac{N}{\sqrt{2 \pi \sigma}\overline{r}}\exp{\Bigg[-\frac{1}{2}\Big(\ln{\big(r/\overline{r}\big)/\sigma^2}\Bigg]}
,\end{equation}
where $n(r)dr$ is the number of cloud particles per kg in the size range $[r,r+dr]$, $N$ the total number of particles per kg and $\sigma$ is the standard deviation of the distribution, fixed at $\sqrt{0.1}$. The assumption of a log-normal size distribution is simplistic, as particle sizes are likely dependent on pressure and locations \citep{parmentier_3d_2013,lee_dynamic_2016}. Thus, particle size distributions might be more complex \citep{lines_simulating_2018,powell_2018} but detailed modelling is out of the scope of this study. The optical depth of a cloud (index $i$) is computed in each layer as 
\begin{equation}
    d\tau_{i} = \frac{3Q_{e,i}}{4\rho_i \overline{r_i}}q_i \frac{dp}{g}
,\end{equation}
with q$_i$ the specific concentration of cloud $i$ in the concerned layer, $dp$ the pressure thickness of the layer and $g$ the gravitational acceleration. It is worth noting that due to our approximations of the Navier-Stokes equations, the gravitational acceleration is fixed at a constant value throughout the atmosphere. \\

\subsection{\silicate clouds and cold trap}\label{subsec: coldtrap}
In our simulations, the temperature profile and \silicate clouds condensation curves (from \citealt{visscher_atmospheric_2010}) shown in Fig.\ref{fig:TP_1D_condensation} lead to the appearance of a cold trap in the deep atmosphere. This cold trap should efficiently remove clouds in the deeper layers of the atmosphere. To confirm our hypothesis, we computed a mixing and sedimentation timescale in the first layer of the model (P = 800 bar), following \cite{charnay_self-consistent_2018}: 
\begin{subequations}
    \begin{align}
        \tau_{\rm mixing} &= \frac{H^2}{K_{\rm zz}} \\ 
        \tau_{\rm sed} &= \frac{H}{V_{\rm sed}},
    \end{align}
\end{subequations}
with H the atmospheric scale height, K$_{\rm zz}$ the eddy diffusion coefficient, computed using the parameterisation of \cite{parmentier_3d_2013}, and V$_{\rm sed}$ the terminal velocity of cloud particles. For particle radius of 1 $\mu$m. the mixing and sedimentation timescale are of the same order of magnitude, leading to a cold trap efficiently removing clouds over a few atmospheric scale heights. For bigger particles, this effect is stronger, but is less efficient for smaller particles. However, microphysical cloud models (see Fig. 3 in \cite{lee_modelling_2023}) show that where the cold trap starts, particles form with a radius of $\approx$1 mm, and only smaller particles are lifted higher up in the atmosphere. Thus, the cold trap should be efficient, removing most of the condensed \silicate particles where they first form. Moreover, this cold-trap depletion impacts the vertical distribution of vapour in the atmosphere, and thus, the amount of available condensable materials to form clouds in the simulations. 

To take this into account, we set the mixing ratio of available condensable vapour in the atmosphere equal to the value at the cold trap, which we take to be the deepest atmospheric layer (see Fig. \ref{fig:profil_clouds}). Since the condensation curve taken from \cite{visscher_atmospheric_2010} yields the condensation temperature assuming that no vapour has condensed yet, the curve is shifted according to Eq. \ref{eq: psat reference} and the abundance of Mg at the top of the clod trap, as depicted in Fig.\ref{fig:TP_1D_condensation}, to account for the fact that there is less vapour in the atmosphere (so a higher `total' pressure is needed at any given temperature for the partial pressure of the condensable vapour to reach the saturation pressure).

\subsection{Model Initialisation}\label{model_init}
\subsubsection{Radiative data computation}
To compute the k-correlated table used by the \texttt{generic PCM}, we started with the k-coefficients computed by \cite{blain_1d_2021}\footnote{available at \url{https://lesia.obspm.fr/exorem/ktables/default/}}, at a spectral resolution of R $\sim500$ at 1 $\mathrm{\mu}$m. We used the 1D radiative-convective code \texttt{Exo-REM} \citep{baudino_interpreting_2015, charnay_self-consistent_2018, blain_1d_2021} with these k-coefficients to simulate a cloud-free atmosphere of our planets. We assumed solar elemental abundances of H$_{\rm 2}$O, CO, CH$_{\rm 4}$, CO$_{\rm 2}$, FeH, HCN, H$_{\rm 2}$S, TiO, VO, Na, K, PH$_{\rm 3}$, and NH$_{\rm 3}$, and out-of-equilibrium chemistry, as we expect chemical quenching  to happen in the atmosphere of highly irradiated, tidally locked planets. Our \texttt{Exo-REM} simulations use 81 equally log-spaced vertical layers between 1000 bars and 0.1 Pa. Using this model, we obtain a temperature profile,
\begin{figure}[h]
    \centering
    {\includegraphics[width=0.5\textwidth]{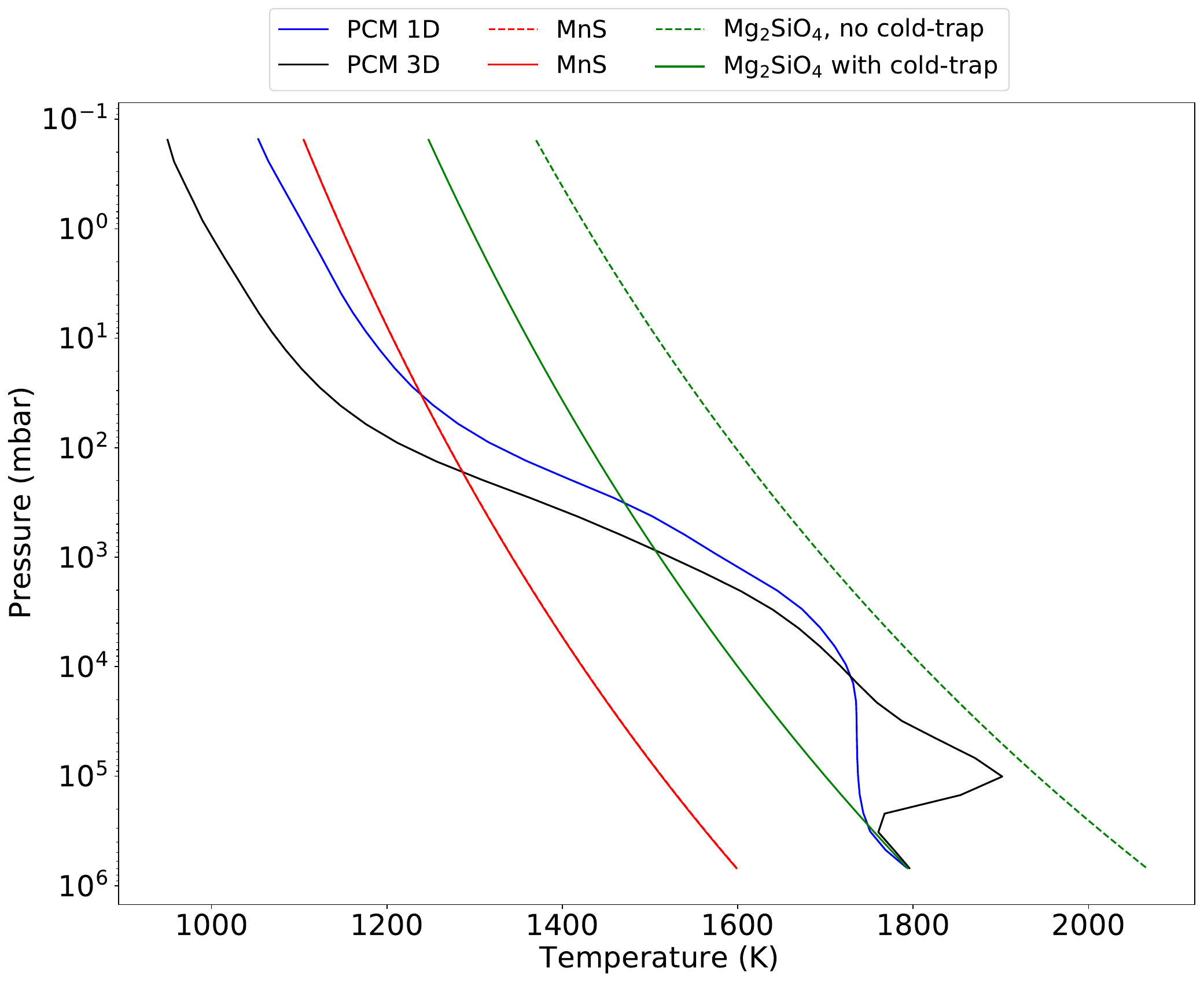}}
    \caption{Temperature profile computed with the 1D version of the \texttt{generic PCM} (blue), 3D globally averaged temperature profiles from the \texttt{generic PCM} (black), and condensation curves of MnS (red curve) and \silicate (green). The dashed lines are from \cite{visscher_atmospheric_2010} and assume that all the \silicate or MnS is in vapour form, and the solid lines are the values assuming that the bottom layer is at saturation after removing \silicate (or MnS) from the cold trap at 800 bars. Only \silicate is affected by the cold trap.}
    \label{fig:TP_1D_condensation}
\end{figure}volume mixing ratio profiles for each of the species mentioned above and a few atmospheric parameters, such as the mean molecular weight and the specific heat capacity of the atmosphere. Simulations are run using a constant eddy-mixing coefficient of $4.5 \times 10^7$ cm$^2$.s$^{-1}$. Vertical chemical profiles computed with \texttt{Exo-REM} are shown in Fig. \ref{fig:exorem_chem}. 

Using these profiles, we created mixed k-tables, depending on pressure and temperature using the \texttt{exo\_k} python package \citep{leconte_spectral_2021}. We used 16 Gauss-Legendre quadrature points. Through these `mixed' k-coefficients, non-equilibrium chemistry is included in the calculation of the radiative forcing, and thus in the radiative feedback on atmospheric dynamics. However, we did not take into account possible horizontal change in chemical composition, especially between the hot dayside and the cold nightside. 
\subsubsection{1D run and 3D shell initialisation}
Using the \texttt{Exo-REM} temperature profile as a starting point, we iterated the 1D version of the \texttt{generic PCM} (i.e. a single-column version of the physical packages with no coupling to the dynamical core, described in Section \ref{physics_package}) until reaching radiative balance. Our diagnostic for reaching radiative balance is for the ratio of the absorbed stellar radiation (ASR) to the outgoing longwave radiation (OLR) to be within 1\%. The 1D temperature profile derived from this procedure is shown in Fig.\ref{fig:TP_1D_condensation}. \\
Finally, we initialised the 3D spherical shell with the temperature profile from the 1D run at radiative equilibrium. Thus, our initial 3D state is horizontally uniform in temperature, and we started the simulation from a rest state (no winds). For all simulations, we used an horizontal grid resolution of 64$\times$48 (longitude $\times$ latitude) and 40 vertical levels, equally log-spaced between 800 bars and 10 Pa. We adjusted the hydrodynamical timestep to be 28.12 seconds for all simulations (2500 steps per \wasp~year), and the radiative and physical timestep to be five times this value (140.6 seconds). We used a dissipation timescale for numerical hyper-diffusion of 2000 seconds, and our simulations use a sponge layer over the topmost 4 atmospheric layers, aiming at reducing spurious wave reflections at the model top.

For all our simulations (\texttt{Exo-REM}, 1D and 3D \texttt{PCM}) the stellar spectrum is taken from the \texttt{BT-NextGen} grid \citep{allard_2012}, assuming a solar metallicity and the respective stellar temperature. As \wasp~is a non-inflated hot Jupiter, we set its interior temperature at 100 K (see Table \ref{num_setup}). 

When clouds are added to the simulations, we initialised them using analytical profiles derived by \cite{visscher_atmospheric_2010} or \cite{morley_neglected_2012}, depending on the species. We started by assuming that only condensable vapour is available in the atmosphere and no condensates have formed. In these profiles, the bottom (deepest) layer is at saturation. For the case of \silicate clouds, this is coherent with a cold trap that would efficiently remove clouds from the deep atmosphere (see Section \ref{subsec: coldtrap}). We used a planetary mean temperature profile from cloudless simulation to compute these vapour profiles. Thus, the analytical profiles give us a maximum amount of clouds that can form in the atmosphere, in a physically motivated manner (see Fig. \ref{fig:profil_clouds}). These profiles are horizontally uniform when initialising the spherical shell.

\section{An investigation of \wasp's atmospheric properties}\label{sec: dynamics}

We integrated the model from the state described in the previous section. Hot Jupiters are known to exhibit a very long convergence time, due to the very long radiative timescale in the deepest levels of the atmosphere \citep{wang_extremely_2020}. To assess if the model has been integrated for long enough, we made use of the super-rotation index \citep{Read_1986,mendonca_angular_2020}, the ratio of the total axial angular momentum of the atmosphere to the total axial angular momentum of the planet assuming a null zonal wind. We considered the model out of the spin-up regime when the variation of the index was less than a few percent over the last 500 days. This number is a mass-weighted metric, meaning that the deepest layers have a strong influence on the variation of the metric as we integrated the model. Thus, we computed this metric below 10 bar, as clouds will mostly affect the upper atmosphere. We also checked the root-mean square of the wind speed as a function of time and pressure to make sure our simulations reached a steady-state. As another diagnostic for reaching an equilibrium state, we also checked the ASR to OLR  ratio, as an indicator of radiative balance above the photosphere. We find that we achieve radiative equilibrium to better than 1.5\% much faster than dynamical equilibrium, as expected \citep{wang_extremely_2020}. \\

\begin{table}
\caption{Opacity bins of the short- and longwave channels (in microns).}             
\label{bins}      
\centering                          
\begin{tabular}{c c |c c}        
\hline\hline                 
\multicolumn{2}{c}{shortwave} & \multicolumn{2}{c}{longwave} \\
\hline
\multicolumn{2}{c}{Wavelength ($\mu$m)} & \multicolumn{2}{c}{Wavelength ($\mu$m)} \\

\hline       \hline                 
   0.261 & 0.400 & 0.612   & 0.678 \\      
   0.400 & 0.495 & 0.678   & 0.745 \\
   0.495 & 0.572 & 0.745   & 0.785 \\
   0.572 & 0.612 & 0.785   & 0.860 \\
   0.612 & 0.675 & 0.860   & 0.910 \\ 
   0.675 & 0.745 & 0.910   & 0.960 \\ 
   0.745 & 0.785 & 0.960   & 1.00 \\ 
   0.785 & 0.860 & 1.00    & 1.10 \\ 
   0.860 & 0.910 & 1.10    & 1.20 \\ 
   0.910 & 0.960 & 1.20    & 1.33 \\ 
   0.960 & 1.00  & 1.33    & 1.50 \\ 
   1.00 & 1.10   & 1.50    & 1.60 \\ 
   1.10 & 1.20   & 1.60    & 1.77 \\ 
   1.20 & 1.33   & 1.77    & 2.02 \\ 
   1.33 & 1.50   & 2.02    & 2.17 \\ 
   1.50 & 1.60   & 2.17    & 2.50 \\ 
   1.60 & 1.77   & 2.50    & 2.99 \\
   1.77 & 2.02   & 2.99    & 3.29 \\ 
   2.02 & 2.17   & 3.29    & 3.80 \\ 
   2.17 & 2.50   & 3.80    & 4.40 \\ 
   2.50 & 2.99   & 4.40    & 5.22 \\ 
   2.99 & 3.29   & 5.22    & 6.45 \\ 
   3.29 & 3.80   & 6.45    & 10.40 \\ 
   3.80 & 4.40   & 10.40   & 20.00 \\ 
   4.40 & 5.22   & 20.00   & 46.00 \\ 
   5.22 & 6.45   & 46.00   & 324.68 \\ 
   6.45 & 10.40  &         &  \\ 
\hline                                   
\end{tabular}
\end{table}
\begin{table}
\caption{Numerical set-up and planetary properties.}             
\label{num_setup}      
\centering          
\begin{tabular}{c c  }     
\hline\hline       
Parameters & \wasp \\ 
\hline                    
   C$_{\rm p}$ (J.K$^{-1}$.kg$^{-1}$) & 12904      \\
   $\mathrm{\mu}$ (g.mol$^{-1}$) & 2.326      \\
   R$_{\rm p}$ (R$_{\rm J}$)     & 1.036    \\
   M$_{\rm p}$  (M$_{\rm J}$)    & 2.052      \\ 
   a (AU)                        & 0.01526    \\
   $\Omega$ (days)               & 0.81347753\\
   g   (m.s$^{-2}$)              & 47.1      \\ 
   T$_{\rm int}$ (K)             & 100      \\
   R$^*$ (R$_{\odot}$)           & 0.667 \\
   T$^*$ (K)                     & 4520  \\
   Integration time (Earth days) & 6508 \\
   Hydrodynamical timestep (s) &28.12\\ 
   Radiative/Physical timestep (s) & 140.6 \\
   Dissipation timescale (s)     &2000 \\
\hline                  
\end{tabular}
\end{table}

\subsection{ Dive into the cloudless dynamics}\label{sec: cloudless_solar}
\begin{figure*}[h]
    \centering
    \subfigure[Mean zonal wind]{\includegraphics[width=0.49\textwidth]{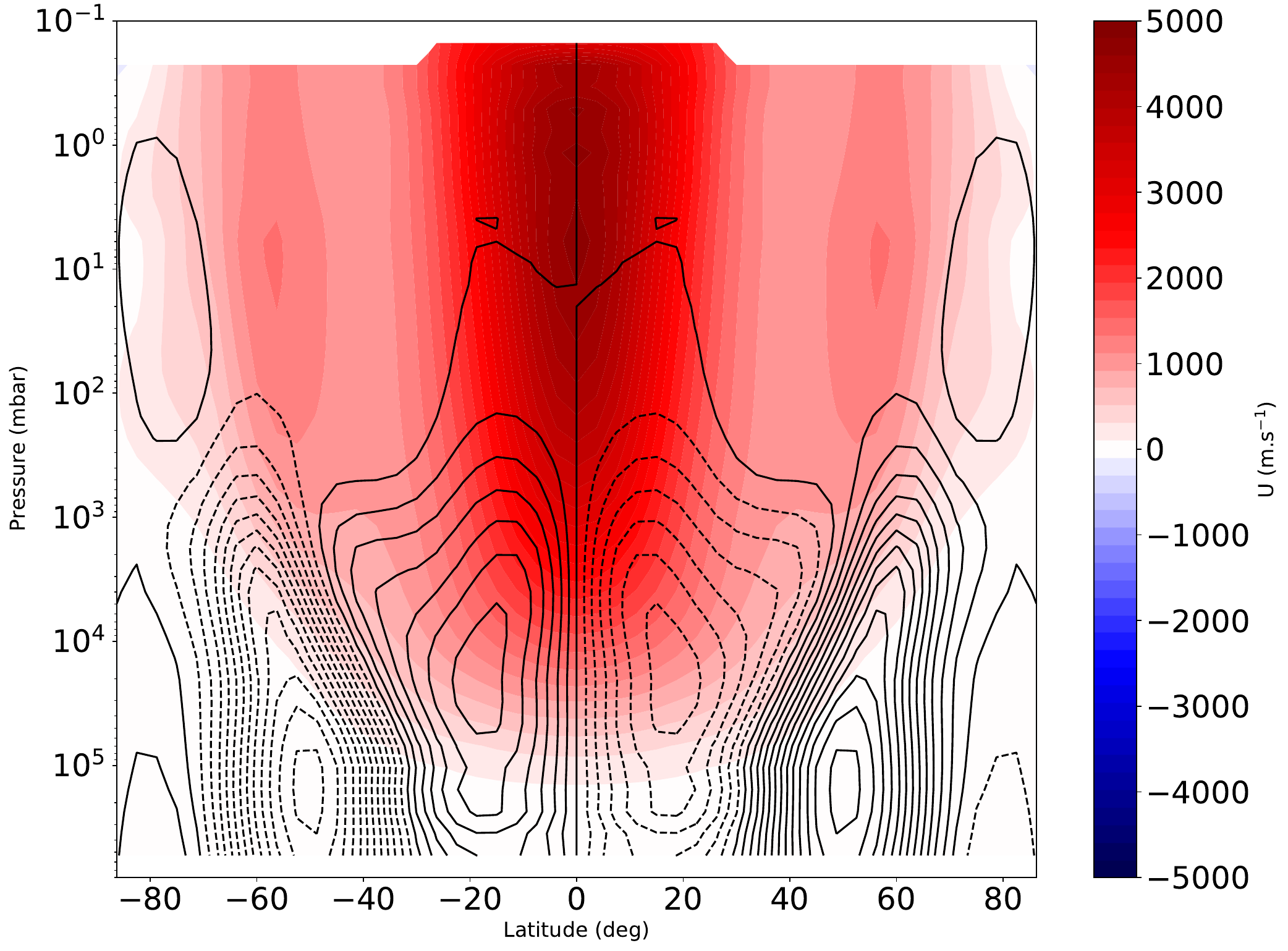}\label{subfig:zmean}}
    \subfigure[Meridional mean of temperature]{\includegraphics[width=0.49\textwidth]{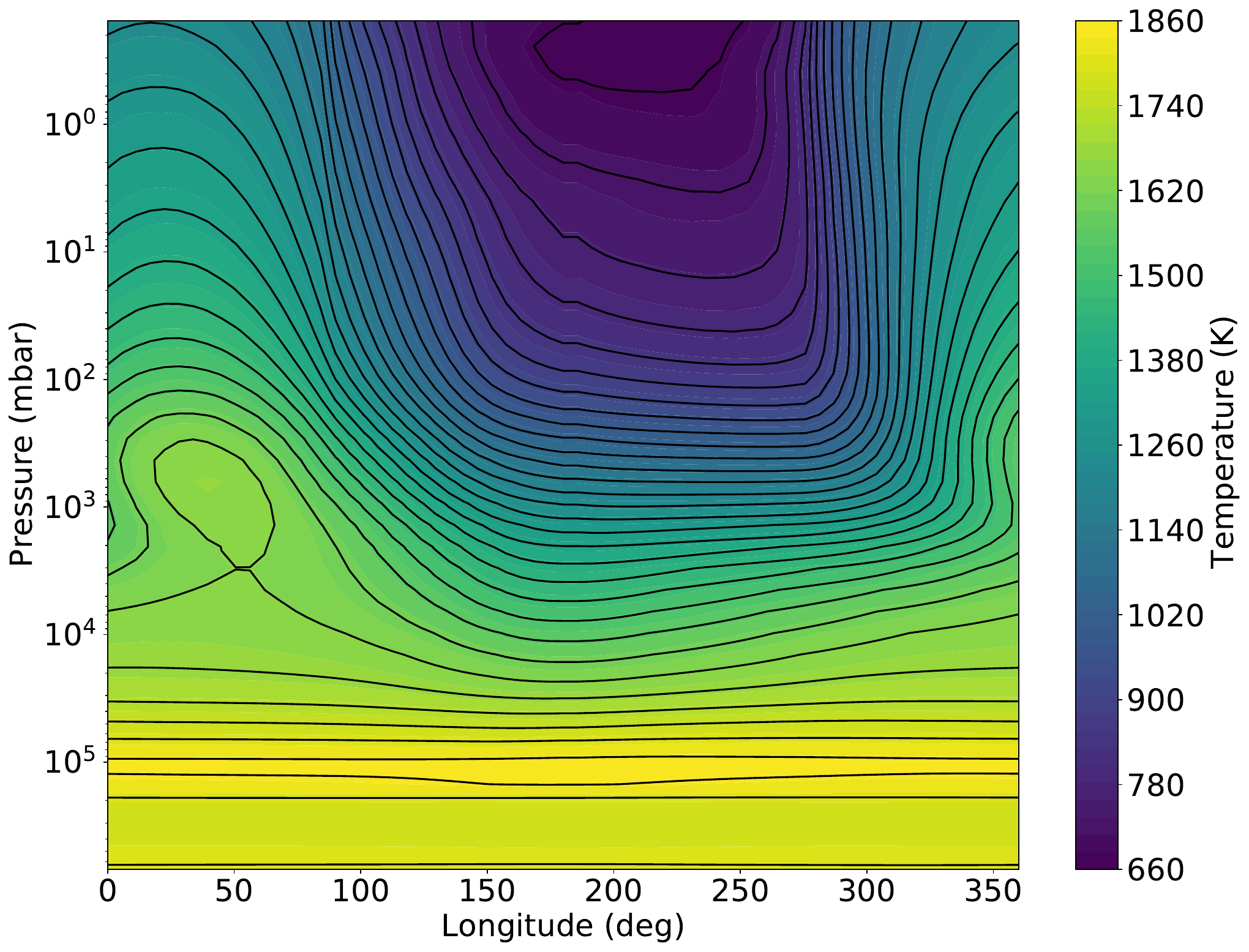}\label{subfig: Tmean_lat}}
    \subfigure[Temperature at 10 mbar]{\includegraphics[width=0.49\textwidth]{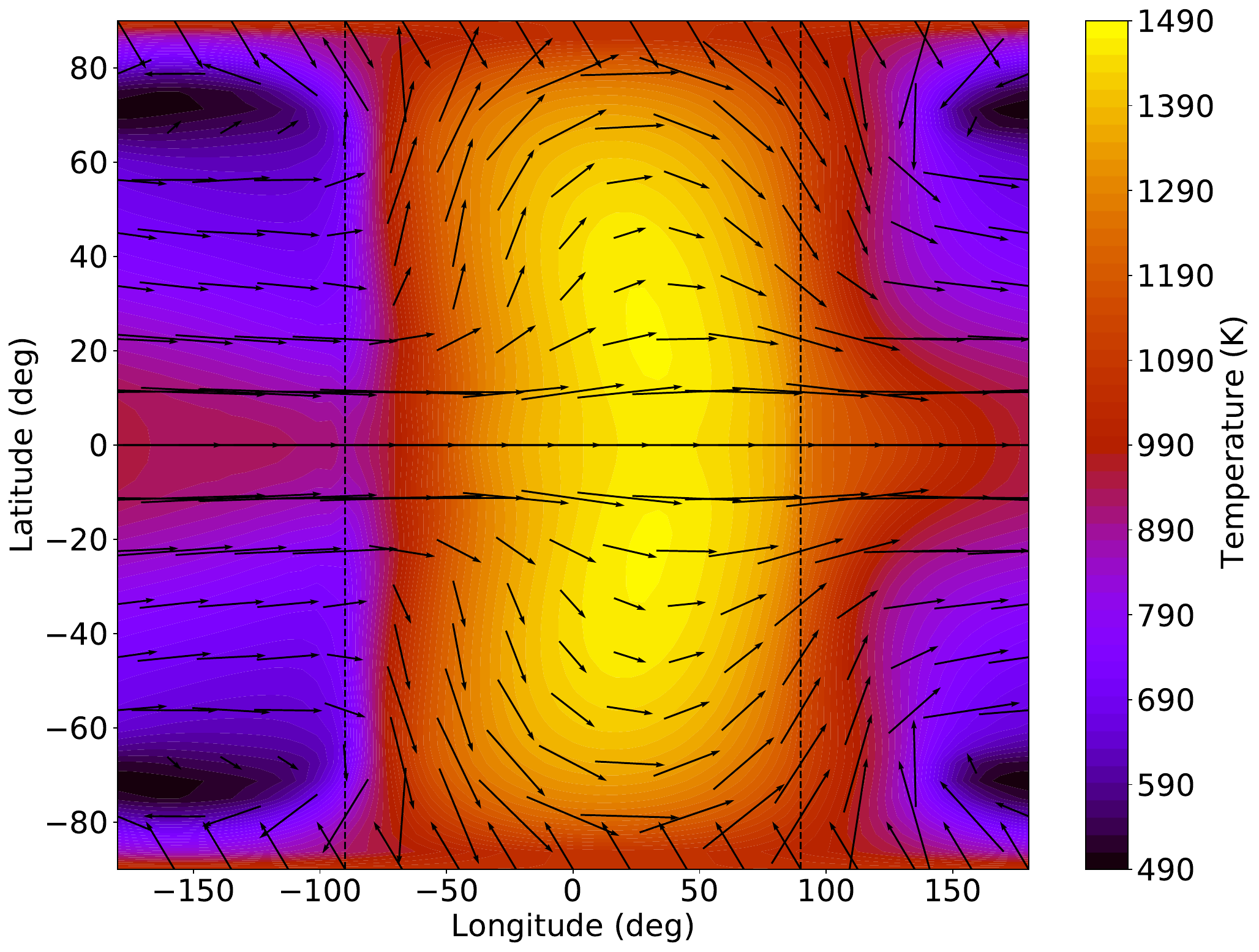}\label{subfig:T10mbar}}
    \subfigure[Outgoing Longwave Radiation]{\includegraphics[width=0.49\textwidth]{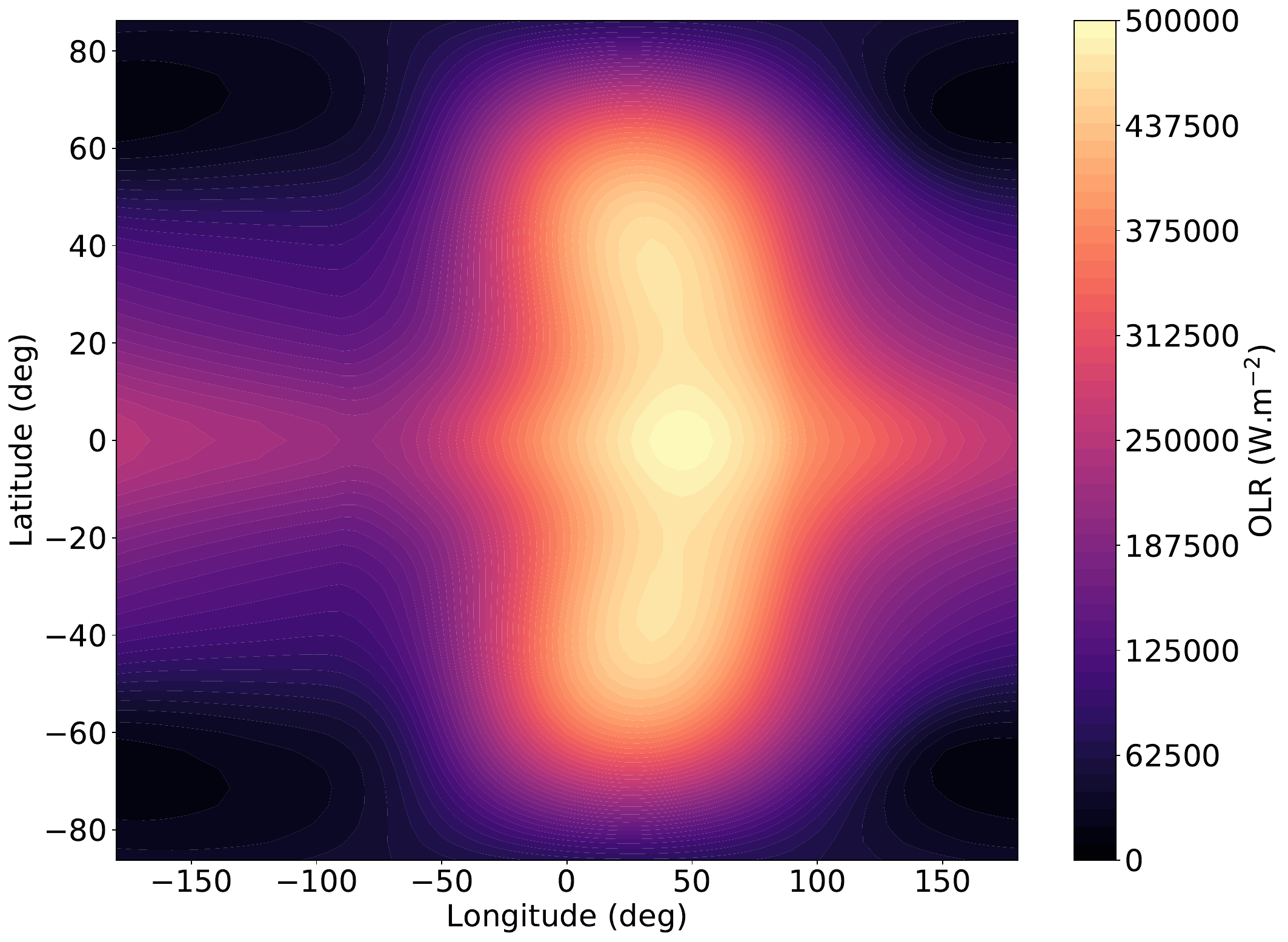}\label{subfig: olr}}
    \caption{Broad dynamical features and outgoing thermal flux for the cloudless simulation of WASP-43 b. (a) Zonal mean and time-averaged zonal wind. The stream functions are plotted in black, with dashed lines representing an anti-clockwise circulation and solid lines a clockwise circulation. (b) Temperature map averaged in time and in latitude. Solid lines are contours corresponding to the colour bar, highlighting the eastward shift of the hotspot. (c) Longitude-latitude temperature map at 10 mbar, averaged in time. Black arrows show the time-averaged wind direction, with the size of the arrow proportional to the magnitude of the wind. Vertical dashed black lines are the two terminators, delimiting the day- and the nightside of the planet. (d) OLR map, averaged in time.}
    \label{fig:cloudless_w43}
\end{figure*}
We integrated our cloudless model for 8000 WASP-43 b years ($\sim$6508 Earth days). All of our results are averaged over the last hundred WASP-43 b years. Our simulation displays the classical broad equatorial super-rotating jet found in other hot Jupiters simulations (see \cite{showman_atmospheric_2020} for a review), as shown in Figure \ref{subfig:zmean}. We also show the large day-night temperature contrast (Fig. \ref{subfig:T10mbar}) and the shift of the hotspot of the atmosphere with regard to the sub-stellar point (Fig. \ref{subfig: Tmean_lat}, \ref{subfig: olr}). We summarise briefly these findings. The strong equatorial jet displays a wind speed as high as 4.7 km.s$^{-1}$, with a latitudinal extent covering $\pm 40^{\degree}$. Because of this strong jet, the hottest regions of the atmosphere are advected to the east of the sub-stellar point by $\approx 30^{\degree}$. We observe the formation of two large-scale cold vortices between 160 and 240$^{\degree}$ east longitude, at $\pm 70 ^{\degree}$  north latitude. In these vortices, the wind speed is very low. Thus, cold air parcels are trapped in these regions and experience radiative cooling. This leads to a cool nightside, whereas the dayside is warm, with air parcels in the equatorial region being advected across one hemisphere on a timescale of $\approx 14$ hours. At 10 mbar, the temperature contrast between the day and nightside reaches $\approx 900$ K and $\approx 1 000$ K at 1 bar. The vertical structure of the atmosphere displays strong coherency between the top of the atmosphere (10 Pa) and 200 mbar with the hotspot location slightly varying in this pressure range. Between 200 mbar and 5 bar, the ratio of the radiative timescale compared to the advective timescale increases. Thus, hot air parcels are further advected to the east before radiative cooling happens, leading to a stronger offset of the hotspot location. \\
\begin{figure*}[h]
    \centering
    {\includegraphics[width=\textwidth]{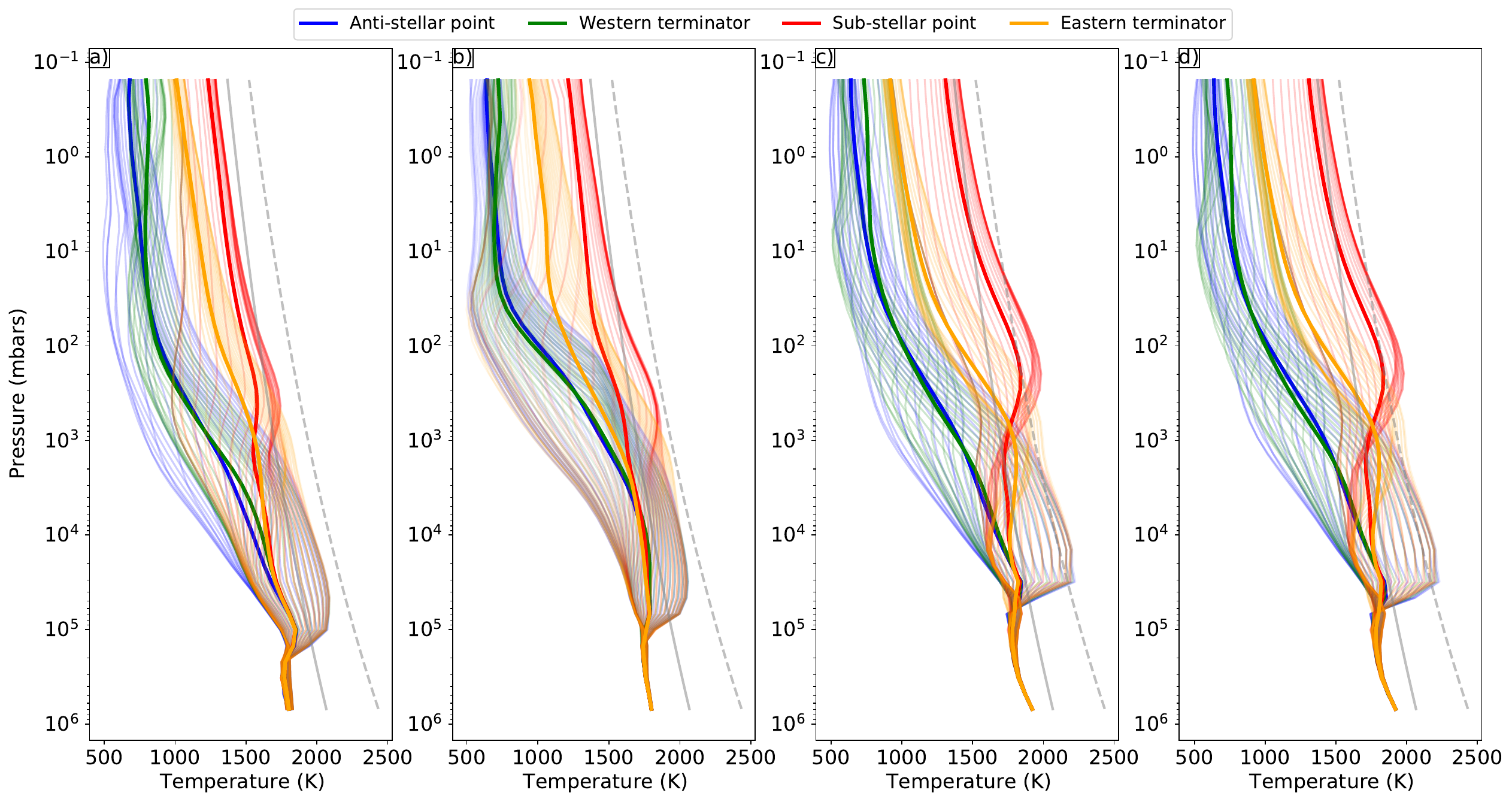}}
    \caption{Modelled temperature profiles at different longitudes, averaged in latitude (thick lines). The light lines represent the temperature profiles at each latitude (spaced every $3.75^{\degree}$). \emph{(a)} Cloudless simulation at solar metallicity. \emph{(b)} With \silicate clouds of 1 $\mu$m at solar metallicity. (\emph{c) }Cloudless at 10x solar metallicity. (\emph{d)} With \silicate clouds of 1 $\mu$m at 10x solar metallicity. The solid grey line represents the condensation curves of \silicate at solar metallicity, and the dashed grey line is the condensation curve of \silicate at 10x solar metallicity.} 
    \label{fig:equatorial_TP}
\end{figure*}
Figure \ref{fig:equatorial_TP} displays temperature profiles at different longitudes of interest, for our cloudless and \silicate clouds simulations at solar metallicity (see Section \ref{sec: dynamic_clouds}), and for atmosphere with 10x solar metallicity (see Section \ref{sec: supersolar }). The thermal contrast between the two sides of the planet is evident in the upper atmosphere, with a warm dayside and east terminator and a cooler nightside and western terminator. Interestingly, the western terminator is cooler than the nightside at the equator. This is understandable by looking at Fig.\ref{subfig:T10mbar}. The equatorial region of the nightside stays warm by means of the advection of heat by the equatorial jet and by adiabatic heating produced by the anti-Hadley circulation \citep{charnay_3d_2015}, but higher latitudes experience a strong cooling. However, the equatorial region of the western terminator is cooler and its meridional temperature gradient is weaker. Thus, the meridionally integrated temperature is cooler on the nightside than in the western terminator. Deeper than 200 bar, the atmosphere starts to become horizontally isothermal. In these regions, the radiative timescale increases again and becomes longer than the advective timescale. Despite the long integration of the model, the thermal structure is still strongly influenced by the horizontally homogeneous initial state and internal heat flux. 

We also observe the formation of two atmospheric cells with an anti-Hadley cell at low latitudes and a clockwise (Hadley-like) cell at mid latitudes, in both hemispheres extending deep into the atmosphere. A change of the circulation between the day and the nightside of the planet is seen, with wide anti-Hadley cells on the nightside from equator to pole and equator-to-mid-latitude anti-Hadley cells followed by Hadley cells on the dayside. The thermal emission of \wasp~closely follows the shift of the hotspot, with a peak of emission located around $45^{\degree}$ at the equator. Most of the emission comes from the dayside of the planets with an eastward shift, and from the equatorial region. Indeed, mid and high latitudes only marginally contribute to the overall emission, especially on the cooler nightside. \\

To investigate the drivers of the atmospheric dynamics, we computed diagnostics on the angular momentum and potential temperature meridional and vertical transport. For two prognostic variables of the simulation, $X$ and $Y$, the decomposition of the total transport of $Y$ by $X$ is given by
\begin{equation} \label{decomposition}
    \Big[\overline{XY}\Big] = \Big[\overline{X}\Big]\Big[\overline{Y}\Big]+\Big[\overline{X^*}   \times \overline{Y^*}\Big]+\Big[\overline{X'Y'}\Big]
,\end{equation}
where 
\begin{subequations}
     \begin{align}
         [X] &=\frac{1}{2 \pi}\int_0^{2 \pi}Xd\lambda \\
         \overline{X}&= \frac{1}{t_2-t_1}\int_{t1}^{t2}Xdt \\
         X^* &= X- [X] \\
         X' & = X-\overline{X}.
     \end{align}
\end{subequations}
In Eq. \ref{decomposition}, the terms on the right respectively represent the contribution from the mean circulation, the stationary waves, and the transient (time-dependent) perturbations (or eddies). For the stationary wave contribution, we added a $\times$ symbol to clearly show that we computed the product of the time-averaged zonal deviations and not the time-averaged of the product of zonal deviations. 
\begin{figure*}[h]
    \centering
    {\includegraphics[width=\textwidth]{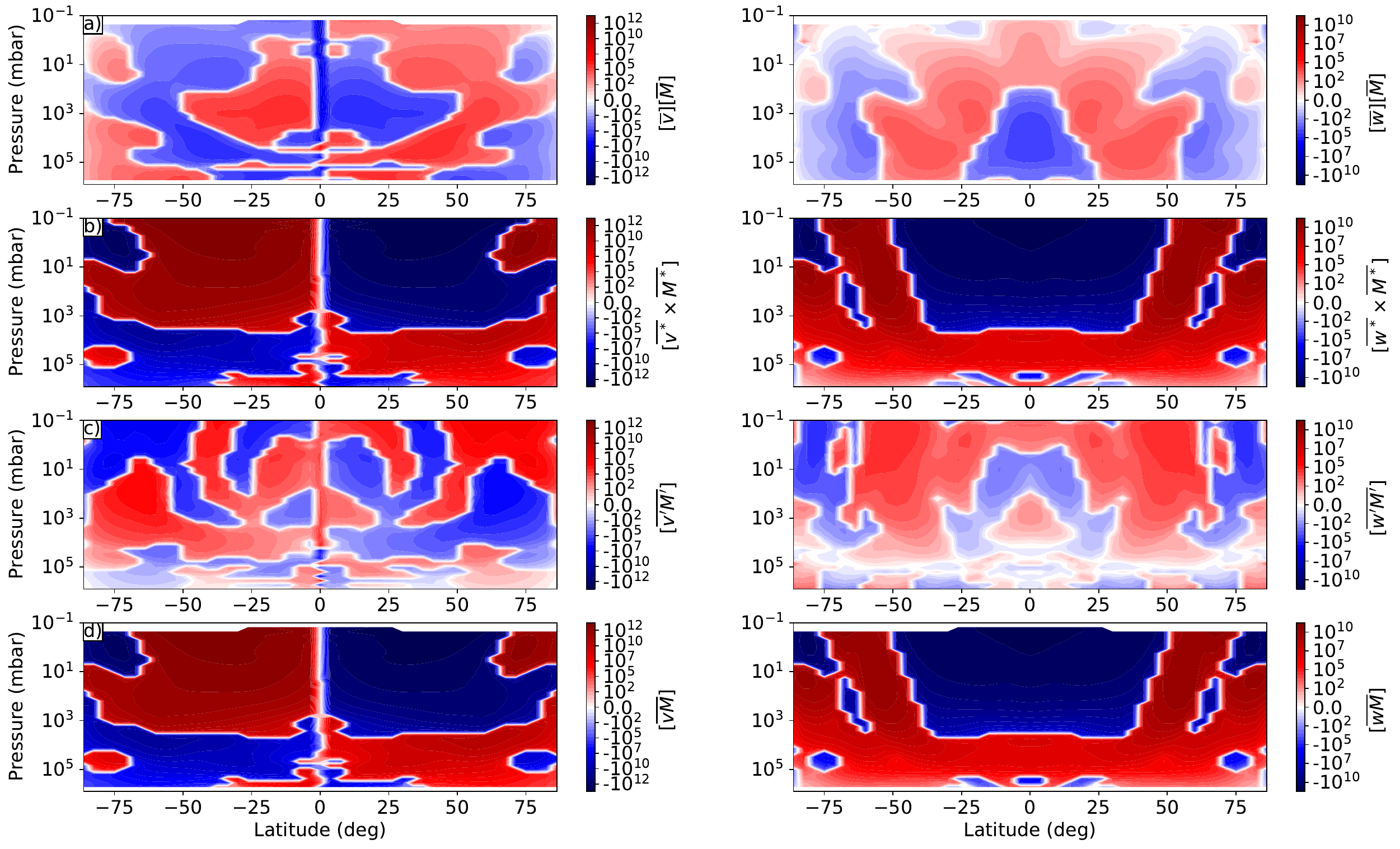}}
    \caption{Axial angular momentum transport for the cloudless simulation of WASP-43b. \emph{Left:} meridional transport of angular momentum. \emph{Right:} vertical transport of angular momentum. Each row displays a different contribution: mean circulation contribution(a), stationary waves contribution (b), transient waves contribution (c), and net transport (d). The units of each plot is $10^{25}$ kg.m$^3$.s$^{-2}$. Positive meridional wind is oriented northward and positive vertical wind is upward.} 
    \label{fig:w43_noclouds_ang_transport}
\end{figure*}

Figure \ref{fig:w43_noclouds_ang_transport} displays the contribution of each term of Eq. \ref{decomposition} to the horizontal (meridional) transport of axial angular momentum (left panels) and to the vertical transport of axial angular momentum (right panels). In agreement with the results of \cite{mendonca_angular_2020}, the horizontal transport in our simulation is dominated by the stationary wave contribution, while transient waves and mean circulation have an equal contribution to the angular momentum redistribution, which is compliant with the findings of \cite{mayne_results_2017}. Stationary waves are expected on hot Jupiters due to the tidally locked rotation and the strong instellation received by the planet \citep{showman_equatorial_2011}. The mean circulation and stationary wave contributions are compliant with the mass stream functions displayed in Fig.\ref{subfig:zmean}, with angular momentum being advected equatorward for pressure lower than a few bars and being advected poleward at deeper pressures. Thus, the equatorial jet is fed by the meridional transport, which strengthens and maintains its speed. The net horizontal transport (last row) confirms this trend, with a maximum of equatorward transport of angular momentum between 1 and $10^3$ mbar, at $\pm 30^{\degree}$ latitude. 

In a similar way, vertical transport of axial angular momentum is dominated by the stationary wave contribution, with a strong downward flux of angular momentum reaching pressure as deep as a few bars between $\pm 40^{\degree}$ latitude. Towards the poles, the transient wave contribution is added to that of the stationary waves to induce an upward transport of angular momentum. It is important to note that the meridional transport is in average stronger than the vertical transport of angular momentum by two order of magnitudes. \cite{carone_equatorial_2020} simulated the atmosphere of WASP-43 b with an extended atmosphere and found retrograde equatorial flows on the dayside, linked to deep wind jets. They link the retrograde flow to a strong upward vertical angular momentum transport at depth $\approx$ 100 bars. However, our simulation do not show this peak in upward vertical angular momentum at $\sim$ 100 bars, nor the retrograde equatorial flow around 10 mbar (see our Fig.\ref{subfig:T10mbar} and their Fig. 2 IIb)). Thus, the axial angular momentum transport in the atmosphere of WASP-43 b is dominated by the contribution of stationary waves. Angular momentum is advected  from the poles to the equator, and downwards for pressure lower than a few bars. In the deep atmosphere, this behaviour reverses, with upward and poleward transport of angular momentum. 
\begin{figure*}[h]
    \centering
    {\includegraphics[width=\textwidth]{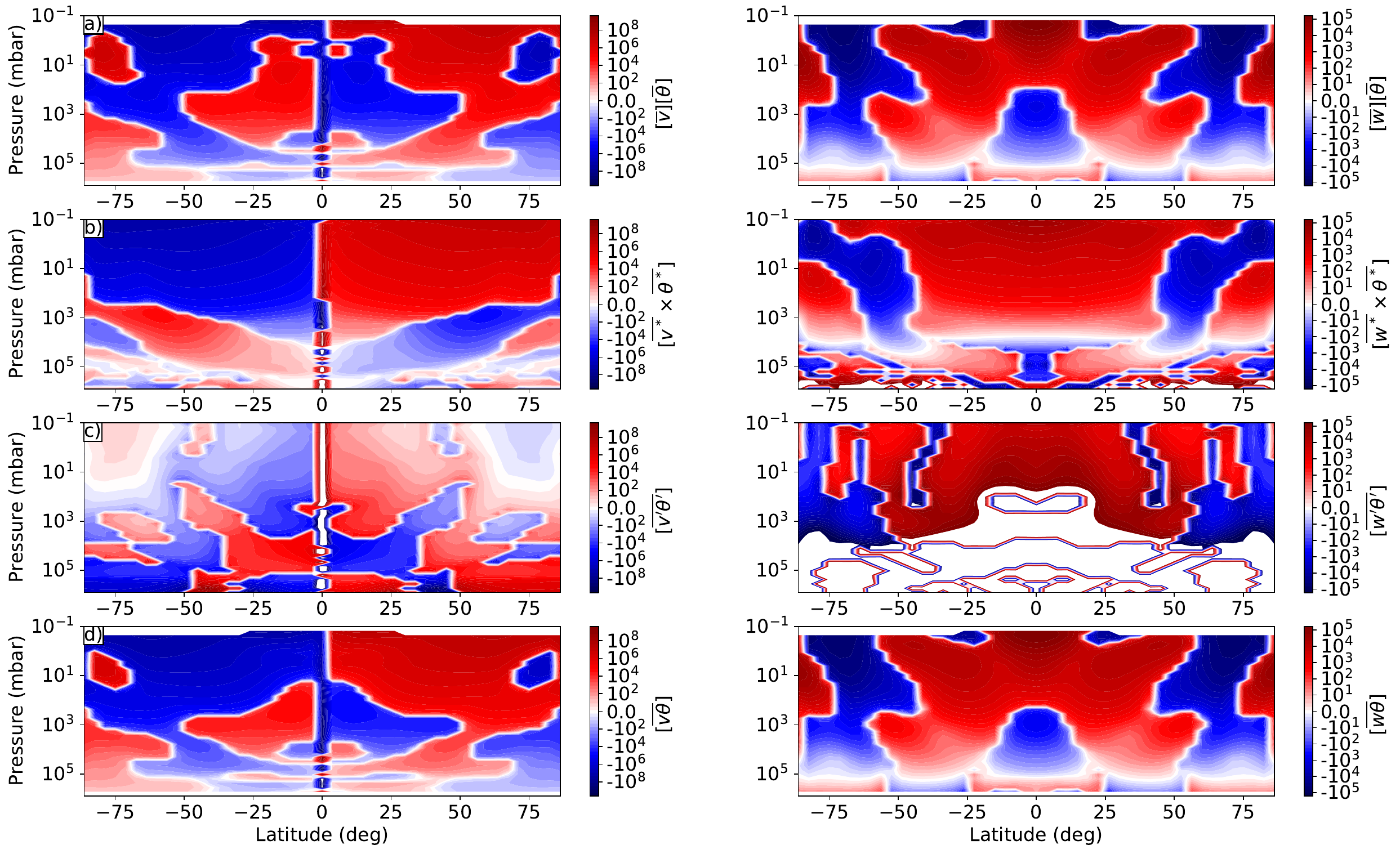}}
    \caption{Potential temperature transport (associated to heat transport) for the cloudless simulation of WASP-43b. \emph{Left:} meridional transport of heat. \emph{Right:} vertical transport of heat. Each row displays a different contribution: mean circulation contribution (a), stationary waves contribution (b), transient waves contribution (c), and net transport (d). The units of each plot is K.m.s$^{-1}$. Positive meridional wind is oriented northward and positive vertical wind is upward.}
    \label{fig:w43_noclouds_heat_transport}
\end{figure*}

We then performed the same computation on the heat transport, by using the potential temperature as a proxy variable of heat. As shown in Fig. \ref{fig:w43_noclouds_heat_transport} left panels, the horizontal transport of heat is dominated by the mean circulation and the stationary waves, similarly to angular momentum transport. As opposed to angular momentum, heat is transported from the equator to the poles by the stationary waves at pressure lower than $\sim$ 1 bar. The mean circulation pattern is somewhat similar to that for angular momentum, with a region of equatorward convergence of heat between $\pm 25^{\degree}$ at pressure levels of $1-10^4$ mbar. Poleward, heat is advected away from the equator. The deep atmosphere transport is dominated by the mean circulation contribution, with alternating equatorward and poleward advection of low intensity as we go to the bottom of the model and its horizontally constant heat flux boundary condition. Overall, for pressure lower than $\sim 1$ bar, the net latitudinal transport of heat is from equator-to-pole, with the exception of an equatorward region of transport between 10 mbars and 1 bars at low latitudes. 

The right panels of Fig. \ref{fig:w43_noclouds_heat_transport} show the different contributions and the total vertical transport of heat. In the same fashion as horizontal heat transport, the vertical transport is dominated by the mean circulation contribution. From equator to mid-latitudes, heat is transported upwards in the upper atmosphere. Deeper than 1 bar, a strong downward transport of heat in the equatorial region is flanked by an upward motion, extending to the top of the atmosphere. Above $\pm 50^{\degree}$, potential temperature is advected from the top to the bottom of the atmosphere. \\

To summarise, we find that meridional heat transport is dominated by the stationary waves and the mean circulation whereas the vertical transport is dominated by the mean circulation. In the upper atmosphere, potential temperature is mixed from equator to poles where the deeper atmosphere behaviour is alternating between poleward and equatorial transport. Vertically, heat is mostly pumped from the deep atmosphere to the upper atmosphere at low to mid-latitudes, with the exception of the equatorial region. At high latitudes, potential temperature is advected from the top of the model to the deep atmosphere. As it is the case for the axial angular momentum transport, meridional transport dominates vertical transport by a few orders of magnitude. Additionally, our simulation ran for around 18 Earth years, which is not enough to reach a hot adiabatic deep atmosphere ($\sim$ 1000 yr), as found by \cite{sainsbury-martinez_idealised_2019}. However, a slight emergence of deep adiabatic warming can be seen for pressure greater than 100 bars when comparing the 3D globally averaged temperature profile to the 1D temperature profile from the \texttt{generic PCM} (Fig. \ref{fig:TP_1D_condensation}). The additional computational cost of multi-wavelength radiative transfer compared to the idealised Newtonian cooling scheme used in their study precludes us from reaching the deep adiabatic steady state.

\subsection{Cloudy simulations} \label{sec: dynamic_clouds}
We ran simulations using the cloud condensation scheme with radiative feedbacks described in Section \ref{RGCS}, for condensates made of \silicate and MnS and cloud particle sizes of 0.1, 0.5, 1, 3, 5 ans 10 $\mathrm{\mu}$m. The choice of MnS and \silicate clouds is driven by the findings of \cite{venot_global_2020}, who state that the nightside of \wasp~could be dominated by MnS, Na$_2$S, MgSiO$_3$ (enstatite) and \silicate (forsterite). They also state that enstatite and forsterite will have similar effect on the nightside spectra. Thus, we only model \silicate clouds and use them as a proxy for all clouds composed of silicate. 

\begin{figure*}[h]
    \centering
    {\includegraphics[width=\textwidth]{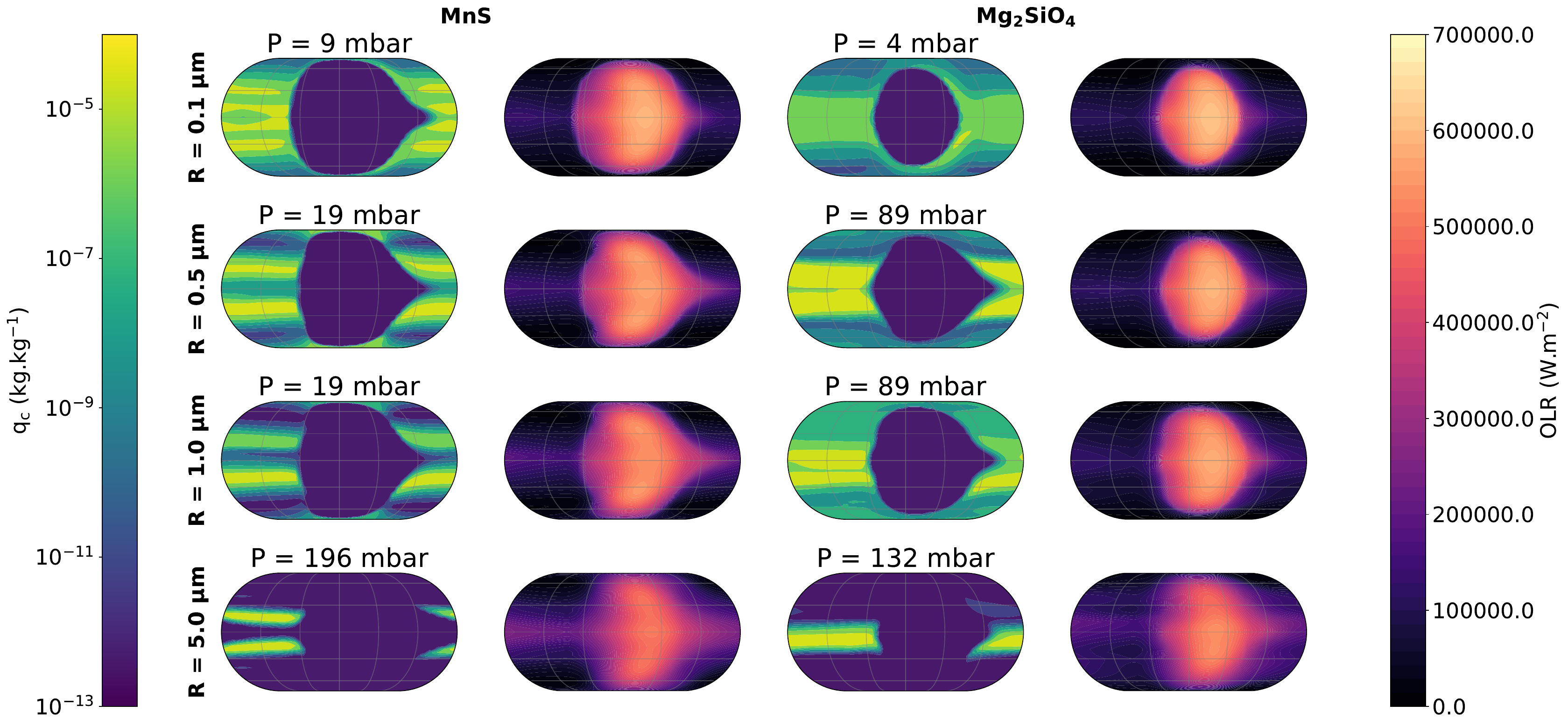}}
    \caption{Isobaric map of the maximum value of the cloud mass density in the atmosphere and OLR for each simulation of WASP-43 b, centred on the dayside. On the left, clouds are composed of MnS, and of \silicate on the right. Each row displays a simulation with a different cloud particle size, from $0.1$ $\mathrm{\mu}m$ to $5$ $\mathrm{\mu}$m. In all our simulations, clouds preferentially form on the nightside and at the western terminator, leaving a cloudless dayside from which most of the thermal flux is emitted.}
    \label{fig:grid_clouds_w43}
\end{figure*}

\
These simulations start from a rest state and are integrated for $2000$ WASP-43 b days. Figure \ref{fig:grid_clouds_w43} displays isobaric maps of the maximum value of the cloud mass density in the atmosphere, for each of these simulations. Regardless of the cloud condensate, the particle size is crucial to determine the depth at which clouds form and settle. Broadly, the bigger the particles, the deeper the cloud deck will settle. In all the simulations, clouds spontaneously form on the cooler nightside of the planet and at the western terminator. The extent of the horizontal cloud coverage west of the sub-stellar point depends on the particle size and the condensate forming, via the thermodynamics and radiative properties of each species. A cloudless dayside for latitudes within $\pm 80^{\degree}$ is also a natural outcome of the simulations. 

The broad pattern of the OLR is similar when including clouds into the simulation. The peak of outgoing flux is located eastwards of the sub-stellar point, with a major contribution from the equatorial region. This pattern closely follows the one of the temperature spatial distribution, as shown in Fig. \ref{fig:w43_clouds_1um}. Indeed, depending on the type of clouds, the peak of emission will change in shape, with the oval-like distribution decreasing in size but increasing in intensity in the case of \silicate clouds with regard to MnS clouds. For a given condensate, the spatial cloud distribution, and accordingly the outgoing radiation,  depends on the particle size (see Section \ref{sec: diff_sizes})  \\

\subsubsection{Impact of the different condensates }
\begin{figure*}[h]
    \centering
    {\includegraphics[width=\textwidth]{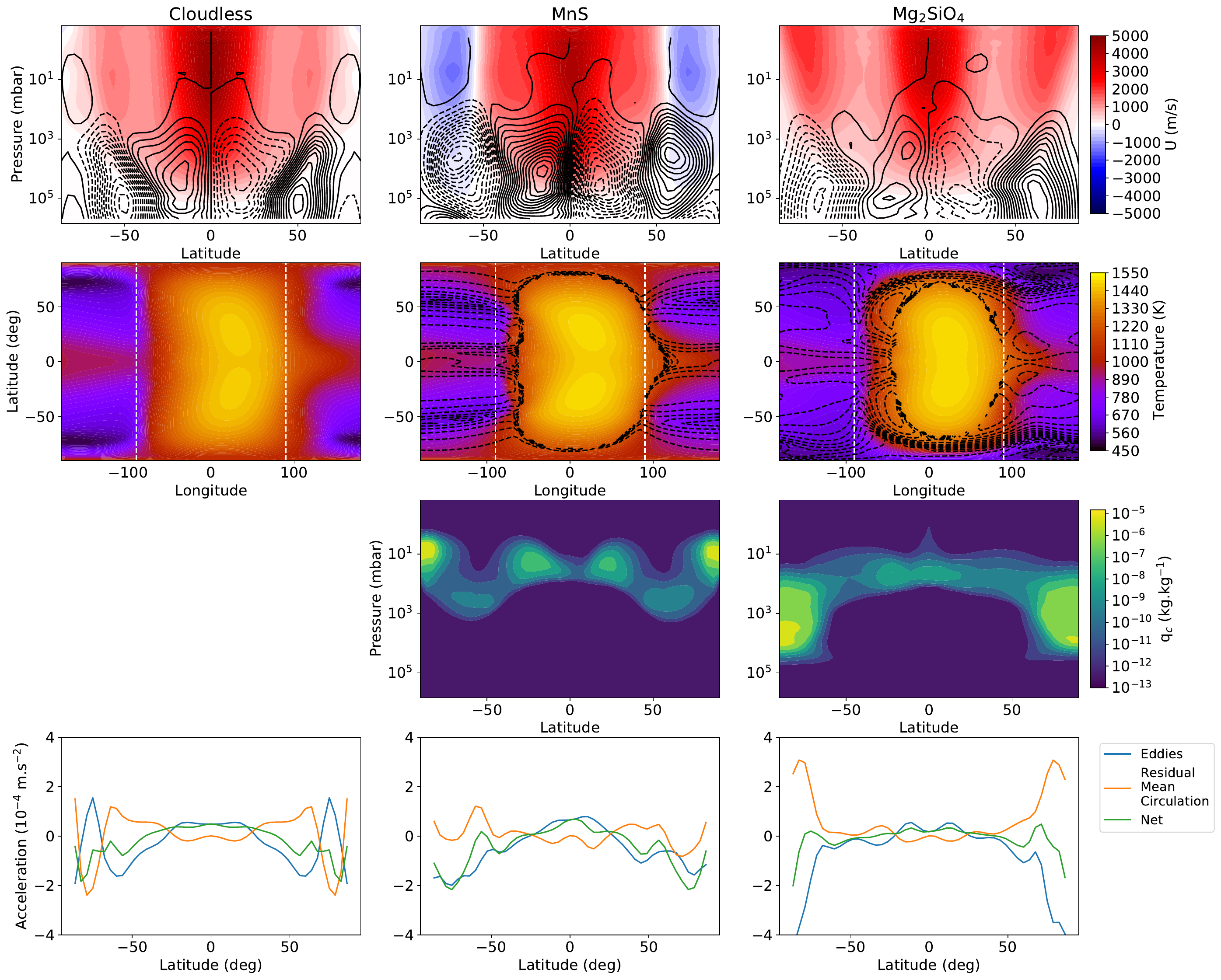}}
    \caption{Comparison of wind patterns, temperature maps at 10 mbar, cloud latitudinal distributions, and acceleration of the jet between a cloudless scenario and 1 $\mu$m size clouds of MnS and Mg$_2$SiO$_4$. Top row: Zonal mean and time-averaged zonal wind. Black contours are the mass stream function, with solid lines indicating a clockwise circulation and dashed lines an anti-clockwise circulation. Second row: Time-averaged temperature map at the 10 mbar pressure level. Black contours denote the location of clouds. White vertical dashed lines are the terminators. Third row: Latitude-pressure map of the cloud distribution, averaged in time and longitude. Bottom row: Vertically integrated jet zonal wind acceleration in zonal and time mean between 5 and 20 mbar. We show the decomposition into residual mean circulation and eddy contributions and the net acceleration. Left panels show the cloudless simulation, middle panels the MnS simulation, and right panels the Mg$_2$SiO$_4$ simulation. The left panels of the first two rows are duplicated from Fig. \ref{fig:cloudless_w43}.}
    \label{fig:w43_clouds_1um}
\end{figure*}

 The thermal structure is directly affected by the radiative feedbacks of clouds. In all cases, a strong temperature contrast between the day and nightside is ubiquitous. However, the shape of the two sides of the thermal structure is widely affected by the clouds. In the MnS simulation, clouds form on the western terminator until $\sim -60^{\degree}$ and the eastern terminator is be cloud-free at 10 mbar. Both terminator regions warm by $\approx$50 K due to the cloud-induced greenhouse effect. For the \silicate simulation, the clouds extend deeper into the dayside at both terminators, with terminators cooling by $\approx$100 K compared to the cloudless case (see Fig.\ref{fig:equatorial_TP}). In contrast, the cloudless part of the dayside is hotter even though smaller. This is easily understandable from the cloud distribution plotted in the third row of Fig.\ref{fig:w43_clouds_1um}. The silicate clouds settle mostly deeper than 10 mbar whereas MnS cloud formation is important at the poles and $\pm 25^{\degree}$ at this pressure. Thus, the cloud deck induces a greenhouse effect below this level, heating the deeper atmosphere in the MnS case, whereas this effect is much less important on the dayside for the silicate simulation. With respect to the cloudless simulation, the clouds also modify the location and shape of the large-scale cold vortices in the nightside. In the MnS simulation, these cold vortices are brought closer to the equator ($\pm 50^{\degree}$) where clouds are not present at 10 mbar (see Fig.\ref{fig:w43_clouds_1um}, third row), but are confined to cloudless regions. This is also true for \silicate  clouds, but as they form deeper in the atmosphere, the nightside 10-mbar  level is uniformly cooler except in the equatorial region, where a small amount of clouds forms. Thus, depending on the diverse assumption on the radiative and thermodynamical properties of the condensates, clouds form at different locations on the planetary sphere. They change the overall thermal structure as the cloud deck induces a greenhouse warming below it, affecting both the day and the nightsides. \\

Figure \ref{fig:w43_clouds_1um} shows the zonal wind, temperature maps and cloud distribution for the cloudless, MnS and \silicate simulations with clouds made of 1 $\mu$m particles. Adding clouds to the simulations has a direct effect on dynamics. Indeed, the cloudless simulation displays a broad equatorial jet that transforms into a narrower equatorial jet with two high latitudes jets at its flanks in the cloudy cases. Depending on the species included in the simulation and thus, its radiative feedback, we can observe the appearance of retrograde or prograde polar jets (see Fig.\ref{fig:w43_clouds_1um}, top row, middle and right panel). The shape of the stream function is broadly identical with alternating anti-Hadley and Hadley cells from equator to poles. However, the strength of the equatorial cells differs from the cloudless case, with an increase in the MnS simulation and a decrease for the silicate one. The altitude of the cells also slightly changes, with complete cells located higher up in the atmosphere for MnS clouds while the two other cases have incomplete deeper cells, cut almost in half by the model's bottom boundary. Also seen is a slight deceleration of the equatorial jet due to clouds, with a maximum wind speed of 4.7 km.s$^{-1}$ in the cloudless case and 4.2 and 3.9 km.s$^{-1}$ for the MnS and \silicate case, respectively. 

We performed the same transport analysis on the cloudy simulations as done on the cloudless simulation in Section \ref{sec: cloudless_solar}. The overall meridional transport of angular momentum is not strongly affected by the addition of clouds (Fig. \ref{fig:ang_v_1um}). However, the contribution from transient waves becomes non-negligible in these cases, and mostly affects the high latitudes. For \silicate clouds, this contribution affects the upper atmosphere with complete pole-to-equator transport, reversing the transport versus the cloudless case at high latitudes. Despite this hemisphere-wide equatorial advection, the counter effects of both the mean circulation and transient waves diminish the intensity of the transport. In the deep atmosphere, the overall pattern of transport is unchanged, with pole-to-equator transport of angular momentum. For MnS clouds, the polar behaviour is a mix between the cloudless and the \silicate cloud cases. Transient waves tend to transport angular momentum to the poles but this contribution hardly affects the overwhelming contribution of stationary waves, advecting polar momentum to the equator. However, a slight decrease in the intensity of the advection is also noticeable. Thus, the effect of clouds is to decrease the equatorward horizontal advection of axial angular momentum that feeds the super-rotating jet, leading to a slight deceleration of the jet's speed. 

Vertical transport of angular momentum is also altered by the addition of clouds. In the same fashion, transient wave contribution becomes non negligible, even if the net transport is still dominated by the stationary waves (see Fig. \ref{fig:ang_w_1um}). From equator to pole, wide downward transport  flanked by streams of upward and downward advection of angular momentum in the upper atmosphere still shape the net vertical transport when adding \silicate clouds. However, the equatorial downward feature extends less deep in the atmosphere than in the cloudless case. Moreover, polar regions of downward motions appear between 10 mbar and a few bars. In the 50-75$^{\degree}$ latitude range, deep condensable vapour is advected upwards and condenses when reaching pressure of 1-80 bars depending on the latitude, while still being advected upwards until a pressure of $\sim$ 100 mbar. There, the clouds are either transported further to the equator and downwards by the mid-latitude anti-Hadley cells, or to the poles and down again to the deep atmosphere by the Hadley cells where they are sublimated. This explains the cloud distribution seen in the third right panel of Fig. \ref{fig:w43_clouds_1um}. Clouds are either located deep in the polar regions or higher up in the equatorial region. As the temperature is cooler at the poles, clouds transported back to the deeper polar regions can settle there whereas the warmer equatorial region will lead to their sublimation. Thus, equatorial clouds will settle higher in the atmosphere than their polar counterpart. Moreover, these alternating upstream and downstream motions combined with the pole-to-equator meridional flow in the upper atmosphere result in a separation of the broad equatorial jet into a narrower equatorial jet and two polar prograde jets tilted from the poles towards the mid-latitude regions. 

MnS clouds have a different impact on the dynamics, with an overall vertical net transport that differs from the cloudless and \silicate cases. At pressure lower than $\sim$ 1 bar, the atmosphere still displays a strong downward transport of angular momentum but with a narrower meridional extension (between $\sim \pm 35^{\degree}$ as opposed to $\sim \pm 50^{\degree}$ latitude). In the $\pm$$40$-$55^{\degree}$ latitude ranges, a region of upward motion is followed by a thin region of downward motion at $\pm 60^{\degree}$ latitude, itself followed by a polar region of upward motion (Fig. \ref{fig:ang_w_1um}). Polar clouds are horizontally advected to the mid-latitudes where they first experience a downward motion due to the Hadley cells in the thin latitudinal region mentioned above. Then, while they are still being advected equatorward, clouds are advected upstream in the $\pm~40-55 ^{\degree}$ latitude ranges. As they reach a pressure of $\sim$ 60 mbar, they can either be brought back to the poles at a pressure of $\sim$ 10 mbar by the Hadley cells, or they can further move to the equatorial region where they experience downward transport by the anti-Hadley cells. As shown in Fig. \ref{fig:w43_clouds_1um}, the nightside is cooler between $\pm 20-60^{\degree}$ than it is in the equatorial region. Thus, clouds will settle there between 20 and 100 mbars and evaporate at the equator or deeper in the atmosphere. 

The jet's acceleration by the mean residual circulation, the eddies and the net total acceleration is displayed in the bottom row of Fig.\ref{fig:w43_clouds_1um}. For both the cloudless and the cloudy cases, the net equatorial acceleration is positive and is mostly driven by the eddy contribution. From the equator to mid-latitudes ($\sim 35^{\degree}$), the eddies tend to decelerate the prograde zonal wind. In the cloudless case, the eddy-induced deceleration is mostly compensated by the acceleration by the residual mean circulation, leading to a deceleration of the equatorial jet when moving away from the equatorial region. Around $\pm 60^{\degree}$, an inversion of the actions of the eddies and the residual mean circulation is observed, although a small deceleration of the zonal wind is maintained. In the MnS case, the behaviour is qualitatively similar but quantitatively different. Indeed, above $\pm 60^{\degree}$, a deceleration occurs in both the eddy and the mean circulation contribution, leading to an overall effect on the net acceleration: the jet is slowed down and reverses, leading to the appearance of two polar retrograde jets. \silicate clouds induce a different behaviour, with a strong acceleration by the residual mean circulation hardly compensated by the eddies. Thus, between $\pm 60$ and $\pm 80^{\degree}$, the net acceleration is positive, leading to the appearance of two prograde polar jets. 

To further investigate the origin of the retrograde polar jets in the MnS simulation, we ran an additional simulation, identical to the MnS one but turning off the radiative effects of clouds. In this non-radiative simulation, we do not observe the formation of polar retrograde jets. Thus, we can confidently say that the radiative effect of MnS clouds leads to a warming of the poles in the upper atmosphere. This warming creates a meridional temperature gradient between the poles and mid-latitudes, which is stronger than in the cloudless scenario, with poles warmer than mid-latitudes. This gradient translates into 
a dynamical equilibrium via the thermal wind equation, 
\begin{equation}\label{eq: thermal_wind}
    \Big[2u\frac{\tan{\phi}}{R_p}+2\Omega \sin{\phi}\Big]\frac{\partial u}{\partial P} = \frac{R}{P}\frac{\partial T}{\partial y}
.\end{equation}
At high-latitudes, the cyclostrophic term dominates the Coriolis term and we can neglect the latter Coriolis term. In our case, in the northern hemisphere, the temperature gradient is positive. Thus, the vertical shear of the zonal wind, $\frac{\partial u}{\partial P}$, should be negative (the zonal wind increases upwards) to allow for the retrograde jet to exist, which is the case in our simulation. In the southern hemisphere, the temperature gradient is negative and thus the vertical shear of the zonal wind is also negative, allowing for a retrograde jet.
 
\subsubsection{Impact of different particle sizes}\label{sec: diff_sizes}

Next we focused on the \silicate simulations and investigated the impact of the particle size on the thermal and dynamical structure of the atmosphere. We focused on \silicate clouds instead of MnS clouds, as \cite{gao_universal_2021} find that they should dominate the cloud composition for hot Jupiter with $T_{\rm eq} \leqslant 2100$ K. For simplicity and clarity, we focused on clouds with particles of 0.1, 1, and 10 $\mu$m in radius. However, the trends found in this section are also valid for intermediate particle sizes simulated in this study.  

Figure \ref{fig:silicate_solar} displays the wind pattern, the 10 mbar temperature map, the zonal mean cloud distribution and the jet acceleration for the three simulations . From left to right, the cloud particles grow. As mentioned before, the clouds settle deeper in the atmosphere when composed of bigger particles. This is illustrated again with an extended cloud layer from the top of the model to $\sim$500 mbars for small particles (0.1 $\mu$m) and a thin meridionally homogeneous cloud layer with intermediate particles (1 $\mu$m) between 10 mbar and 1 bar and abundant deep polar clouds. However, the large particle case (10 $\mu$m) displays a unique behaviour. Clouds only settle in the equatorial region, with a thin vertical extent centred around 500 mbars, with hints of deeper clouds at high latitudes. Indeed, deeper than 500 mbar in the equatorial region, the temperature rises above the condensation temperature, leading to the evaporation of the clouds. Above this patchy cloud, the temperature is colder than the condensation temperature. Thus, clouds form higher up in the atmosphere where they experience sedimentation, until reaching their condensation temperature. At mid to high latitudes, the temperature is always below the condensation temperature, above 100 bars. Thus, clouds form and then experience both sedimentation and advection by the atmospheric cells, bringing them back to the equator, or deep in the polar regions where condensation happens. For smaller cloud particles, the terminal velocity of Eq. \ref{eq: terminal_velocity} is small compared to the vertical wind speed, leading to clouds higher up in the atmosphere.  

As seen in the previous section, one of the effects of clouds on atmospheric dynamics is to reduce the speed of the equatorial jet. As cloud particles are smaller and higher up in the atmosphere, this effect gets stronger. Indeed, the jet celerity reaches 3.5 km.s$^{-1}$ if \silicate particles of 0.1 $\mu$m are included in the simulation, and 4.6 km.s$^{-1}$ for 10 $\mu$m particles. A clear correlation between the wind speed and the size of the particles is seen, with a stronger deceleration of the jet when an extended cloud layer exists in the upper atmosphere (i.e. for smaller particles). Indeed, the acceleration by eddies and the residual mean circulation are shown on the last row of Fig.\ref{fig:silicate_solar}. Overall, the net equatorial acceleration is smaller for smaller particles. At the poles, the 1-$\mu$m simulation displays an acceleration that corresponds to the prograde polar jets mentioned previously. However, this is not the case when adding bigger cloud particles. Indeed, as bigger particles will lead to a patchy equatorial cloud, the polar regions remain almost unaffected by their radiative effect and the wind structure closely resembles the ones of the cloudless case, as the eddies counterbalance the acceleration by the mean residual circulation. \\

To summarise our findings, clouds have multiple effects on the atmospheric dynamics. Regardless of the condensate, if able to form in the atmosphere, clouds mostly form on the cooler nightside and terminator regions of the planet. Once formed, they influence the vertical and meridional transport of axial angular momentum. In particular, the transient waves contribution becomes non-negligible in the cloudy case, and influences the wind and thermal structure. Overall, the equatorial jet slows down and breaks into a narrower jet flanked by two polar jets. The shape and strength of the polar jet  depends on the radiative feedbacks of the clouds generating them, for a solar-like atmosphere. In the next section, we investigate the effect of a super-solar metallicity on the atmospheric states.

\subsection{Impact of metallicity} \label{sec: supersolar }
In this section, we explore the impact of a super-solar metallicity on the thermal and wind structure of \wasp. To do so, we ran a cloudless simulation at 10x solar metallicity, using the method described in Section \ref{model_init}. The only change to our set-up is in the \texttt{Exo-REM} simulation where we increased the metallicity, effectively changing the vertical chemical profile of the atmosphere and increasing the opacity. Thus, the radiative data computed during our initialisation are different from those used in the previous section. Appendix \ref{fig:exorem_chem} displays these profiles for the solar and super-solar case. We then computed a grid of simulation including \silicate clouds, with the same radii as described above. \\
\subsubsection{Simulation without clouds}
The broad thermal and dynamical pattern of the atmosphere of \wasp~do not qualitatively change with metallicity. However, quantitative differences emerge. Indeed, the thermal structure is overall warmer at higher metallicity on the dayside and cooler on the nightside (see Fig. \ref{fig:supersolar}). As the opacities increase, heating rates and cooling rates also increase, leading to shorter radiative timescale. The photosphere rises with increasing opacities, as shown for the nightside on Fig.\ref{fig:tau_silicate} and predicted by \cite{kataria_atmospheric_2015}. The equatorial jet is accelerated by this increased longitudinal temperature contrast, reaching 5.8 km.s$^{-1}$ and is flanked by two polar retrograde jets. Moreover, the vertical extent of the jet is narrower, extending to $\sim$ 50 bars instead of $\sim$ 200 bars before reaching zero. The circulation is characterised by one large anti-Hadley cell in each hemisphere instead of the two cells previously discussed in Section \ref{sec: cloudless_solar}. This is consistent with previous studies on the effect of metallicity \citep{showman_atmospheric_2009,kataria_atmospheric_2015}. Interestingly, we note that similarly to our simulations with MnS clouds, a strong meridional temperature gradient between the poles and mid-latitudes emerges. This is understandable from the thermal wind equation (Eq. \ref{eq: thermal_wind}). Retrograde polar jets and negative upward vertical zonal wind shear are balanced by a positive mid-latitude to pole meridional gradient of temperature. 

Thus, increasing the metallicity has a different effect on the thermal and wind structure of the atmosphere of \wasp~compared to having clouds, as jet speed increases instead of diminishing. This jet increase is also found by \cite{kataria_atmospheric_2015} and is linked to the greater longitudinal thermal forcing. In the next section, we investigate the joint effect of clouds and super-solar metallicity on the dynamical structure of the atmosphere.
\subsubsection{Simulation with \silicate clouds}
Starting from the previously computed  cloudless simulation at super-solar metallicity, we added \silicate clouds of different particle radius and ran the simulation for 1000 \wasp~years. The initial profile of condensable vapour is computed as described in section \ref{model_init}. However, as metallicity increases, the condensation curve of \silicate shifts to higher temperature for a given pressure \citep{visscher_atmospheric_2010}. The initial profile of condensable vapour is then lower than the one used in the solar metallicity case because most of the vapour has condensed in layers below the lower boundary of our model. As less material is available for condensation, clouds are thus thinner and less abundant at higher metallicity. 

The cloud distribution closely follows our findings of section \ref{sec: dynamic_clouds}. As cloud particles are larger, the terminal velocity of sedimentation increases, leading to clouds layers located deeper in the atmosphere. Moreover, deeper than 1 bar, anti-Hadley cells dominate the circulation (see Fig. \ref{fig:supersolar}), effectively transporting polar clouds upwards and equatorward, where the temperature is above the condensation temperature. Thus, large cloud particles settle deeper in the atmosphere. In contrary to the solar metallicity case where large cloud particles mostly affect the equatorial region, super-solar metallicity impacts the thermal structure at all latitudes. The equatorial region is too warm for clouds to form and the cooler mid-latitudes are preferred for cloud formation and settling. Moreover, as the equatorial region is almost not affected by clouds, the jet speed is unaltered when adding \silicate clouds to the simulations. This is an entirely different behaviour from the one found in solar metallicity atmospheres. However, it is not entirely clear how super-solar metallicity and \silicate clouds globally impact the dynamical and thermal structure of \wasp. Hence, in the next section, we study the albedo and cloud radiative forcing for solar and super-solar metallicity. \\

\subsection{Albedo and cloud radiative forcing} \label{sec:albedo_CRF}
\begin{figure}[h]
    \centering
    {\includegraphics[width=0.5\textwidth]{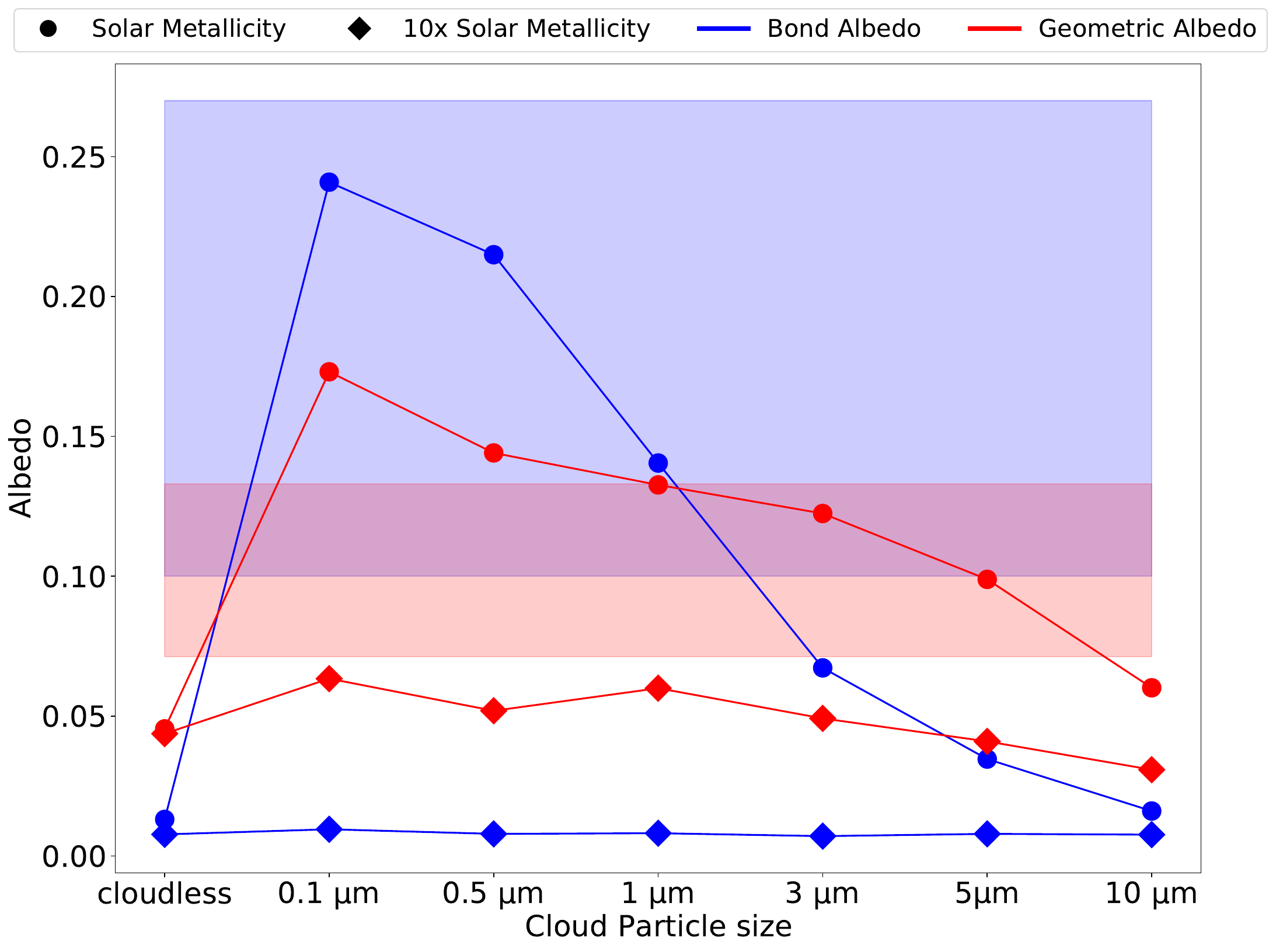}}
    \caption{Disk-integrated Bond albedo and geometric albedo in the CHEOPS band (0.33-1.1 $\mu$m) for the cloudless and \silicate cloud simulations at solar (circles) and 10x solar (diamond) metallicity. The red shaded area corresponds to the constraint on the geometric albedo derived by \cite{scandariato_phase_2022}. The blue shaded area corresponds to the estimated range of Bond albedo derived by \cite{stevenson_spitzer_2017}.}
    \label{fig:bond_albedo}
\end{figure}
In this section, we compute the Bond and geometric albedo of our simulations, along with the cloud radiative forcing \citep{Ramanathan_1989}. As pointed out in the last section, discriminating between a solar cloudy atmosphere and a super-solar cloudy or cloud-free atmosphere is complicated by looking at the thermal structure only. We show the Bond albedo (integrated between 0.612 and 325 $\mu$m) and the geometric albedo in the CHEOPS band (0.33-1.1 $\mu$m) in Fig. \ref{fig:bond_albedo}. For the geometric albedo, we only considered the reflected light component for the computation. As the CHEOPS filter leaks into the infrared, the thermal emission is non-negligible and needs to be removed to accurately compute the geometric albedo. We find that the cloudy solar metallicity simulations always have a higher Bond albedo than the super-solar atmospheres and the cloudy atmospheres as a result of the increasing amount of scattering provided by the increasing cloud content. Moreover, for solar metallicity, the size of the cloud particles has a detectable impact on the albedo, and the sub-micron to micron size simulations are compatible with the Bond albedo derived by \cite{stevenson_spitzer_2017}. This is easily understandable from the cloud distributions shown in Fig. \ref{fig:silicate_solar}. The smaller the cloud particles, the easier they are lofted above the photosphere and impact the measured Bond albedo. The geometric albedo in the CHEOPS band is similar between the cloudless cases of different metallicity, as no clouds induce a strong reflective component and the flux is dominated by thermal emission for wavelength above 0.7 $\mathrm{\mu}$m. In the cloudy cases, the solar metallicity simulations always have a higher bond albedo due to the reflective clouds than the super-solar simulations. Scenario with particle sizes ranging from 1 to 10 $\mu$m are compatible with the derived estimates from \cite{scandariato_phase_2022}. However, super-solar atmospheres constantly under-predict the measured geometric albedo in the CHEOPS band. Thus, precise measurements of the Bond and the geometric albedo should help disentangle the degeneracy between super-solar and cloudy solar metallicity atmospheres, and help constrain the size of the cloud particles. 

\begin{figure}[h]
    \centering
    \subfigure[cloud radiative forcing]{\includegraphics[width=0.5\textwidth]{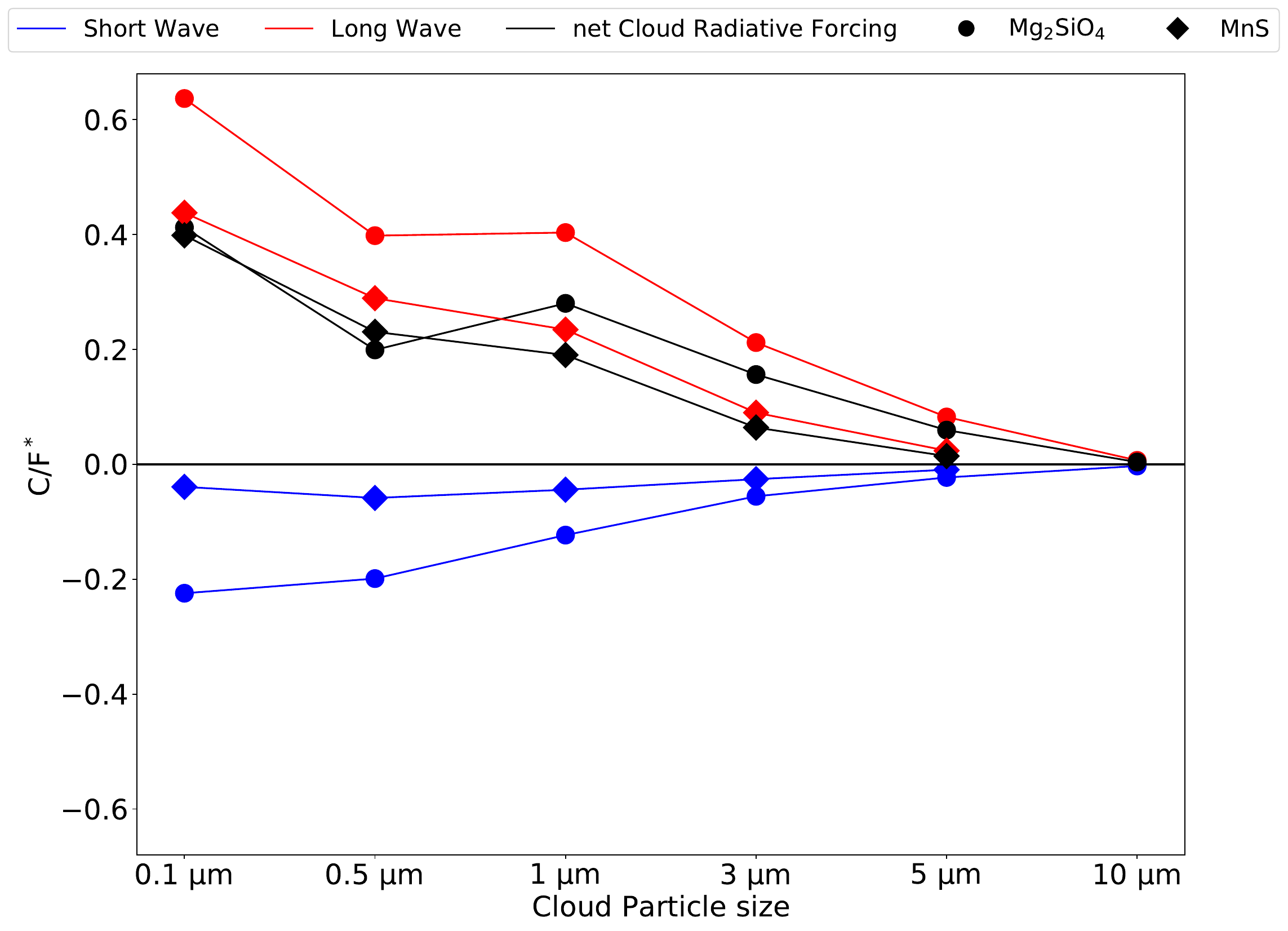}\label{subfig:CRF}}
    \subfigure[Globally averaged Temperature profile]{\includegraphics[width=0.5\textwidth]{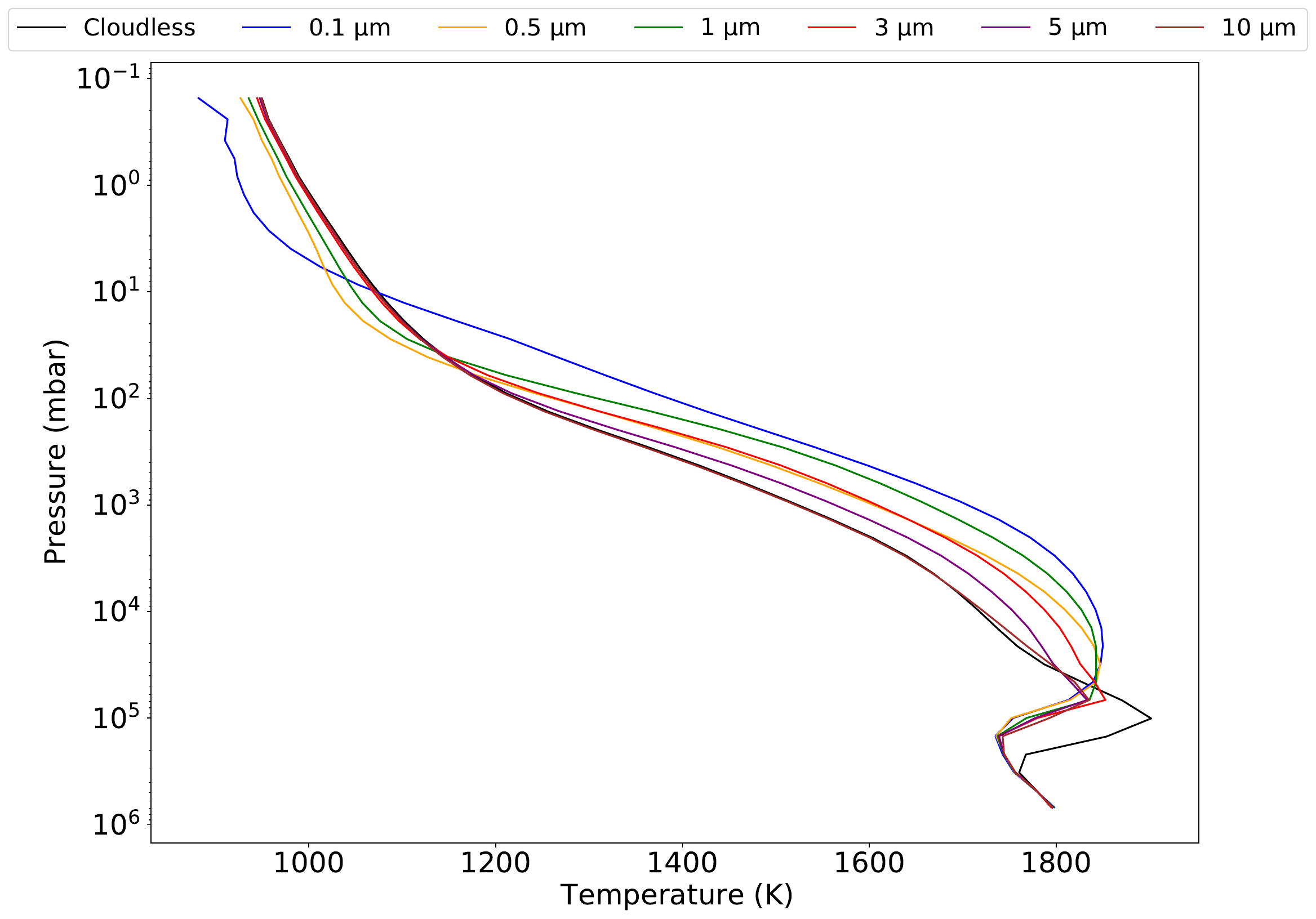}\label{subfig:globalTP}}
    \caption{Normalised cloud radiative forcing and globally averaged temperature profiles. \emph{Top: }Cloud radiative forcing (normalised by the stellar irradiance) as a function of cloud particle sizes for solar simulations, and clouds composed of \silicate and MnS. Blue lines denote the shortwave cloud radiative forcing, the red line the longwave cloud radiative forcing, and the black line the net, total cloud radiative forcing. Circles denote \silicate clouds, and diamonds are MnS simulations. \emph{Bottom:} Globally averaged temperature profiles for our cloudless and solar metallicity \silicate cloud simulations.}
    \label{fig:CRF_solar}
\end{figure}

To understand the global effect of cloudiness on the atmosphere of \wasp, we computed the cloud radiative forcing. Following \cite{Ramanathan_1989}, the net radiative heating $H$ of an atmospheric column is\begin{equation}
    H = (1-\rm ASR)-\rm OLR
,\end{equation}
with OLR the thermal flux radiated to space. Thus, the cloud forcing can be written as\begin{equation}
    C = H_{\rm cloud}-H_{\rm clear}
.\end{equation}
Finally, the cloud forcing can be decomposed into a shortwave and longwave component, yielding
\begin{equation}\label{eq: CRF_decomposed}
    C = C_{\rm sw} + C_{\rm lw} = (\rm ASR_{\rm cloud}-\rm ASR_{\rm clear}) + (\rm OLR_{\rm clear}- \rm OLR_{\rm cloud})
.\end{equation}
We computed this value directly from our cloudy simulations, by turning off the radiative feedbacks of clouds and running an additional time-step of the models. In this way, the clear models have the same thermal structure ias the cloudy models and only differ by the lack of clouds opacities. Figure \ref{subfig:CRF} displays the shortwave, longwave and cloud radiative forcing components normalised by the stellar irradiance at the top of the atmosphere for our solar simulations with \silicate and MnS clouds. 
It is clear that radiative clouds induce a positive forcing in all of our simulations, regardless of the size of the cloud particles and the type of condensates. Shortwave forcing corresponds to the dayside albedo effect of the cloud, which translates into a cooling effect. On the other side of the spectrum, longwave forcing corresponds to the greenhouse effect of clouds, mostly on the nightside, which leads to a warming effect. Thus, clouds have a net warming effect on the atmosphere of \wasp, as the longwave forcing is $\sim 3$ greater than the shortwave forcing (see the globally averaged temperature profile of Fig. \ref{subfig:globalTP}). This is in contrast with the global net cloud radiative effect on Earth, where low-altitude clouds have a cooling effect, high-altitude clouds a warming effect, with a net global cooling effect \citep{IPCC_chap7}. Both the shortwave and longwave components displays a trend with particle sizes. The smaller the cloud particles, the stronger the impact on the shortwave and longwave components, as they are more abundant and yield a greater optical depth. Interestingly, the longwave radiative forcing for \silicate does not follow that trend for cloud particles of $0.5$ and $1$ $\mathrm{\mu}$m. This is due to the peak emission wavelength of \wasp, located around 2 $\mu$m, which corresponds to the peak in extinction efficiency of \silicate clouds for $1$ $\mathrm{\mu}m$ particles whereas the peak of extinction efficiency for $0.5$ $\mathrm{\mu}$m particle is located around 0.9 $\mathrm{\mu}$m. Thus, the net cloud radiative forcing is slightly stronger for $1$ $\mu$m \silicate clouds than for $0.5$ $\mu$m particles, as the shortwave component is not affected by this thermal amplification. Comparing simulations with MnS and \silicate clouds, the cloud radiative forcing is stronger for \silicate clouds. The cloud radiative forcing for simulations with super-solar metallicity is shown in Fig. \ref{fig:CR_10x}. Because of the low amount of clouds in these simulations, the net cloud radiative forcing is unsurprisingly weaker than in the solar metallicity atmosphere by a factor of $\sim$700-5000 depending on the cloud particle size. Both the long and shortwave component follow the previously identified trend with particle size, but the net overall forcing is in favour of a cooling instead of a warming. The driver of this cooling is the Bond albedo of the planet. Despite its low value, the longwave optical depth induced by clouds is not enough to counteract the shortwave radiative cooling. 

As the temperature contrast between the day and the nightside increases with metallicity, and the offset of the hotspot is smaller with clouds, spectral phase curves should encode all the necessary information to discriminate between a cloud-free super-solar atmosphere and a cloudy solar one. In the next section, we study the impact of clouds on spectral phase curves in the infrared.

\section{Spectral phase curves and the impact of clouds}\label{sec: observations}

 In this section, we compare our grid of models to phase curves of \wasp~obtained with HST \citep{stevenson_thermal_2014}, the \textit{Spitzer} Space Telescope \citep{morello_2019} and JWST MIRI-LRS \citep{ERSpaper}. For the MIRI-LRS data reduction we used the Eureka! \citep{Bell2022Eureka} pipeline. Stage 1 and 2 use the \texttt{jwst} pipeline \citep{stsci2022jwst} with cosmic ray detection threshold set to 5, calibration files from commissioning and ramp fitting weighting set to uniform.  For stage 3 we defined a rectangular selection to extract the flux of the source (y window of 140-393px and an x window of 11-61px) and then performed background subtraction on each integration using a column away from the trace. In Stage 4 we generated the spectroscopic light curves with bins of 0.5~$\mu$m, spanning 5-12~$\mu$m, for a total of 14 spectral channels. Then, the light curves are sigma-clipped and the first 779 integrations removed. For the fit, we modelled the phase curve with a second order sinusoidal function and the eclipse with the \texttt{batman} package \citep{kreidberg_batman}, to model the instrumental systematics we used a linear polynomial model in time, an exponential ramp, a first order polynomial in $y$ position and a first order polynomial in PSF width in the $s_y$ direction. We fitted the data using the \texttt{emcee} sampler \citep{emcee_2021} with 500 walkers and 1500 steps with the following jump parameters (randomly perturbed at each step): $R_{\rm p}/R_{\rm \star}$, $F_{\rm p}/F_{\rm \star}$, $u_1$, $u_2$, AmpCos1, AmpSin1, AmpCos2, AmpSin2, $c_0$, $c_1$, $r_0$, $r_1$, $y_{pos}$, and $y_{width}$. The convergence of the fit was then assessed using the Gelman and Rubin test. Finally, from the best fit we extracted the spectra of the planet at different phases (nightside, day-side and the two terminators) of its orbit.\\
 
 To produce phase curves from the \texttt{generic PCM}, we used the \texttt{Pytmosph3R} software \citep{falco_toward_2022}. Our wavelength coverage is identical to the one used in \texttt{generic PCM} (see Table \ref{bins}) but at a resolution of R$\sim$500 and the same opacities are used.\\ 

\subsection{Spectral phase curves at solar metallicity}\label{sec: spectral_solar}
\begin{figure*}[h]
    \centering
    {\includegraphics[width=\textwidth]{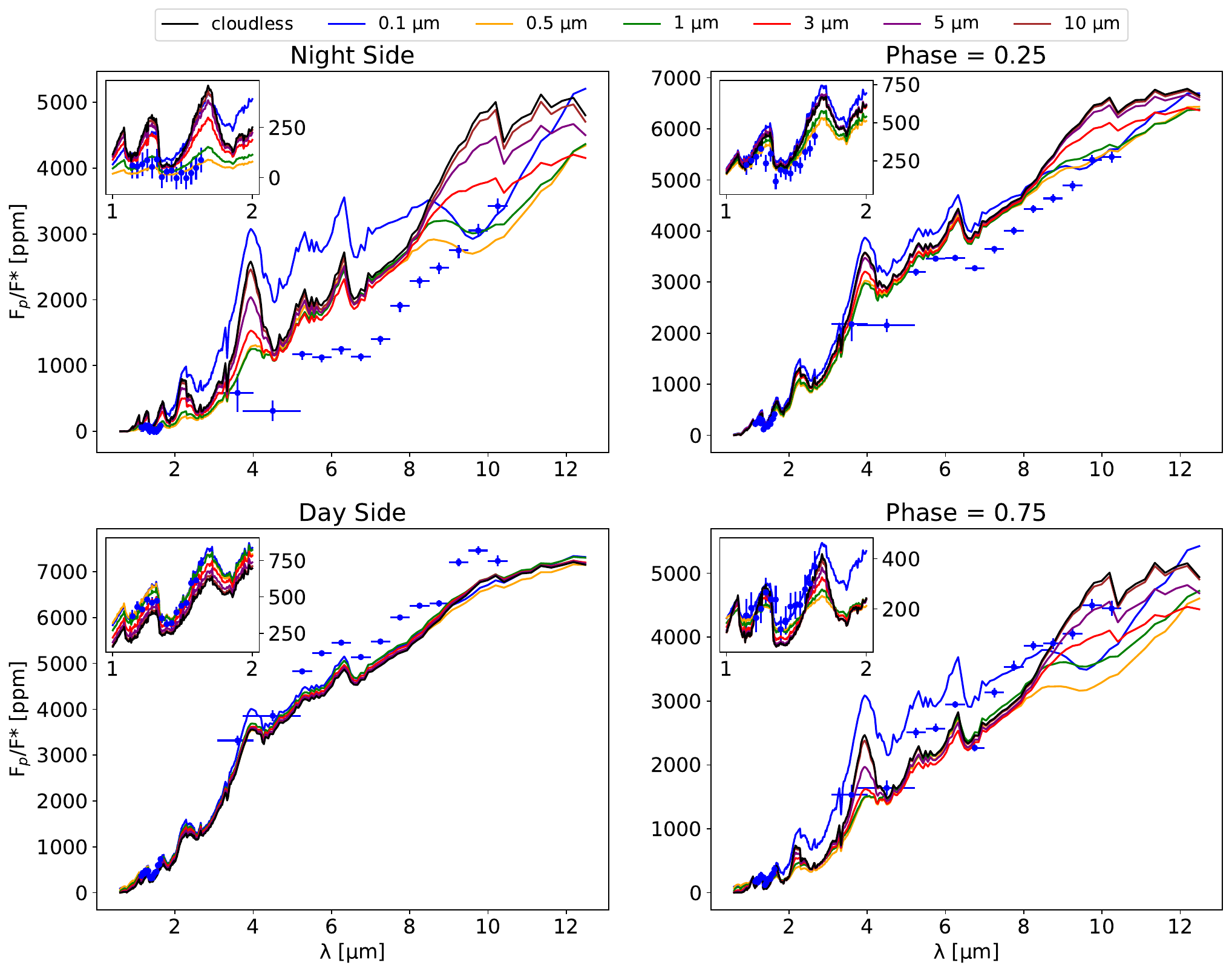}}
    \caption{Spectral phase curves of \wasp~at four given phases, from 0.6 to 12.5 $\mu$m. The black line is the cloudless simulation. Each colour corresponds to a simulation with a different \silicate cloud particle size. Blue data points and error bars from HST are taken from Table 5 of \cite{stevenson_spitzer_2017}. \textit{Spitzer} data points are from \cite{morello_2019}, and MIRI-LRS data points are from the \texttt{Eureka! v2 Reduction}, as explained above. In each panel, the subplot is a zoomed-in view of the HST data, between 1.1 and 1.7 $\mu$m. Dayside corresponds to the planet passing behind its star and nightside to the planet passing in front of its star. The 0.25 phase is the quarter phase between the night and day phases, and the 0.75 phase is the quarter phase between the day and night phases. Thus, the eastern terminator is seen at phase 0.25 and the western terminator at phase 0.75. }
    \label{fig:spectral_w43b}
\end{figure*}

\begin{figure*}[h]
    \centering
    {\includegraphics[width=\textwidth]{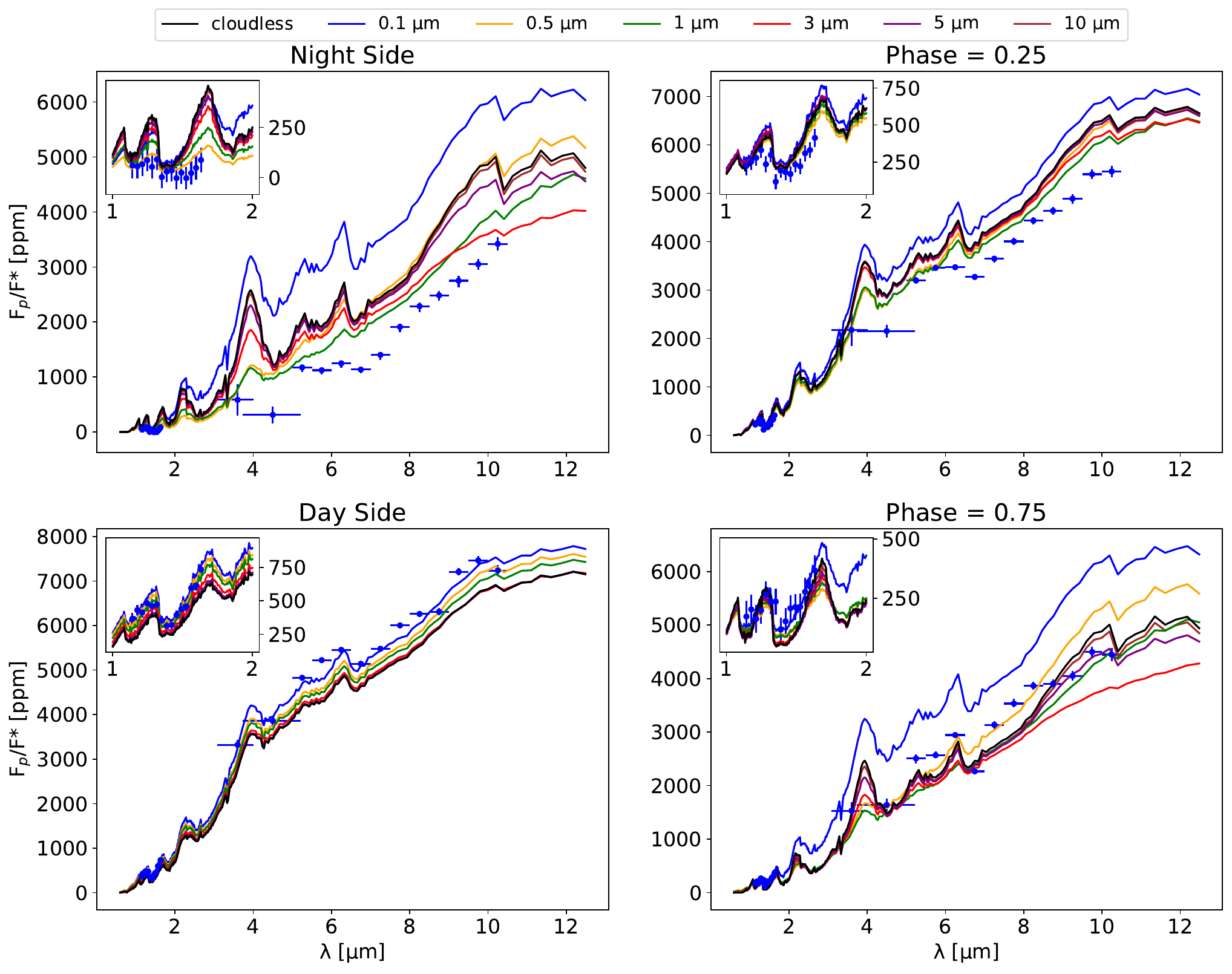}}
    \caption{Spectral phase curves of \wasp~at four given phases, from 0.6 to 12.5 $\mu$m. The black line is the cloudless simulation. Each colour corresponds to a simulation with a different MnS cloud particle size. Blue data points and error bars from HST are taken from Table 5 of \cite{stevenson_spitzer_2017}. \textit{Spitzer} data points are from \cite{morello_2019}, and MIRI-LRS data points are from the \texttt{Eureka! v2 Reduction}, as explained above. In each panel, the subplot is a zoomed-in view of the HST data, between 1.1 and 1.7 $\mu$m. Dayside corresponds to the planet passing behind its star and nightside to the planet passing in front of its star. The 0.25 phase is the quarter phase between the night and day phases, and the 0.75 phase is the quarter phase between the day and night phases. Thus, the eastern terminator is seen at phase 0.25 and the western terminator at phase 0.75. }
    \label{fig:spectral_w43b_MnS}
\end{figure*}

Our simulations are post-processed at four orbital phases, the nightside emission, the dayside emission and the two intermediate quarter phases for our cloudless and cloudy simulations that we show in Fig.\ref{fig:spectral_w43b} for cloudless and \silicate simulations and in Fig. \ref{fig:spectral_w43b_MnS} for MnS simulations. It is straightforward from the plots that clouds are a plausible explanation for the observed spectra. Indeed, the cloudless simulation strongly overestimates the flux in the nightside, and slightly underestimates it on the dayside.
At quarter phases, the cloudless model overestimates the flux at phase 0.25 (when the eastern terminator is visible by the observer) whereas the agreement with the phase 0.75 (when the western terminator is visible by the observer) is quite good, despite a slight overestimation of the water band depth in the HST wavelength coverage and an underestimation of the flux in the MIRI range. Adding clouds to the simulations drastically changes the nightside spectrum. As clouds preferentially form on the cooler nightside, they reduce the amount of emitted flux on this side of the planet. Moreover, the clouds warm the atmosphere below them and this additional heat is transported back to the dayside by the equatorial jet. Thus, the dayside warms up in the cloudy cases with regards to the cloudless simulation. This redistribution of heat from the night to the dayside is not clearly seen in the case of \wasp, as seen on the dayside panel. Indeed, depending on the size of the particles forming the clouds and the type of condensates, dayside heating will be more or less efficient. Small particle of MnS clouds are the most efficient to produce a night-to-day advection of heat. A possible explanation of this phenomenon is because of the cooler condensation curves of MnS, clouds will tend to form higher up in the atmosphere (see Fig.\ref{fig:grid_clouds_w43}), inducing a stronger greenhouse effect, allowing for more heat to be available for horizontal advection by the jet. At phase 0.25, the effect of clouds should be noticeable, as we probe the eastern terminator and parts of the nightside. Even though the eastern terminator is mostly cloud-free in our simulations, micron-sized clouds allow for a reduction of the flux consistent with HST and MIRI-LRS data but not consistent with the 4.5 $\mu$m \textit{Spitzer} point. The 0.75 phase is interesting, as no cloudy case perfectly matches the near infrared data. The cloudless case reproduces  the data fairly well until 5 $\mu$m, even though it slightly overestimates the depth of the 1.4 and 6.1 $\mu$m water bands. It seems that \silicate clouds have disappeared from the observations at this phase, which could potentially be explained by microphysical effects, not taken into account in our approach. However, MnS clouds composed of sub-micron particles are in good agreement with this phase. In the mid-infrared, the nightside shows a disagreement with the cloudy simulations, as the emitted flux is overestimated. Moreover, the strong \silicate absorption feature around 9.6 $\mu$m is not seen in the data and is present in our simulations with sub-microns cloud particles. Thus, we can advocate with a certain degree of confidence that if \silicate clouds play a role in the observable flux of \wasp, the cloud particles are bigger than 1 $\mu$m. Also, it seems difficult to disentangle a contribution from MnS clouds from one of \silicate clouds with the current observations, despite the findings of \cite{gao_universal_2021} and \cite{helling_cloud_2021}, who claim that MnS should not play a major role in the nightside spectra of \wasp. Thus, we only investigate \silicate clouds in the following.

\begin{figure}[h]
    \centering
    {\includegraphics[width=0.47\textwidth]{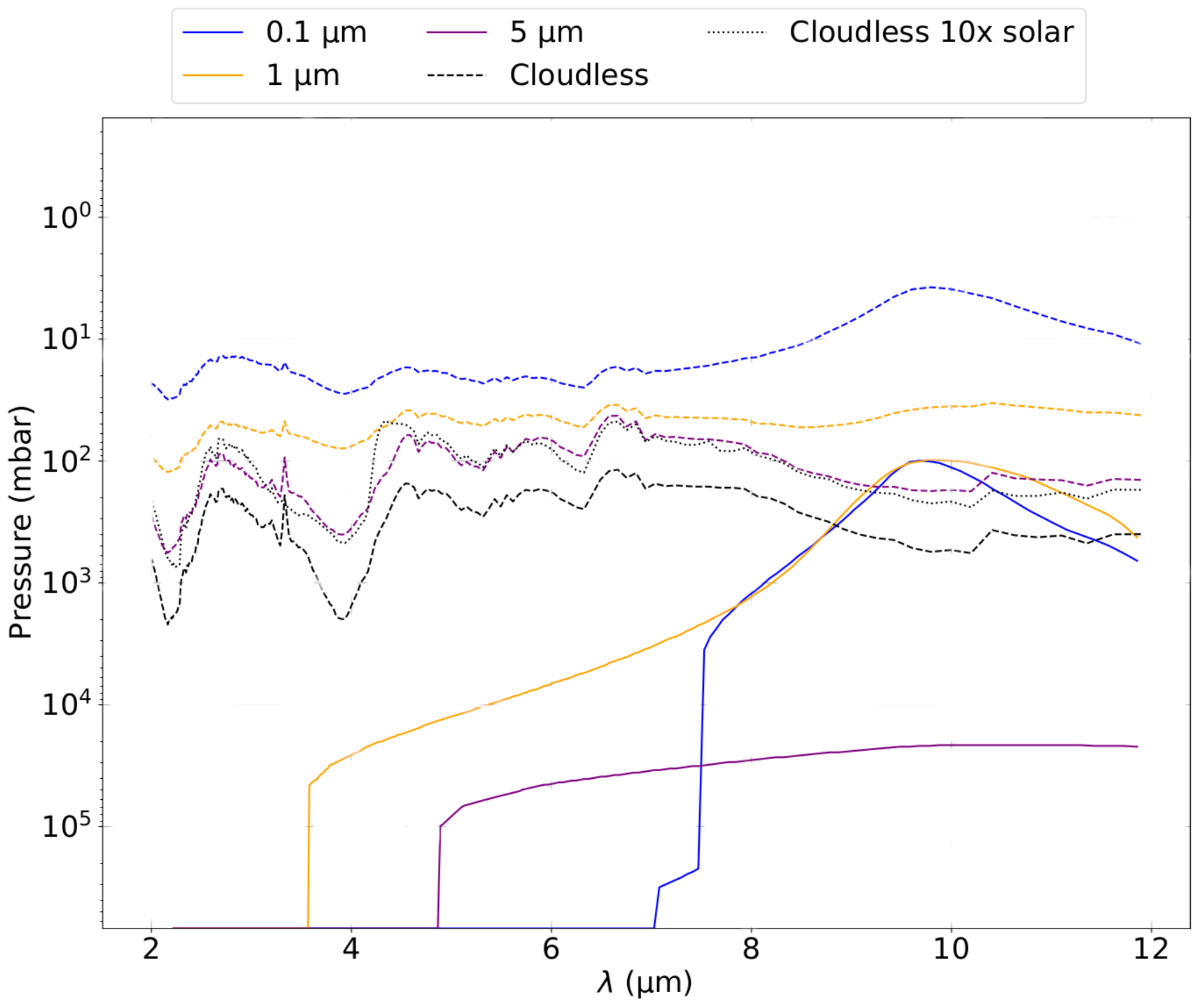}}
    \caption{Nightside pressure at which \silicate clouds becomes optically thick as a function of wavelength, for simulations with different particle sizes (solid lines) and atmospheric photospheric pressure (dashed lines) with solar metallicity atmospheres. Solid lines denotes the pressure at which the optical depth of \silicate clouds is equal to 1. For all simulations, \silicate clouds become optically thick below the photospheric pressure. We  added the atmospheric photospheric pressure for the cloudless case for reference in black for the solar metallicity (dashed line) and 10x solar metallicity (dotted line).}
    \label{fig:tau_silicate}
\end{figure}

As mentioned above, the 10 $\mu$m \silicate absorption feature is not seen in the MIRI-LRS data and in our simulation with cloud particles bigger than 1 $\mu$m. Indeed, this is expected for large particles ($\geq 10$ $\mu$m) as the absorption coefficient does not peak. For smaller particle sizes, we computed the optical depth of \silicate clouds around 10 $\mu$m as a function of pressure and the photospheric pressure at the same wavelength (see Fig.\ref{fig:tau_silicate}). Here, the photospheric pressure is defined as the pressure  at which the atmospheric temperature is equal to the brightness temperature. For all simulations, the \silicate clouds become optically thick at a pressure greater than the photospheric pressure of the nightside. However, optically thin clouds still contribute to the rising of the photosphere, via their thermal feedback. This translates to an effect in the nightside spectra located around 9.6 $\mu$m clearly detectable for 0.1 $\mu$m cloud particles.
Thus, we can only conclude that if \silicate clouds are present, their spectral contributions are shaped by particles bigger than 1 $\mu$m.
Except on the cloud-free dayside, our cloudy simulations always overestimate the flux in the mid-infrared. Our explanation for this behaviour is that our models lack a source of longwave opacity. A few possible culprits for this additional opacity could be linked to a lack of clouds due to a too strong cold trap in our models or horizontal variations of the chemical composition (possibly linked to the CO-CH$_{\rm 4}$ equilibrium), which we did not account for \citep{venot_global_2020}. Super-solar metallicity could also be responsible for this additional heating, as we investigate in the next section.

\subsection{The effect of super-solar metallicity on spectral phase curves}
\begin{figure*}[h]
    \centering
    {\includegraphics[width=\textwidth]{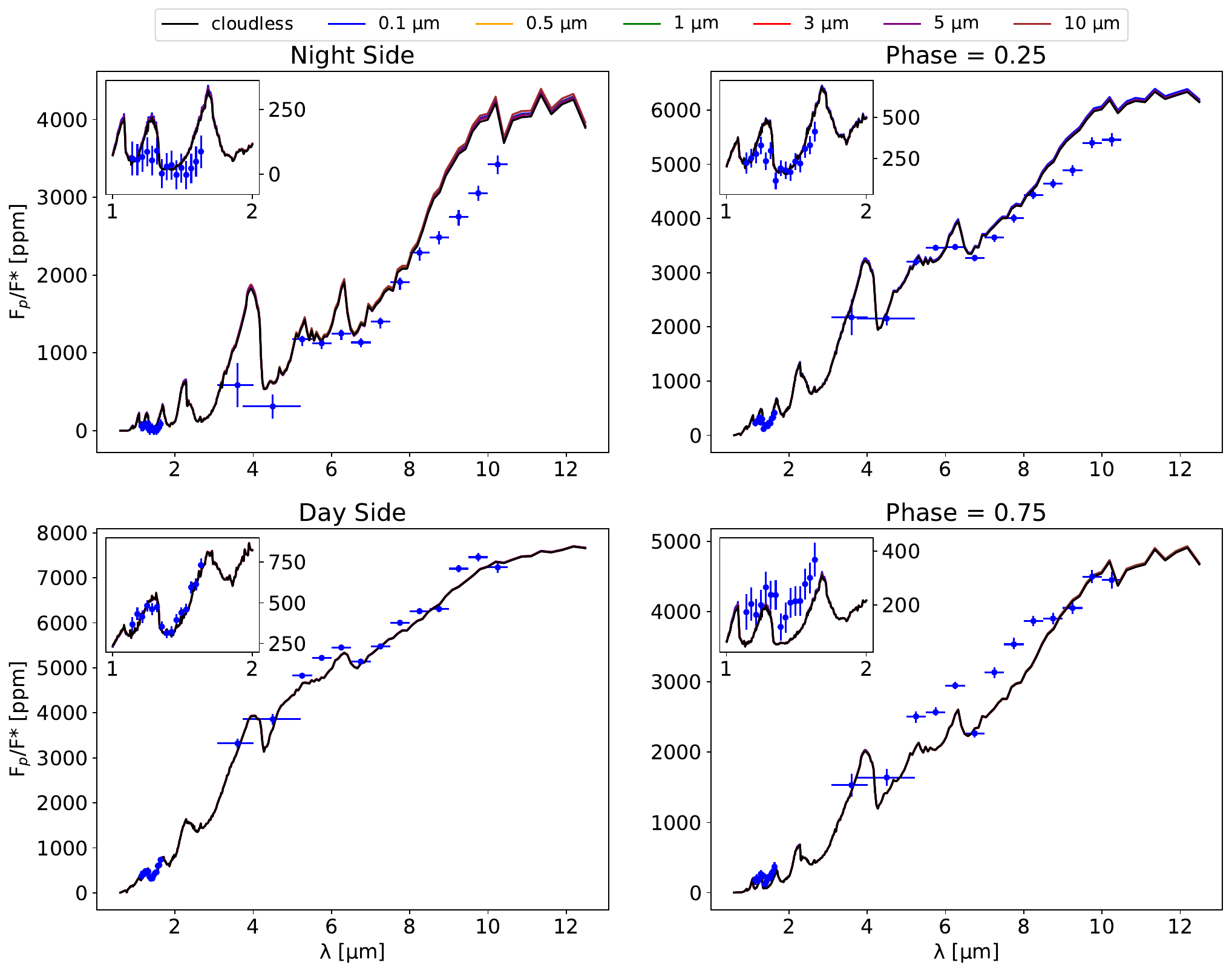}}
    \caption{Spectral phase curves of \wasp~at four given phases, from 0.6 to 12.5 $\mu$m, for a super-solar metallicity atmosphere. The black line is the cloudless simulation. Each colour corresponds to a simulation with a different cloud particle size. Each coloured line is a simulation with \silicate clouds and a different radius. Blue data points and error bars  from HST are taken from Table 5 of \cite{stevenson_spitzer_2017}. \textit{Spitzer} data point are from \cite{morello_2019}, and MIRI-LRS data points are from the \texttt{Eureka! v2 Reduction}, as explained above. In each panel, the subplot is a zoomed-in view of the HST data, between 1.1 and 1.7 $\mu$m. Dayside corresponds to the planet passing behind its star and nightside to the planet passing in front of its star. The 0.25 phase is the quarter phase between the night and day phases, and the 0.75 phase is the quarter phase between the day and night phases. Thus, the eastern terminator is seen at phase 0.25 and the western terminator at phase 0.75.}
    \label{fig:spectre_10x}
\end{figure*}

We computed spectral phase curves from the \texttt{generic PCM} in the same fashion as done for the solar metallicity case (see Fig.\ref{fig:spectre_10x}). As opacity increases, the model dayside is warmer than in the solar case and perfectly fits the data at all wavelengths, and for all simulations (clear and cloudy). This is understandable by the fact that the amount of clouds forming in the atmosphere is very low and the dayside is always cloudless, leading to very small differences in the eclipse spectra between the clear and cloudy cases. 
On the nightside, as the temperature is lower than in the solar metallicity simulations, all spectra have a lower flux. The 4.5 $\mu$m \textit{Spitzer} point is better fitted than in the solar metallicity case, even if the simulated feature is not deep enough. This is also the case at the other phases, with a better agreement with this data point. This is due to the enhance CO$_{\rm 2}$ abundance in the 10x solar case, as shown in Fig. \ref{fig:exorem_chem}.

Interestingly, the 6.1 $\mu$m water feature is overestimated on the nightside case and  underestimated at phase 0.75, as the water bands in the HST range. This seems to indicate that a super-solar metallicity is needed to understand the spectra at different phases, but clouds are also mandatory on the nightside. If so, an intermediate metallicity, allowing for more cloud formation could be an answer to our current mismatch between simulations and observations. Moreover, we again see a flux overestimation at all phases for wavelength greater than 8 $\mu$m, indicating again a potential lack of opacity in our simulations (except on the cloud-free dayside).

\begin{figure*}[h]
    \centering
    {\includegraphics[width=\textwidth]{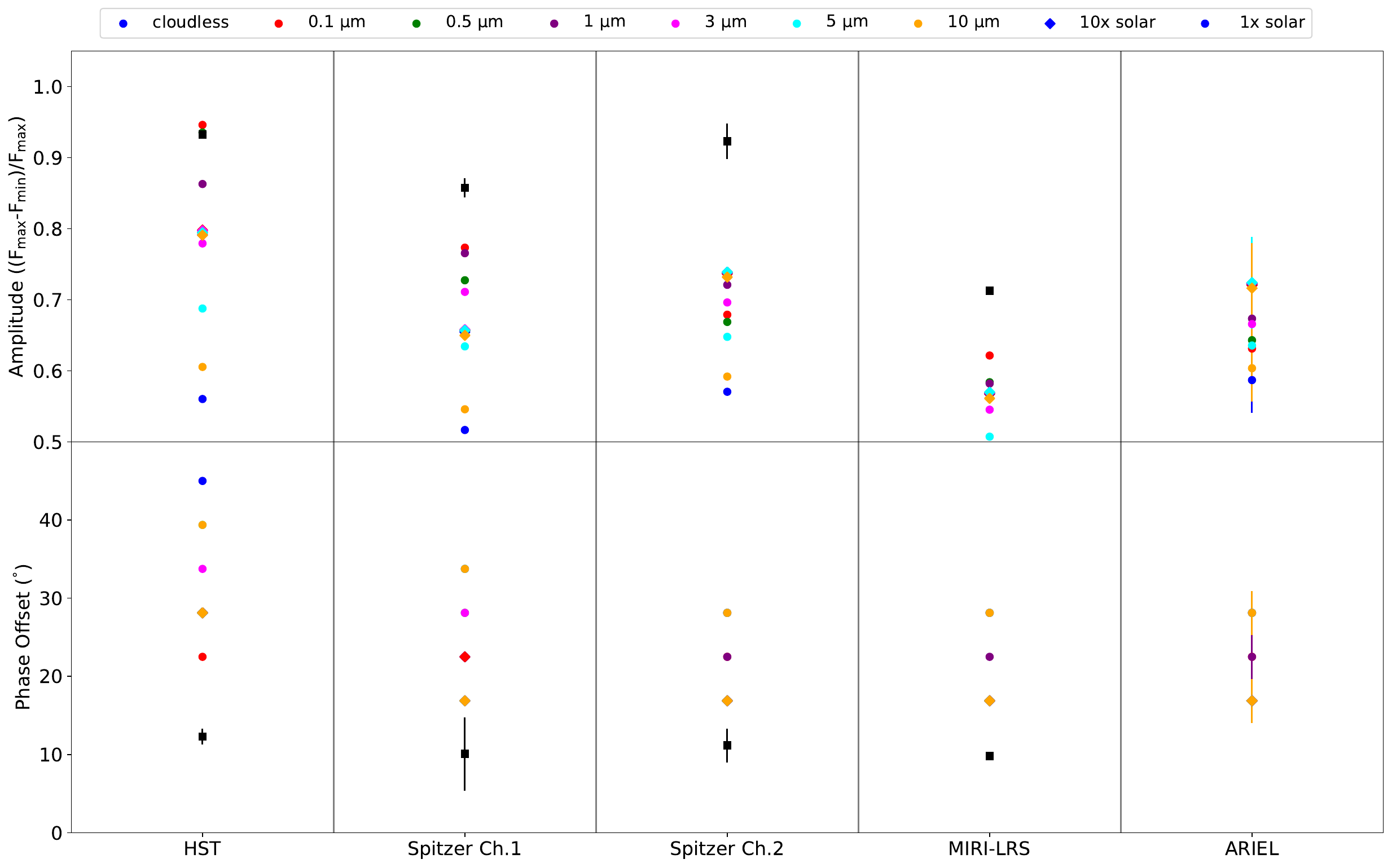}}
    \caption{Amplitude and offset of the simulated broadband phase curves in the HST, \textit{Spitzer}, and MIRI-LRS bands. Predictions for the amplitude and offset in \textit{Ariel}'s band is also added to the plots. The black square represents data points from \cite{stevenson_thermal_2014} and \cite{morello_2019}, or our MIRI-LRS reduction. Each colour is associated with a cloudy or cloud-free scenario. Circles represent the solar metallicity simulations and diamonds the 10x solar metallicity simulations.}
    \label{fig:amplitude_offset}
\end{figure*}

To discriminate between the different scenarios and better understand the data at our disposal, we computed the amplitude and the offset of the spectral phase curves for the cloudless and \silicate simulations at solar and super-solar metallicity, in each spectral band of the three instruments used in this study (see Fig. \ref{fig:amplitude_offset}). We also added the amplitude and offset as would be seen by \textit{Ariel} (see Section \ref{subsec: spectral_ariel}). For all instruments, the offset and amplitude of the phase curves are better matched by cloudy simulations rather than by cloud-free ones. In particular, the amplitude of the phase curves are best represented by solar metallicity simulations with cloud particles ranging from 0.5 to 3 $\mu$m, to the exception of \textit{Spitzer}'s Channel 2 data, which is always best represented by 10x solar metallicity simulations. The phase offsets are always overestimated in our simulations and favour the super-solar metallicity scenarios, to the exception of the HST data. Indeed, these observations favours sub-microns particles of \silicate at solar metallicity. This overall behaviour is not surprising as the effect of both clouds and higher metallicity is to reduce the phase offset \citep{parmentier_cloudy_2020,kataria_atmospheric_2015}. Moreover, it is also seen in all simulations that clouds significantly increase the phase curve amplitude, as found by \cite{parmentier_cloudy_2020}.  Interestingly, in \textit{Spitzer}'s and MIRI-LRS's bands, the modelled phase offset  for cloudy simulations is always higher at solar metallicity than in the corresponding super-solar metallicity case, but this is not the case in the HST band. Indeed, in the near infrared ($\leq 2$ $\mathrm{\mu}$m), reflected light leaks into the HST band. As shown by \cite{parmentier_transitions_2016}, phase curves in reflected light are dominated by the clouds contribution and peak after the secondary eclipse. In our solar metallicity simulations, the reflected light component induced by the clouds compete with thermal emission in HST's band, leading to a smaller phase offset than in the super-solar metallicity simulations with less cloudy atmospheres and thus, a smaller reflected light component.

To summarise, clouds have a major impact on spectral phase curves of \wasp. Their main effect is to strongly reduce the nightside thermal flux and to mask or reduce the intensity of spectral features that would otherwise be observable. Depending on the type and size of the condensates forming the clouds, this flux reduction is mitigated. On the dayside, equatorial and mid-latitudes temperatures are too warm for clouds to form, and nightside clouds will tend to warm this side of the planet by redistributing heat trapped below the cloud deck on the nightside. Using observations from HST, \textit{Spitzer,} and MIRI-LRS, we show that \silicate clouds composed of sub-micron particles cannot explain the nightside mid-infrared spectra and that MnS clouds cannot be distinguish from \silicate clouds if the particles are bigger than 1$\mu$m.

Differentiating between a solar or super-solar cloudy atmosphere is not straightforward. Our models broadly match the spectra on the dayside, nightside and terminators (phase 0.25 and 0.75). However, no model reproduces both the amplitude and the phase offset of the observations. As super-solar atmospheres have a low amount of \silicate clouds in their atmosphere, constraining the abundance of this condensate with a reflected light phase curve could help us discriminate between the solar and super-solar metallicity scenarios presented here.
Furthermore, other condensates could contribute to the observed phase curves, such as Na$_2$S, Al$_2$O$_3$, KCl, or ZnS, and explain the lacking nightside opacity needed to fit the nightside mid-infrared spectrum.

In the next section, we simulate what the \textit{Ariel} Space Telescope could see if observing a phase curve of \wasp.
\subsection{Preparing for Ariel} \label{subsec: spectral_ariel}
\begin{figure*}[h]
    \centering
    {\includegraphics[width=\textwidth]{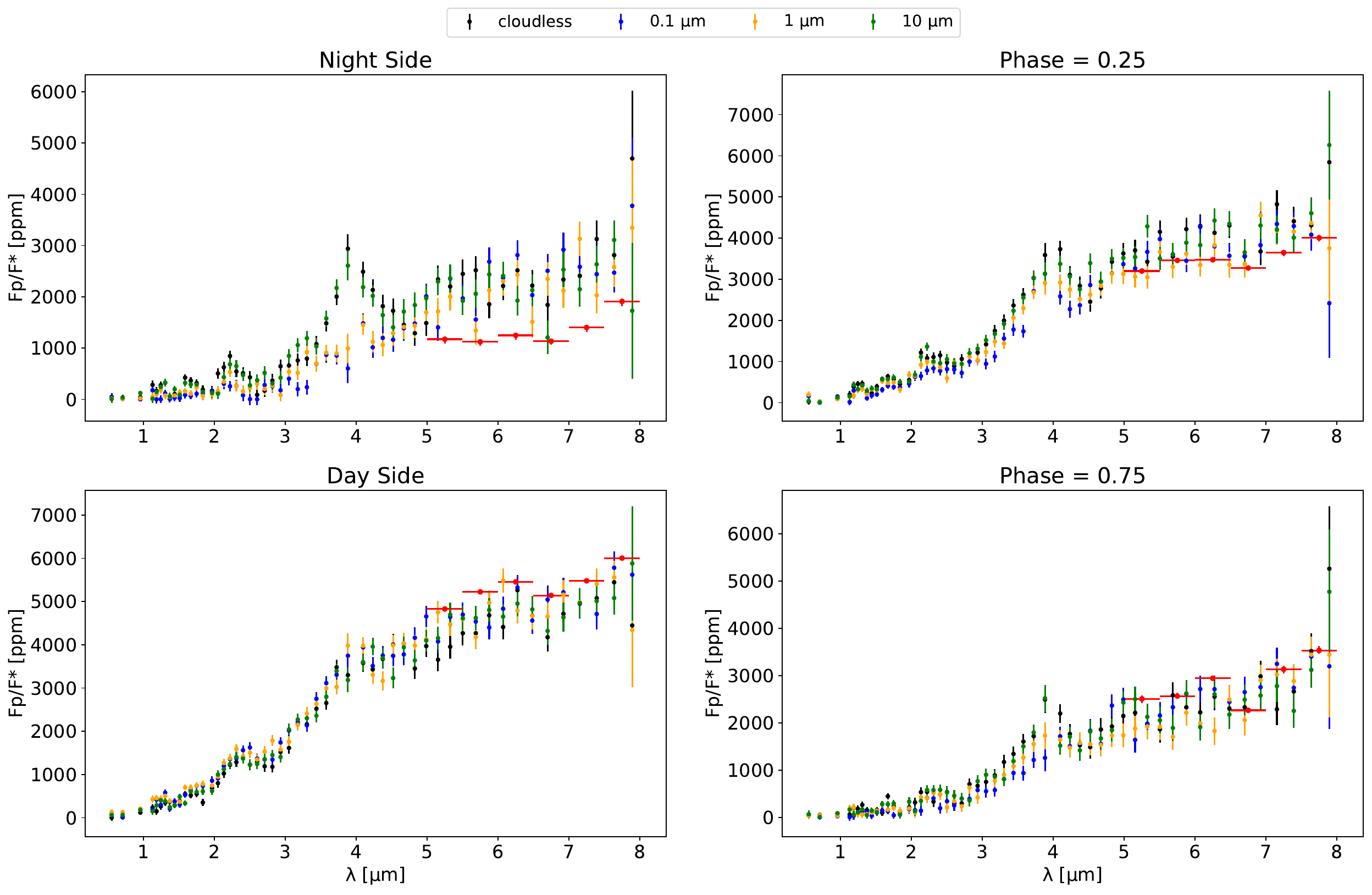}}
    \caption{Estimated phase curves as seen by \textit{Ariel} with the estimated noise. We considered that each phase has been integrated for twice the transit duration and took that into account for scaling the noise. Simulations used here are the cloudless and \silicate simulation with solar metallicity with particle sizes of $0.1$, $1,$ and $10$ $\mathrm{\mu}$m. Red points are the MIRI-LRS data for comparison. }
    \label{fig:ariel}
\end{figure*}
As described in \cite{charnay_survey_2021}, the \textit{Ariel} mission will dedicate one tier to exoplanet phase curves, with the objective of constraining atmospheric dynamics, composition, thermal structure and cloud distribution and composition. Here, we used our simulations to predict the spectral phase curves that \textit{Ariel} could yield for \wasp, considering the spectral resolution and estimated noise. We performed this computation on our cloudless and \silicate simulations and summarised our findings in Fig.\ref{fig:ariel}. The noise is estimated using \texttt{ArielRad} \citep{Mugnai_2020}. For each phase, we considered an integration time of twice the transit duration (2.32 hours). Distinguishing between a clear and a cloudy atmosphere of small particles should be straightforward using the nightside flux above 2 $\mu$m. The dayside is fairly identical between our cloudless and cloudy simulations, as stated in Section \ref{sec: spectral_solar}. 

It is clearly noticeable that \textit{Ariel} will not be able to put stronger constraints on the flux than what can be done using MIRI-LRS. However, the advantages of \textit{Ariel} will reside in its slightly higher spectral resolution in the mid-infrared, compared to what is currently achieved with MIRI-LRS (see the Methods section of \citep{ERSpaper}) and its wide spectral coverage. Thus, spectral features could enable us to measure chemical abundances, which are greatly needed to constrain the metallicity of \wasp.

Most importantly, \textit{Ariel} will allow for a simultaneous characterisation of both reflected light and thermal emission. Thus, it will constrain the overall heat budget by precise measurements of the Bond and further constrain the geometric albedo.

\section{Discussion} \label{sec: discussion}

In this study, we have introduced the adaptation of the \texttt{generic PCM} to the context of hot gaseous giant exoplanets, such as hot Jupiters. Motivated by previous studies and observations, we developed a simple cloud scheme to try to understand the dynamical and radiative impact of clouds in these atmospheres, and we focused on the special case of \wasp. We postpone a broader analysis of the impact of clouds as a function of different parameters of the system (such as surface gravity, planetary radius, and instellation) to a future study. Numerous general circulation models of hot Jupiters tackled the clouds issue, using different assumptions and modelling techniques \citep{kataria_atmospheric_2016, oreshenko_optical_2016,parmentier_transitions_2016,parmentier_cloudy_2020,roman_modeling_2017,roman_modeled_2019,roman_clouds_2021,lee_dynamic_2016,mendonca_revisiting_2018,komacek_patchy_2022,lines_simulating_2018}. The cloud models range from simple parameterisations, to detailed cloud microphysics and chemistry. We discuss our findings with regard to others and how our model fits into the diversity of existing models in the next section.
\subsection{Cloud modelling assumptions}

The different cloud modelling techniques in global circulation models can be sorted into four different categories of increasing complexity. The simplest way to take the impact of clouds into account is to add them as post-processed clouds. This approach neglects the opacities of clouds during run-time and only incorporates them when creating observables, using the thermal structure computed without the radiative feedbacks of clouds. When using {post-processed}  clouds, the model either does not include them at all in the computation or includes them as a passive tracer field, meaning that the clouds can be transported by the dynamics but cannot have an impact on the thermal structure of the atmosphere. This technique is widely used as it reduces the computation time, and still yields a pretty good match to observations \citep{parmentier_transitions_2016}. However, the thermal structure of the simulation is lacking the cloud contribution. Thus, when post-processing the simulations, the cloud location may vary significantly from the one expected with the cloud contribution, yielding an incorrect thermal flux radiated by the atmosphere \citep{roman_modeled_2019}. 

Adding complexity to the model, the cloud location can be prescribed at the start of the simulation, and different physical phenomena can be modelled to take into account cloud formation. In these models, the radiative feedbacks of clouds is taken into account, using grey or double-grey radiative transfer \citep{mendonca_revisiting_2018,roman_modeling_2017,roman_modeled_2019,roman_clouds_2021,parmentier_cloudy_2020}. In these studies, the clouds are treated as (i) a fixed, constant source of opacity such as a nightside cloud deck at a given pressure with grey radiative transfer \citep{mendonca_revisiting_2018}, (ii) as a static distribution of scatterers with varying scattering properties in double-grey radiative transfer \citep{roman_modeling_2017}, or (iii) as a temperature-dependent source of opacity, using double-grey radiative transfer \citep{roman_modeled_2019,roman_clouds_2021} or k-correlated radiative transfer \citep{parmentier_cloudy_2020}. These approaches are more physically motivated than the {post-processed} ones, even though they require a certain amount of tuning of the model for the prescription of the cloud locations. For instance, \cite{mendonca_revisiting_2018} managed to reproduce the HST and \textit{Spitzer} phase curve of \wasp~with their grey opacity nightside clouds. However, their model does not inform on how clouds interact with the atmospheric flow and does not adequately take into account the heating rates associated with the thermal effect of clouds. As found by \cite{roman_modeled_2019}, radiative feedbacks of clouds strongly alter the temperature field, and the alteration depends on the aerosol scattering properties and abundances. However, their treatment of the aerosol scattering properties is also incomplete, as they only use two values (one in visible and one in infrared light) to represent the single scattering albedo, the asymmetry parameter and the extinction efficiency of their aerosols. Thus, their modelling is not sensitive to the strong absorption bands of condensates, such as the 10-$\mu$m silicate feature, detected for instance in the atmosphere of the planetary-mass companion VHS 1256-1257 b \citep{Miles_jwst_2023} . As JWST is enabling multi-phase emission spectra at higher spectral resolution and with a broader spectral coverage, accurate spectral modelling of clouds is needed to predict and understand the forthcoming observations. 

Clouds are not static in 3D atmospheres, as can be daily experienced on Earth. Although the models described above are a good compromise between accurate cloud physics and computational time, they lack the interaction between clouds and the atmospheric flow. This interaction is taken into account in models treating clouds as active tracers. Active refers to the consideration of the optical properties of clouds and thus their radiative feedbacks while tracers are quantities that are advected with the flow, at each dynamical time-step of the simulation. Hence, the third category of models describes clouds as active tracers. These models provide information on the location at which cloud form and how they interact with the atmospheric flow. This is commonly used in Solar System GCMs, for instance on Mars, Venus and Titan \citep{forget_2006,lefevre_2018,schneider_2012}. \cite{komacek_patchy_2022} applied such a model to the case of ultra-hot Jupiters and found that clouds are patchy in these very warm atmospheres. \cite{tan_atmospheric_2021-1} used a similar parameterisation for the radiative effects of clouds in brown dwarf atmospheres. However, their models use the double-grey approximation and thus lack the physical insights needed to yield quantitative predictions. Indeed, \cite{Lee_2021} showed that grey models can only qualitatively reproduce the results of correlated-k schemes. 

Despite the growing complexity and physics put in the previous models, cloud physics encompasses a range of phenomena that can strongly impact the formation and evaporation of clouds. For instance, none of the above models takes into account nucleation, the chemical interplay happening at the surface of the grain, condensation growth, desorption energy of the condensates etc... All these can be grouped into the term microphysics, which more accurately describes cloud formation, growth, sedimentation and sublimation. Coupling microphysical cloud models to GCMs has been done for particular studies \citep{lee_dynamic_2016,lines_simulating_2018,christie_impact_2021}. However, the additional computation cost of this coupling makes it unpractical for global studies, even though necessary to more thoroughly understand cloud behaviour in these hot atmospheres. One recent study from \cite{lee_modelling_2023} coupled a GCM with a microphysical cloud model tailored for fast computation and found a good match to observations of HAT-P-1b. \\ 

Hence, our model lies as a compromise between accurate modelling of the cloud properties and too simplistic parameterisations. The aim behind our work is to understand the broad behaviour of clouds in the atmosphere of hot giant gaseous exoplanets and their impact on the thermal, dynamical and observable properties. As stated in Section \ref{sec: spectral_solar}, our modelling is not sufficient to explain the observed flux at phase 0.75. Microphysical modelling of clouds might help explain how clouds behave during their advection from the nightside to the dayside. Moreover, we only included one type of condensates in our simulations. As pointed out by \cite{helling_cloud_2021}, the upper atmosphere of \wasp~is dominated by silicates, with the nightside also containing metal oxides and clouds composed of carbon-bearing materials. The addition of these clouds to the simulation, along with \silicate clouds might help us better understand the observed phase curves. Also, chemical interactions between the different clouds that are not taken into account in our model could have an impact on the cloud distribution and thus, on the thermal emission of the atmosphere.
\subsection{Clouds and metallicity}

Clouds and metallicity induce similar behaviours on phase curves of hot Jupiters. They both tend to reduce the phase offset and amplify the amplitude of the phase curves. In the special case of \wasp, clouds are needed to reproduce the nightside flux. However, the dayside spectrum is compliant with super-solar metallicity in all bands. As we show, super-solar metallicity atmospheres contain less dense \silicate clouds than their solar metallicity counterpart. This is due to the metallicity-dependence of the condensation curve \citep{visscher_atmospheric_2010}. Thus, we propose two ideas to break this observed degeneracy in \wasp's atmosphere. The first one would be to measure a reflected light phase curve of the planet. Following \cite{oreshenko_optical_2016}, such phase curves encode all the needed information about cloud distribution in cloudy atmospheres. Assessing the influence of \silicate clouds would thus indirectly inform on the atmospheric metallicity. We argue that observations with the CHEOPS facility should yield interesting constraints on the longitudinal distribution of clouds. The second idea would be to perform a precise measurement of the water abundance on the nightside. To date, such measurements are available with HST data only and have been difficult to carry out due to the dim nightside \citep{stevenson_thermal_2014}. The MIRI-LRS phase curve is a step ahead towards more accurate measurements but the low spectral resolution (R $\sim$100) makes accurate derivation of molecular abundances difficult. Thus, a mid-resolution phase curve is needed to assess the water amount on the nightside. Using the combination G395H/F290LP (3-5 $\mu$m, R $\sim$1000) disperser and filter of {JWST-NIRSpec} could provide this measurement \citep{alderson_early_2023},  as could also be done using the NIRISS-SOSS instrument \citep{feinstein_early_2023}. However, as the NIRISS-SOSS wavelength coverage ranges from 0.6 to 2.8 $\mu$m, reflected light could complicate the inference of the water budget on the cloudy nightside of \wasp. Moreover, the higher spectral resolution of {NIRSpec-G395H} should help accurately constrain water.

\subsection{Other limitations}

In this work, we did not investigate the impact of chemistry on the atmospheric circulation and on the observables. It is expected that chemical abundances vary with longitude, as hot Jupiters exhibit a strong temperature contrast. In particular, thermochemical equilibrium predicts CH$_4$ to be the dominant carbon-bearing constituent in the nightside of \wasp, but not on the warmer dayside where CO should be dominant. Using 1D and pseudo-2D kinetic models, \cite{venot_global_2020} showed that vertical and horizontal quenching will partially damp these heterogeneities. Moreover, HCN could be photochemically produced and be more abundant than methane at all longitudes at the infrared photosphere. In our model, as all longitudes and latitudes are initialised with the same vertically quenched chemical profiles, we are in the extreme case of uniform horizontal quenching. This is reasonably not the case and relaxing this assumption should be done, in line with coherent chemical network predictions. We also neglected the potential photochemical production of molecules such as SO$_2$ in our model, which we now know to be happening in hot Jupiter atmospheres \citep{tsai_photochem_2022}. \\ 

Our model simulates an extended atmosphere as we set the inner boundary at 800 bars. This choice is motivated by a few studies that highlight the connection between the upper and deep atmosphere, shaping the overall climatology of hot Jupiters \citep{carone_equatorial_2020,schneider_exploring_2022}. However, these deep layers have a long  numerical convergence time, as the radiative timescale becomes very long \citep{wang_extremely_2020,showman_atmospheric_2020}. An other difficulty arising with the simulation of deep atmospheres is the accurate representation of the thermal properties of these layers. At these temperature and pressure, the ideal gas law that is commonly used in GCM is not valid anymore and the assumption of constant thermal capacity is also greatly challenged. Finally, the boundary condition is set to a constant interior temperature based on what is commonly used in the literature for Jupiter and Saturn. More accurate constraints on the internal flux are thus needed to accurately model the deep layers and their interaction with the upper atmosphere dynamics. \\

In this study, we neglected the potential radiative feedback of photochemical hazes that could alter both the thermal and dynamical structure of hot Jupiters \citep{steinrueck_2023}. Indeed, haze radiative feedbacks can increase the day-night temperature contrast and potentially mask cloud effect. If photochemical hazes are produced and/or advected to the western terminator of \wasp, they could potentially be responsible for the discrepancy between the modelled and observed spectrum for $\lambda \geq 4$  $\mu$m. \\

Finally, the \texttt{generic PCM} solves the primitive hydrostatic equations of meteorology. Despite this set of equations being commonly used in the exoplanet atmosphere modelling community, \cite{mayne_unified_2014} used a fully compressible, non-hydrostatic, deep atmosphere global circulation model to simulation the atmosphere of hot Jupiters. Their main findings are that their more complete set of equations allow for an exchange between the vertical and horizontal angular momentum, and a stronger exchange between the deeper and upper atmosphere, leading to the degradation of the prograde super-rotating jet. However, the general atmospheric structure they derive is quite similar to that obtained with the primitive hydrostatic simulation. This finding is consistent with a study from \cite{noti_examining_2023} that uses the non-hydrostatic deep equations and compare them to simulations using the quasi-hydrostatic deep equations. For the case of \wasp, they find that the choice of equations set does not have a strong impact on the dynamical and thermal state.

\section{Conclusions}\label{sec: conclusion}

We modelled the hot Jupiter \wasp~using the \texttt{generic PCM} with and without including clouds. Our cloud model is based on the assumption that clouds form or do not form depending on the local thermodynamical conditions. We also added the radiative feedbacks of clouds on the thermal structure. Once formed, clouds are horizontally advected and vertically sedimented as tracers of fixed radius. We ran multiple scenarios with different assumptions on the type of clouds, the size of the cloud's particles, and different atmospheric metallicities. We investigated the mechanisms responsible for the transport of heat and axial angular momentum. We post-processed our simulations to produce spectral phase curves and compared them to datasets obtained by HST, \textit{Spitzer,} and the {JWST's MIRI-LRS} instrument. We computed estimations of the spectral phase curves as will be observed by \textit{Ariel}. Our conclusions are the following: 
\begin{enumerate}
    \item In the cloudless case, our model reproduces the broad dynamical findings of previous studies, such as an equatorial super-rotation prograde jet, an eastward shift of the hotspot with regard to the substellar point, and a strong temperature contrast between the irradiated dayside and the dark nightside.\\
    \item In the cloudless case, the horizontal and vertical transport of axial angular momentum is dominated by the stationary waves' contribution, and the transport of heat is dominated by the mean circulation and the stationary waves' contribution (for horizontal transport). \\ 
    \item Clouds have multiple effects on the thermal and dynamical structure. One common feature of clouds is to reduce the overall jet speed. Depending on the thermal and optical properties of the condensates and the size of the cloud particles, the cloud distribution varies, leading to different atmospheric structures. Both the planetary and visible geometric albedo increase with clouds. \\
    \item Clouds have a net positive radiative feedback, warming the atmosphere. \\
    \item Super-solar metallicity leads to stronger longitudinal thermal contrasts and to fewer clouds in the upper atmosphere. \\ 
    \item Clouds are necessary to explain the spectra of the dark nightside of \wasp. The mid-infrared MIRI-LRS data allowed us to rule out sub-micron \silicate cloud particles as the main opacity source in the atmosphere of hot Jupiters. With current observations, our model is not able to discriminate between cloudy solar and cloudy super-solar atmospheres. \\ 
    \item Further phase curves at different wavelengths are needed to constrain the atmospheric metallicity, the size of the cloud particles, and to further constrain the cloud physics in warm giant planet's atmospheres. In particular, reflected light and thermal phase curves are needed to constrain the location of clouds and to constrain the overall energy budget of the planet.
\end{enumerate}
Though our cloud modelling ignores the microphysical processes inherently shaping cloud formation, it allows us to derive constraints on the atmospheric state induced by the radiative feedbacks of clouds, and to constrain the cloud settling locations without further assumptions. In the era of JWST and future space-based observatories such as \textit{Ariel} and {PLATO}, the observational constraints of increasing quality must be met by atmospheric models that take the full 3D picture and feedbacks between its constituents  into account.   

\begin{acknowledgements}
This work was granted access to the HPC resources of MesoPSL financed by the Region Ile de France and the project Equip@Meso (reference
ANR-10-EQPX-29-01) of the programme Investissements d’Avenir supervised by the Agence Nationale pour la Recherche. 
This work was supported by CNES, focused on AIRS on Ariel.
This work is partly based on observations made with the NASA/ESA/CSA JWST. The data were obtained from the Mikulski Archive for Space Telescopes at the Space Telescope Science Institute, which is operated by the Association of Universities for Research in Astronomy, Inc., under NASA contract NAS 5-03127 for JWST. These observations are associated with program JWST-ERS-01366. Support for program JWST-ERS-01366 was provided by NASA through a grant from the Space Telescope Science Institute. We thank C. Wilkinson for constructive discussions about the deep atmosphere and the equations of state, and for moral support during this work. We thank G. Morello for kindly providing his Spitzer data. E. D acknowledges support from the innovation and research Horizon 2020 program in the context of the  Marie Sklodowska-Curie subvention 945298. We thank the anonymous referee for constructive comments that strongly enhanced the quality of this manuscript.
\end{acknowledgements}


\bibliography{biblio}
\bibliographystyle{aa}

\begin{appendix}
\section{Vertical chemical  and cloud profiles}
\begin{figure}[h]
    \centering
    {\includegraphics[width=0.5\textwidth]{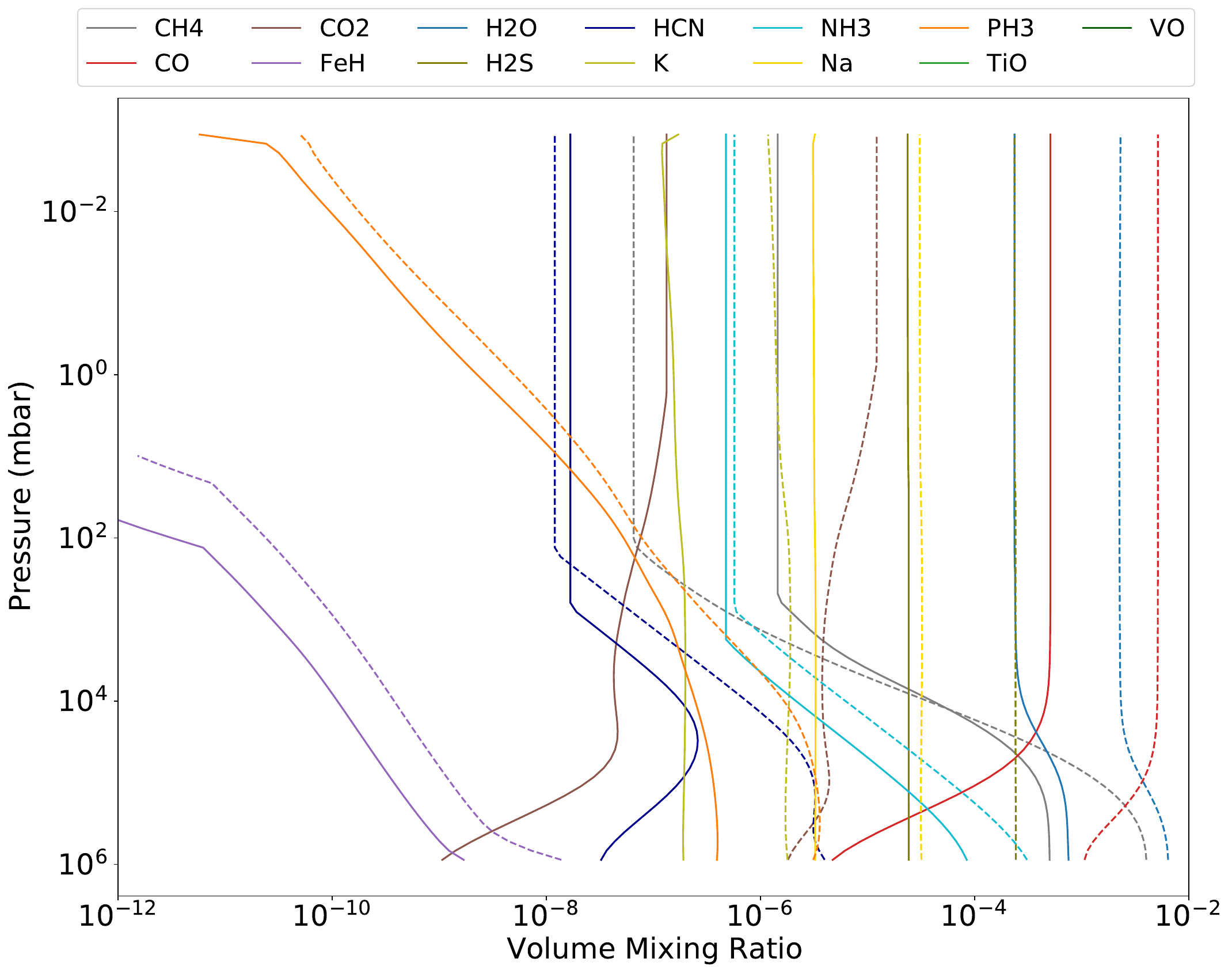}}
    \caption{Vertical abundances profiles of radiatively active species computed with \texttt{Exo-REM}. Solid lines assume solar metallicity, and dashed lines have 10x solar metallicity. Disequilibrium chemistry is assumed in both cases. The upper atmosphere is dominated by CO, with a strong abundance of H$_2$O. TiO and VO are negligible at these temperatures even though they are included in the computation.}
    \label{fig:exorem_chem}
\end{figure}
\begin{figure}[h]
    \centering
    {\includegraphics[width=0.5\textwidth]{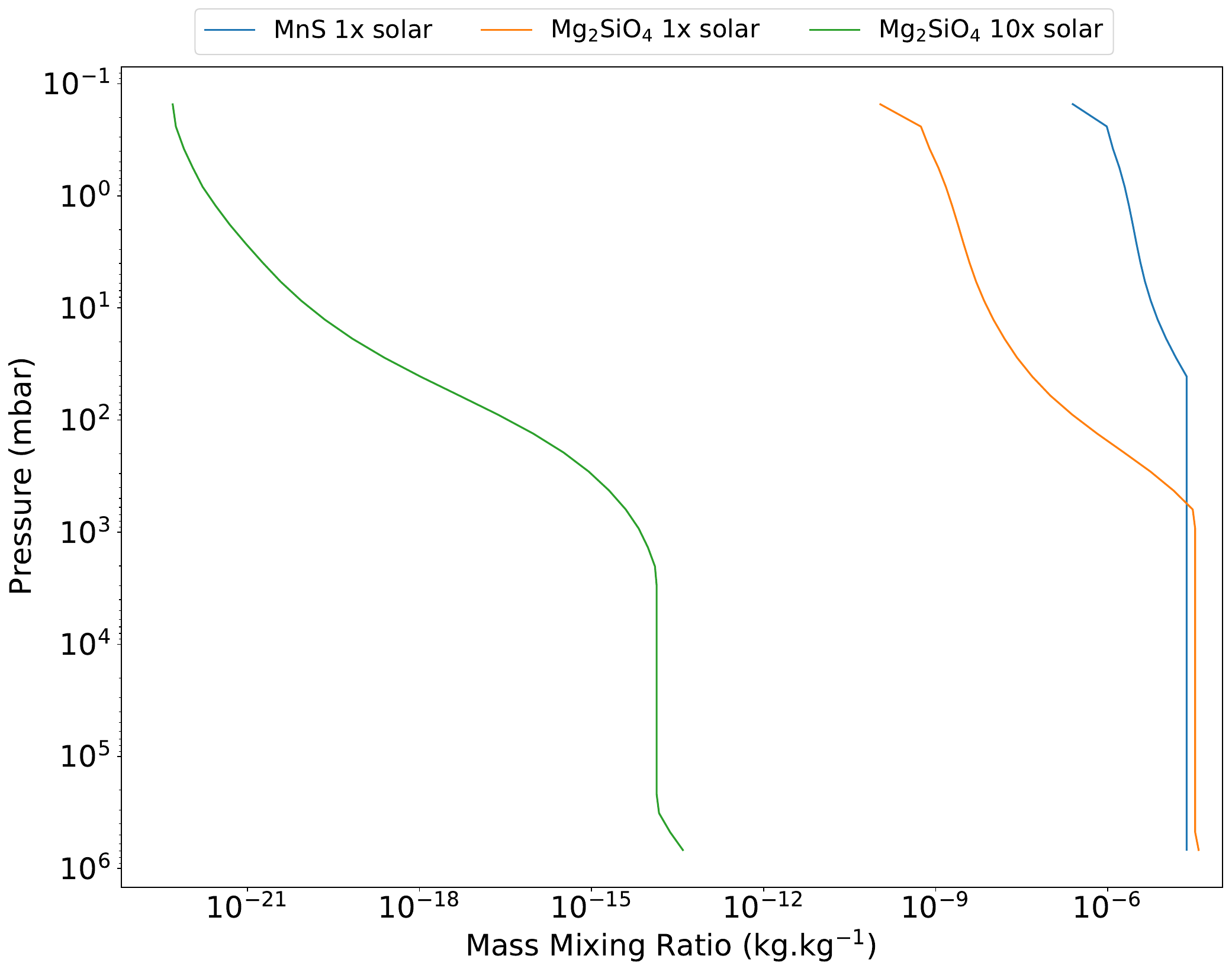}}
    \caption{Initial condensable vapour profile for MnS and \silicate clouds. These profiles are computed using analytical profiles from \cite{visscher_atmospheric_2010} and \cite{morley_neglected_2012}, with a planetary averaged temperature profile from the cloudless simulation.}
    \label{fig:profil_clouds}
\end{figure}

To derive the radiative data used in our 3D simulations, we computed vertical abundances of typical absorbers using the 1D radiative convective code \texttt{Exo-REM} \citep{baudino_interpreting_2015,charnay_self-consistent_2018,blain_1d_2021} with disequilibrium chemistry. For each planet, we assumed that H$_{\rm 2}$O, CO, CH$_{\rm 4}$, CO$_{\rm 2}$, FeH, HCN, H$_{\rm 2}$S, TiO, VO, Na, K, PH$_{\rm 3}$, and NH$_{\rm 3}$ will be radiatively active. The vertical profiles of the absorbers are presented in Fig.\ref{fig:exorem_chem} for a solar and a 10x solar metallicity. In both cases, CO and H$_{\rm 2}$O dominates the atmosphere below 10 bars. TiO and VO are negligible for \wasp. \\ 
Figure \ref{fig:profil_clouds} displays the initial condensable vapour mass mixing ratio used to initial the 3D spherical shell of the \texttt{generic PCM}. These profiles are based on analytical derivations from \citep{visscher_atmospheric_2010} for \silicate and \citep{morley_neglected_2012} for MnS.

\newpage
\section{Axial angular momentum transport induced by 1 $\mathbf{\mu}$m clouds}
Here are plotted the horizontal and vertical transport of angular momentum contribution as wall as the net transport, as described in section \ref{sec: dynamic_clouds}.
\begin{figure*}[h]
    \centering
    {\includegraphics[width=\textwidth]{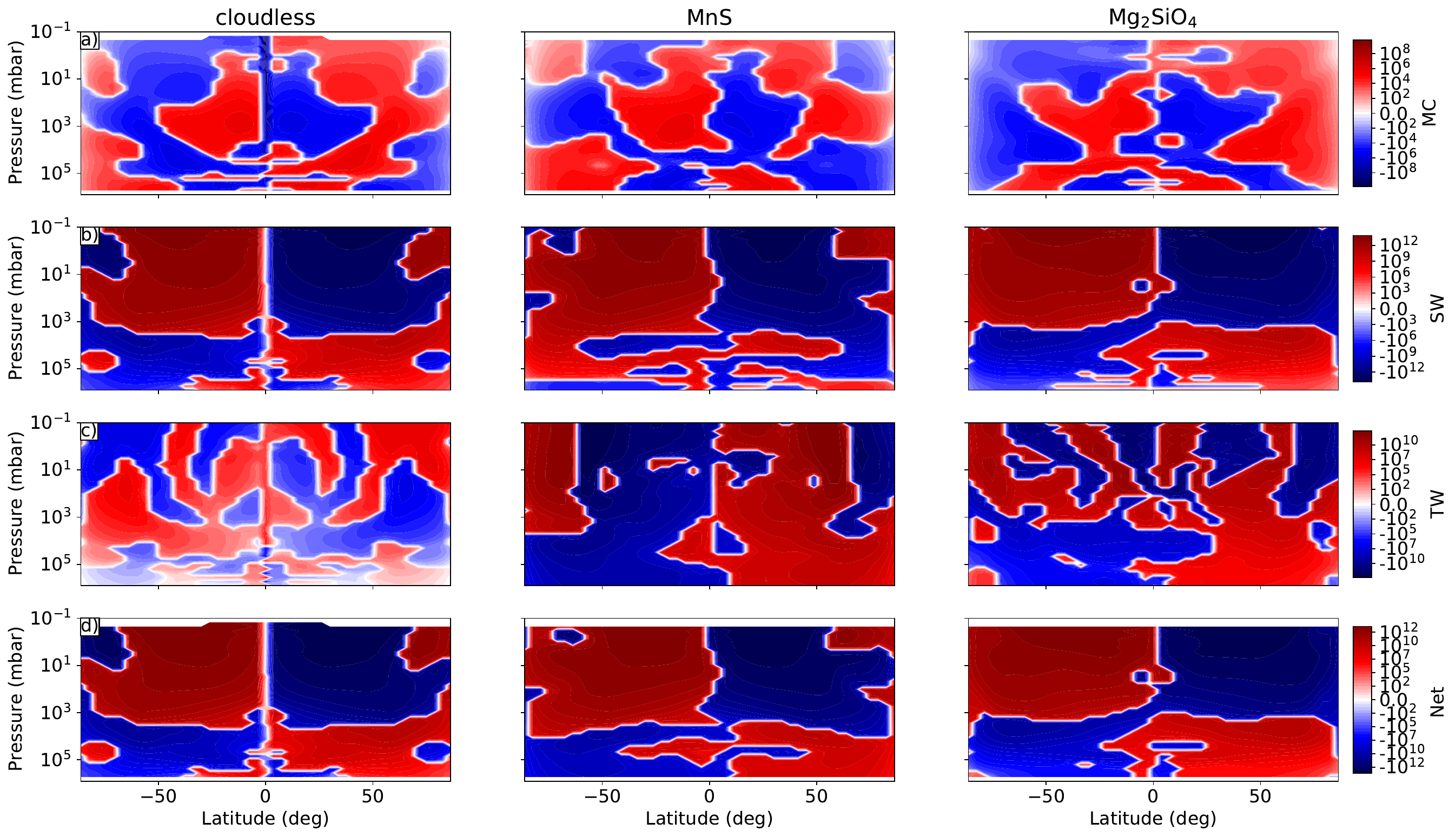}}
    \caption{Axial angular momentum horizontal transport for the cloudless simulation of \wasp~(\emph{left}) and 1$\rm \mu$m clouds of MnS (\emph{middle}) and \silicate (\emph{right}). Each row displays a different contribution: mean circulation contribution (a) stationary wave contribution (b), transient wave contribution (c), and net transport (d). Each row is plotted with a different range to show the amplitude of each contribution to the net transport.}
    \label{fig:ang_v_1um}
\end{figure*}
\begin{figure*}[h]
    \centering
    {\includegraphics[width=\textwidth]{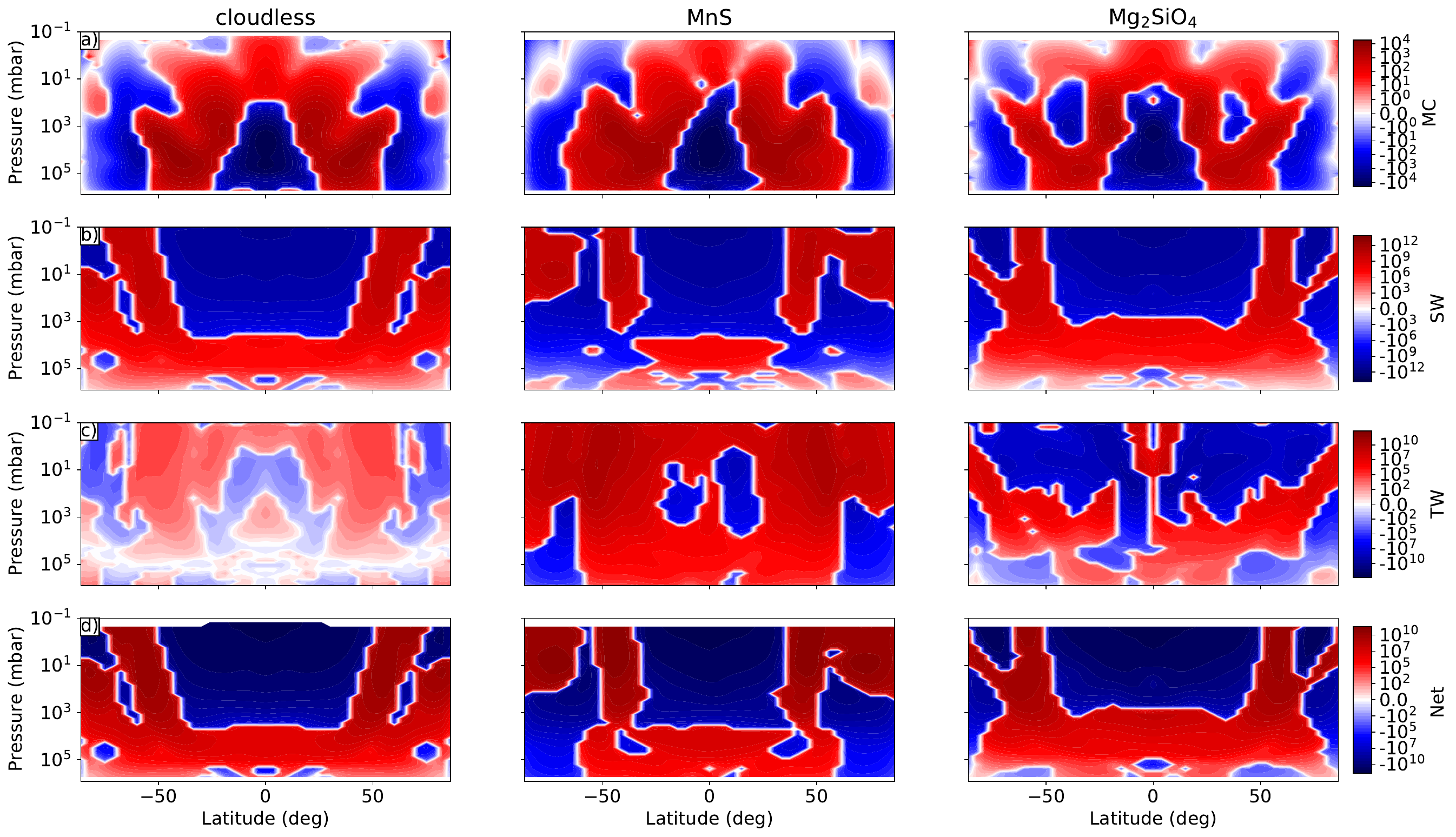}}
    \caption{Axial angular momentum vertical transport for the cloudless simulation of \wasp~(\emph{left}) and 1$\rm \mu$m clouds of MnS (\emph{middle}) and \silicate (\emph{right}). Each row displays a different contribution:  mean circulation contribution (a), stationary wave contribution (b), transient wave contribution (c), and net transport (d). Each row is plotted with a different range to show the amplitude of each contribution to the net transport.}
    \label{fig:ang_w_1um}
\end{figure*}
\newpage
\section{Solar and super-solar simulation: Dynamical and thermal structure of the atmosphere}
\begin{figure*}[h]
    \centering
    {\includegraphics[width=\textwidth]{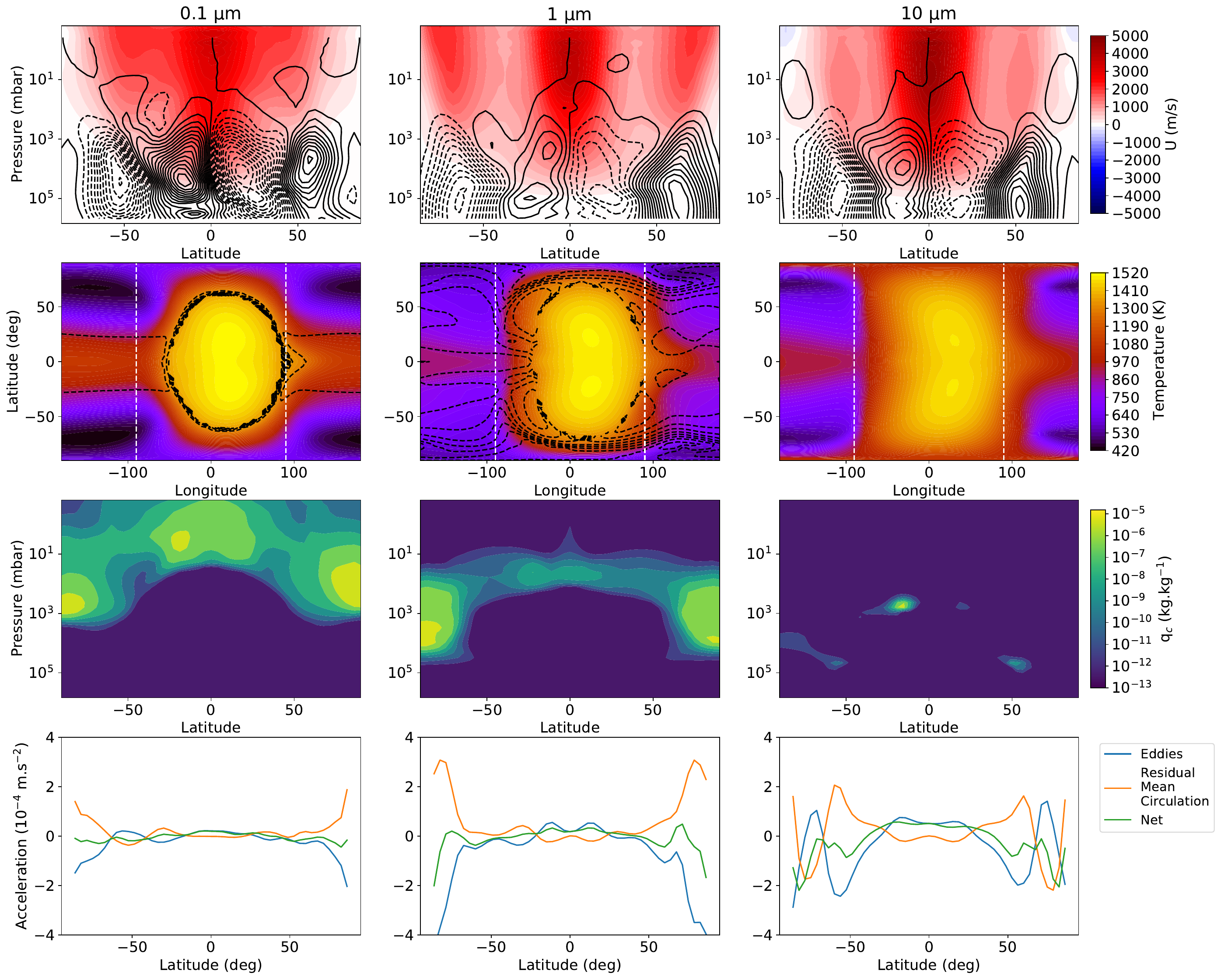}}
    \caption{Comparison of wind patterns, temperature maps at 10 mbar, cloud latitudinal distributions, and jet acceleration for simulations with \silicate clouds and particle sizes of $0.1$, $1$ and $10$ $\mathrm{\mu}$m. Top row: Zonal mean and time-averaged zonal wind. Black contours are the mass stream function, with solid lines indicating a clockwise circulation and dashed lines an anti-clockwise circulation. Second row: Time-averaged temperature map at the 10 mbar isobaric level. Black contours denote the location of clouds. White vertical dashed lines are the terminators. Third row: Latitude-pressure map of the cloud distribution, averaged in time and longitude. Bottom panel: Vertically integrated jet zonal wind acceleration in zonal and time mean between 5 and 20 mbar. We show the decomposition into residual mean circulation, the eddy contributions, and the net acceleration. From left to right, cloud particles increase in size. }
    \label{fig:silicate_solar}
\end{figure*}

\begin{figure*}[h]
    \centering
    {\includegraphics[width=\textwidth]{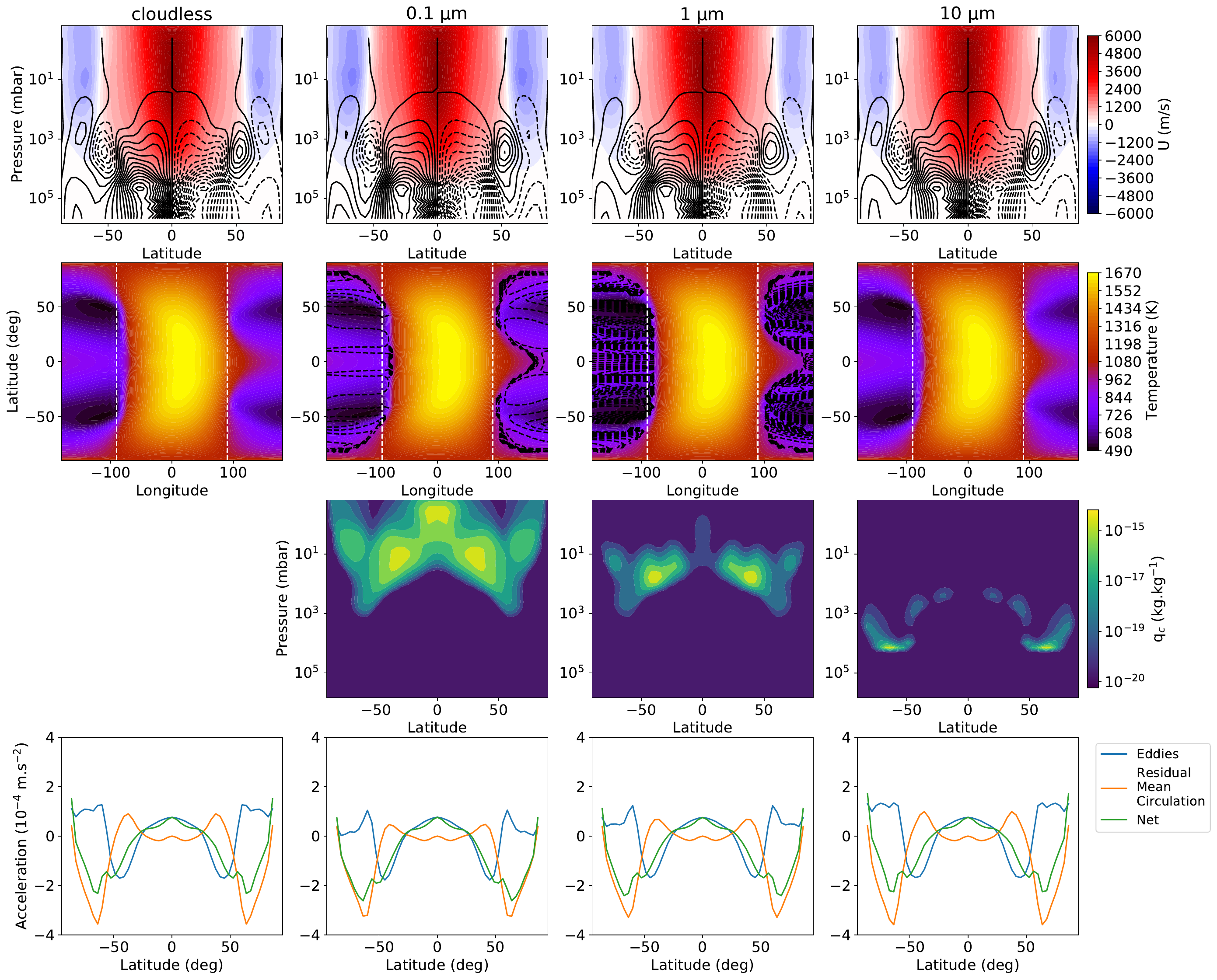}}
    \caption{Comparison of wind patterns, temperature maps at 10 mbar, cloud latitudinal distributions, and jet acceleration for simulations with \silicate clouds at super-solar metallicity and particle sizes of $0.1$, $1,$ and $10$ $\mathrm{\mu}$m. The first column is the cloudless simulation, and the following are the cloudy ones. Top row: Zonal mean and time-averaged zonal wind. Black contours are the mass stream function, with solid lines indicating a clockwise circulation and dashed lines an anti-clockwise circulation. Second row: Time-averaged temperature map at the 10 mbar isobaric level. Black contours denote the location of clouds. White vertical dashed lines are the terminators. Third row: Latitude-pressure map of the cloud distribution, averaged in time and longitude. Bottom panel: Vertically integrated jet zonal wind acceleration in zonal and time mean between 5 and 20 mbar. We show the decomposition into residual mean circulation, the eddy contributions, and the net acceleration.}
    \label{fig:supersolar}
\end{figure*}

Here are plotted the broad dynamical and thermal pattern of the atmosphere. Figure \ref{fig:silicate_solar} is for the solar metallicity simulations with \silicate clouds. Figure \ref{fig:supersolar} is the same for super-solar metallicity, with the first column being the cloudless simulation and the three other panels cloudy simulations with different particle sizes. In all plots where clouds are included, clouds are composed of \silicate.

\begin{figure}[h]
    \centering
    {\includegraphics[width=0.47\textwidth]{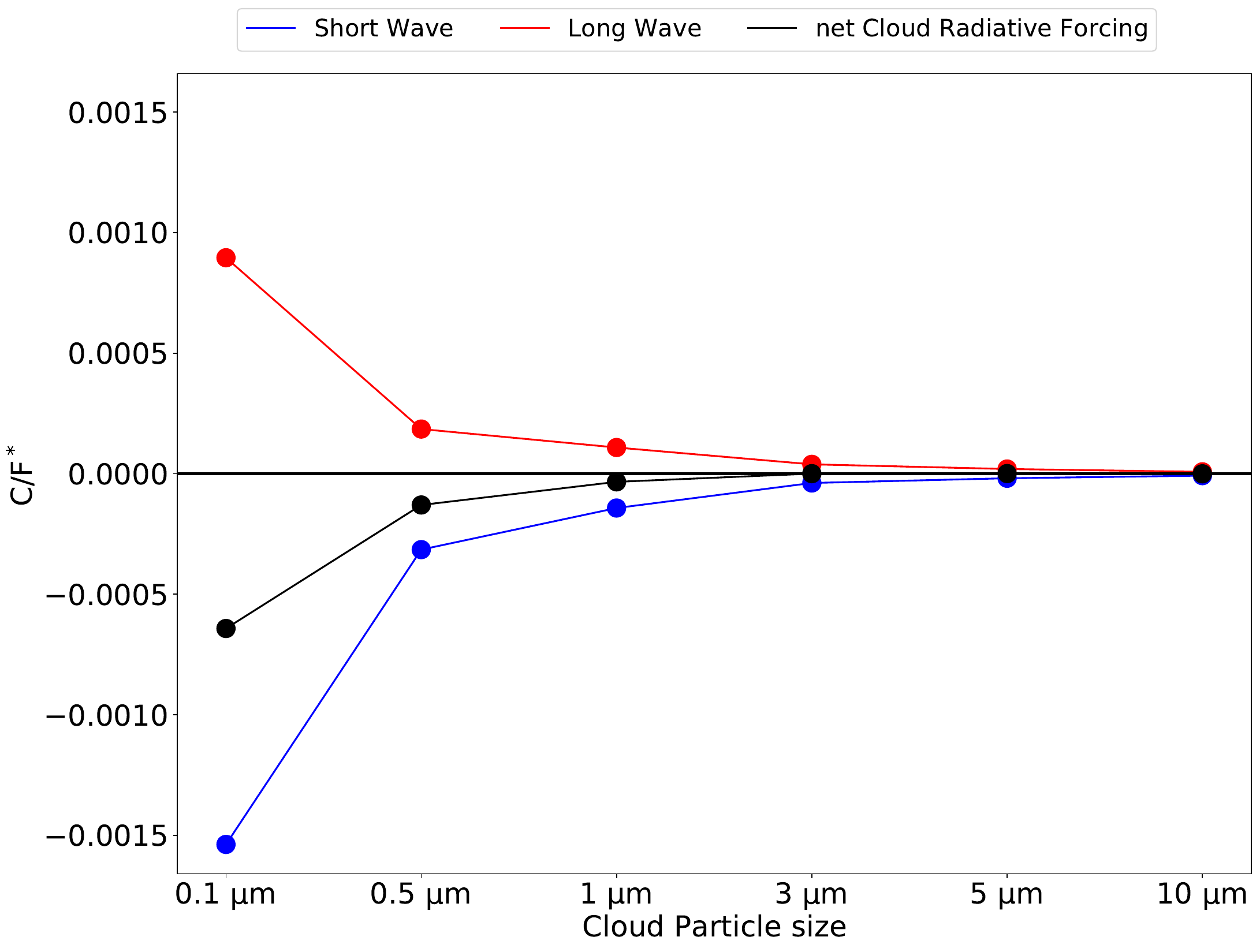}}
    \caption{Cloud radiative forcing (normalised by the stellar irradiance) for super-solar metallicity atmospheres and \silicate clouds. See Section \ref{sec:albedo_CRF} for a definition.}
    \label{fig:CR_10x}
\end{figure}

We also plot the cloud radiative forcing for the super-solar metallicity simulations with \silicate clouds, in Fig. \ref{fig:CR_10x}

\end{appendix}
\end{document}